\documentclass[useAMS,usenatbib]{mn2e}

\usepackage{aas_macros}
\usepackage{graphicx}
\usepackage{float}

\graphicspath{{figures/}}

\newcommand{\Ks}{K_s}
\newcommand{\ab}{}
\newcommand{\ud}{\mathrm{d}}
\newcommand{\simgt}{\lower.5ex\hbox{$\; \buildrel > \over \sim \;$}}
\newcommand{\simlt}{\lower.5ex\hbox{$\; \buildrel < \over \sim \;$}}
\newcommand{\rs}[1]{{#1}}

\newcommand{\Mstar}{\hbox{{${M}_{\rm \star}$}}}
\newcommand{\Mstart}[1]{\hbox{{${M}_{\rm \star}^{#1}$}}}
\newcommand{\Mstaro}{\hbox{{${M}_{\rm \star, 0}$}}}

\newcommand{\Mh}{\hbox{{${M}_{\rm h}$}}}
\newcommand{\Msun}{\hbox{${M}_{\odot}$}}

\title[HOD in CFHTLenS/VIPERS]{The galaxy-halo
  connection from a joint lensing, clustering and abundance analysis
  in the CFHTLenS/VIPERS field}

\author[J. Coupon et al.]
{J.~Coupon,$^1$\thanks{jean.coupon@unige.ch}
S.~Arnouts,$^2$ 
L.~van~Waerbeke,$^3$ 
T.~Moutard,$^2$ 
O.~Ilbert,$^2$\newauthor 
E.~van~Uitert,$^{4}$ 
T.~Erben,$^4$ 
B.~Garilli,$^5$ 
L.~Guzzo,$^{6,7}$ 
C.~Heymans,$^{8}$ 
H.~Hildebrandt,$^4$\newauthor 
H.~Hoekstra,$^9$ 
M.~Kilbinger,$^{10}$ 
T.~Kitching,$^{11}$ 
Y.~Mellier,$^{12,10}$ 
L.~Miller,$^{13}$\newauthor  
M.~Scodeggio,$^5$
C.~Bonnett,$^{13}$ 
E.~Branchini,$^{15,16,17}$ 
I.~Davidzon,$^{18,19}$ 
G.~De Lucia,$^{20}$ \newauthor
A.~Fritz,$^5$ 
L.~Fu,$^{21}$ 
P.~Hudelot,$^{12}$ 
M.~J.~Hudson,$^{22,23}$
K.~Kuijken,$^9$
A.~Leauthaud,$^{24}$\newauthor 
O.~Le~F\`{e}vre,$^2$ 
H.~J.~McCracken,$^{12}$ 
L.~Moscardini,$^{18,19,23}$ 
B.~T.~P.~Rowe,$^{25}$\newauthor  
T.~Schrabback,$^{4}$ 
E.~Semboloni,$^4$ 
and M.~Velander$^{13}$\\
Affiliations can be found after the references
}

\begin{document}

\date{}

\pagerange{\pageref{firstpage}--\pageref{lastpage}} \pubyear{2014}

\maketitle

\label{firstpage}

\begin{abstract}
  We present new constraints on the relationship between galaxies and
  their host dark matter halos, measured from the location of the peak
  of the stellar-to-halo mass ratio (SHMR), up to the most massive
  galaxy clusters at redshift $z\sim0.8$ and over a volume of nearly
  0.1~Gpc$^3$. We use a unique combination of deep observations in the
  CFHTLenS/VIPERS field from the near-UV to the near-IR, supplemented
  by $\sim60\,000$ secure spectroscopic redshifts, analysing galaxy
  clustering, galaxy-galaxy lensing and the stellar mass function.  We
  interpret our measurements within the halo occupation distribution
  (HOD) framework, separating the contributions from central and
  satellite galaxies.  We find that the SHMR for the central galaxies
  peaks at $M_{\rm h, peak} = 1.9^{+0.2}_{-0.1}\times10^{12}\Msun$
  with an amplitude of $0.025$, which decreases to $\sim0.001$ for
  massive halos ($\Mh > 10^{14} \Msun$).  Compared to central galaxies
  only, the total SHMR (including satellites) is boosted by a factor
  10 in the high-mass regime (cluster-size halos), a result consistent
  with cluster analyses from the literature based on fully independent
  methods.  After properly accounting for differences in modelling, we
  have compared our results with a large number of results from the
  literature up to $z=1$: we find good general agreement,
  independently of the method used, within the typical stellar-mass
  systematic errors at low to intermediate mass ($\Mstar < 10^{11}
  \Msun$) and the statistical errors above.  We have also compared our
  SHMR results to semi-analytic simulations and found that the SHMR is
  tilted compared to our measurements in such a way that they over-
  (under-) predict star formation efficiency in central (satellite)
  galaxies.
\end{abstract}

\begin{keywords}
  cosmology: observations -- dark matter -- galaxies: clusters:
  general -- gravitational lensing: weak.
\end{keywords}

\section{Introduction}
\label{sec:intro}

The last few years have seen an increasing interest in statistical
methods linking observed galaxy properties to their dark matter halos,
owing to the availability of numerous large scale multi-wavelength
surveys. Those techniques are based on the assumption that the spatial
distribution of dark matter is predictable and one is able to match
its statistical properties with those of the galaxies.  The halo model
\citep[see][]{Cooray:2002aa} is a quantitative representation of the
distribution of dark matter, characterised by three main ingredients:
the halo mass function describing the number density of halos per
mass, the halo bias tracing the clustering amplitude, and the halo
density profile.

Galaxies are born and evolve in individual halos where the baryonic
gas condensates, cools and forms stars. Galaxies are gravitationally
bound to dark matter and share a common fate with their host, e.g.
during mergers. Although we understand qualitatively individual
physical processes likely to be involved in galaxy evolution, a number
of key answers are missing.

Observations show that a fraction of galaxies experienced star
formation quenching and have become passive, shaping the galaxy
population into a bimodal blue/red distribution
\citep{Faber:2007aa,Ilbert:2013dq}.  The number of these passive
galaxies is higher today than in the past and increases with
increasing halo mass. Might feedback processes in massive halos be
responsible for this, or is there a universal critical stellar mass
above which star formation ceases, independently of the halo mass?
Studying the connection between galaxies and their host halos is
crucial to answer these questions.

Another enigmatic question is the low stellar mass fraction in low
mass halos, seen in early studies connecting galaxies to their host
halos \citep{Yang:2003aa,Vale:2006fm,Zheng:2007ac}. In fact, when
measuring the stellar-to-halo mass ratio (SHMR) as a function of time,
we observe that stellar mass is building up asymmetrically, first in
massive halos, later on in low-mass halos
\citep{Conroy:2006aa,Behroozi:2013aa}.  This asymmetry in the SHMR is
one corollary of the so-called galaxy \emph{downsizing} effect
\citep{Cowie:1996aa}.  In low-mass halos, stellar winds and supernovae
may slow down star formation until the potential well grows deep
enough to retain the gas and increase the star formation rate. Again,
it becomes necessary to relate galaxy properties to their host halo
mass.

A number of studies have related galaxy properties to dark matter
halos using the \emph{abundance matching} technique
\citep{Marinoni:2002aa, Conroy:2006aa, Behroozi:2010ja, Guo:2010do,
  Moster:2010ep}, which employs the stellar mass (or luminosity)
function and the halo mass function to match halo-galaxy properties
based on their cumulative abundances.  The \emph{conditional
  luminosity function} technique proposed by \cite{Yang:2003aa}
includes a parameterised $\Mstar-\Mh$ relationship whose parameters
are fitted to the luminosity function. Both this formalism and recent
abundance matching studies feature a scatter in $\Mstar$ at fixed
$\Mh$, which is an important ingredient to account for, given the
steep relation between the two quantities at high mass.

More recently, models adopting a similar approach to abundance
matching consist of directly populating dark matter halos from N-body
simulations, to reproduce the observed stellar mass functions as a
function of redshift, using a parameterised star formation rate (SFR)
model to account for redshift evolution
\citep{Behroozi:2013aa,Moster:2012dl}.

Except in some rare cases where central or satellite galaxies can be
individually identified \citep[e.g.][]{More:2011aa,George:2011kv}, in
studies based on luminosity or stellar mass distributions, the
satellite galaxies' properties cannot be disentangled from those of
the central galaxies. To remedy the problem, abundance matching
techniques either assume an ad-hoc fraction of satellites or use a
sub-halo mass function estimated from numerical simulations.
Unfortunately, as sub-halos may be stripped and disappear after being
accreted onto larger halos, the subhalo mass function at the time
considered might not correspond to the distribution of satellites, and
one must consider the mass of sub-halos at the time of accretion,
further extrapolated to the time considered.  Obviously these
complications limit the amount of information one can extract about
galaxy satellites.

Galaxy clustering, on the other hand, allows separation of the
contributions from central and satellite galaxies due to the different
typical clustering scales. To model the clustering signal of a given
galaxy population, the halo occupation distribution (HOD) formalism
assumes that the galaxy number per halo is solely a function of halo
mass and that the galaxy satellite distribution \rs{is correlated to} that of the
dark matter \citep{Berlind:2002aa,Kravtsov:2004aa}.

One achievement of HOD modelling was to demonstrate from simulations
\citep{Berlind:2003aa, Moster:2010ep} that only a handful of
parameters was necessary to fully describe galaxy halo occupation.
This parametric HOD was fitted to a number of observations over a
large range of redshifts and galaxy properties. Among the more
remarkable results are the local Universe galaxy clustering and
abundance matching studies performed on the Sloan Digital Sky Survey
\cite[see e.g.][]{Zehavi:2011hl} and at higher redshifts
\citep{Foucaud:2010gb,Wake:2011cn,Coupon:2012aa,de-la-Torre:2013aa,Martinez-Manso:2014aa}.

However, some underlying assumptions on the distribution of dark
matter halos implied in the HOD formalism are observationally
challenging to confirm and one has to rely on N-body simulations.
Fortunately, additional techniques may be used to relate galaxy
properties to halo masses, among which gravitational lensing is one of
the most powerful probes: by evaluating the distortion and
magnification of background sources, one is able to perform a direct
estimation of the dark matter halo profile \citep[for a review,
see][]{Bartelmann:2001aa}.  The low signal-to-noise ratio associated
with individual galaxies, however, forces us to ``stack'' them
(e.g. binned together within narrow stellar mass ranges), using a
technique known as \emph{galaxy-galaxy lensing}
\citep{Brainerd:1996aa,Hudson:1998aa,Hoekstra:2004aa,mandelbaum:2005aa,Yoo:2006aa,Uitert:2011aa,Velander:2014aa,Cacciato:2014aa,Hudson:2015aa}.

Clearly, each of the above methods brings a different piece of
information and combining all observables together is particularly
interesting, although doing so properly is challenging. In a recent
study using COSMOS data, \cite{Leauthaud:2012aa} have successfully
combined galaxy clustering, galaxy-galaxy lensing and the stellar mass
function, fitted jointly and interpreted within the HOD framework: the
authors have used a global central galaxy $\Mstar-\Mh$ relationship
(as opposed to measuring the mean $\Mh$ per bin of stellar mass) and
extended it in a consistent way to satellite galaxies.

In this paper, we apply this advanced formalism using a new dataset
covering a uniquely large area of $\sim$25~deg$^2$ with accurate
photometric redshifts in the redshift range $0.5< z<1$ and stellar
masses $> 10^{10} \Msun$. Our galaxy properties' measurements are
calibrated and tested with $70\,000$ spectroscopic redshifts from the
VIPERS survey and a number of publicly available datasets.  Our data
span a wide wavelength range of ultra-violet (UV) deep data from
GALEX, optical data from the CFHT Legacy Survey, and $\Ks$-band
observations with the CFHT WIRCam instrument. This large statistical
sample allows us to measure with high precision the stellar mass
function, the galaxy clustering, and we use the CFHTLenS shear
catalogue to measure galaxy-galaxy lensing signals.  The galaxy
clustering is measured on the projected sky for the photometric sample
and in real space for the spectroscopic sample.

This paper is organised as follows: in Section \ref{sec:data}, we
describe the observations, the photometric redshift and stellar mass
estimates. In Section \ref{sec:measurements} we present the
measurements of the stellar mass function, the galaxy clustering (both
from the photometric and spectroscopic samples) and galaxy-galaxy
lensing signals. In Section~\ref{sec:model} we describe the HOD model,
and the MCMC model fitting results are given in
Section~\ref{sec:results}.  In Section~\ref{sec:conclusions}, we
discuss our results and conclude.  Throughout the paper we adopt the
following cosmology: $H_0$=72\ km s$^{-1}$ Mpc$^{-1}$ and $\Omega_{\rm
  m}=0.258$, $\Omega_{\Lambda}=0.742$ \citep{Hinshaw:2009jq} unless
otherwise stated. To compute stellar masses we adopt the initial mass
function (IMF) of \cite{Chabrier:2003ki} truncated at 0.1 and
$100\Msun$, and the stellar population synthesis (SPS) models of
\cite{Bruzual:2003aa}.  All magnitudes are given in the AB system. The
dark matter halo masses are denoted as $\Mh$ and defined within the
virial radius enclosing a mean overdensity $\Delta_{\rm vir}$ compared
to the mean density background, taking the formula from
\cite{Weinberg:2002aa}. At $z=0.8$, $\Delta_{\rm vir}=215$. All masses
are expressed in unit of $\Msun$.  Measured quantities are denoted as
$\widetilde{X}$ and theoretical quantities as $X$. We call
\emph{cosmic variance} the statistical uncertainties caused by the
density fluctuations of dark matter and we define the \emph{sample
  variance} as the sum of the cosmic variance and Poisson noise
variance.

\section{Data}
\label{sec:data}

In this work, we combine several datasets to build a volume-limited
sample of galaxies more massive than $\Mstar = 10^{10}\Msun$ in the
redshift range $0.5 < z < 1$.  Our galaxy selection is based on NIR
($\Ks\ab<22$) observations, collected in the two fields of the VIMOS
Public Extragalactic Redshift Survey (``VIPERS-W1'' and
``VIPERS-W4''), overlapping the CFHTLS-Wide imaging survey, and
covering a total unmasked area of 23.1 deg$^2$.  We refer to
\cite{Arnouts:2014aa} for a complete description of the
multi-wavelength UV and NIR observations, reduction and photometry.

Our background galaxy selection used for the measurement of the
lensing signal is based on the CFHTLS-Wide $i$-band selection in the
area that overlaps with the NIR observations.

\subsection{The CFHTLS-Wide survey}

The Canada-France-Hawaii Telescope Legacy
Survey\footnote{http://www.cfht.hawaii.edu/Science/CFHTLS/} (CFHTLS)
is a photometric survey performed with MegaCam \citep{Boulade:2003aa}
on the CFHT telescope in five optical bands $u^{\star}, g, r, i, z$
($i\ab < 24.5-25$, 5$\sigma$ detection in 2$\arcsec$ apertures) and
covering four independent patches in the sky over a total area of
154~deg$^2$. In this analysis, we use the photometric and shear
catalogues produced by the CFHTLenS\footnote{http://cfhtlens.org/}
team \citep{Heymans:2012aa}.  The CFHTLenS photometry is performed
with \texttt{SExtractor} \citep{Bertin:1996aa} on the PSF-homogenised
images \citep{Erben:2013aa,Hildebrandt:2012aa}.  Magnitudes are based
on the \texttt{MAG\_ISO} estimator where the isophotal apertures are
derived from the $i$-band detection image. This approach optimises the
colour measurements and leads to an improvement in the photometric
redshift accuracy \citep{Hildebrandt:2012aa}. 
\rs{To estimate the total magnitude of each source, a global shift is
  applied to the \texttt{MAG\_ISO} magnitude in all the bands based on the
  difference between \texttt{MAG\_ISO} and \texttt{MAG\_AUTO} magnitudes, as measured
  in the $i$-band detection image \citep{Hildebrandt:2012aa}.}

As the magnitude errors are measured with \texttt{SExtractor} directly
from the local background in the PSF-homogenised image, we need to
correct for the noise correlation introduced by the convolution
process. To do so we multiply the CFHTLenS magnitude errors in all
bands by the ratio of the $i$-band detection image errors to the
$i$-band PSF-homogenised image errors. The correction factor ranges
from 3 to 5, where the strongest correction occurs when the seeing
difference between the $i$-band and the worse-seeing image is the
largest. As the $i$-band image is usually the best-seeing image, this
procedure may slightly overestimate the correction in the other bands,
however we neglect it here.

In addition, magnitude errors must be rescaled to account for image
resampling. Two independent tests have been performed to accurately
estimate the correction factor: we measured the dispersion of
magnitudes between the $i$-band detection (un-convolved) magnitudes
and the CFHTLS-Deep magnitudes, and between duplicated observations of
the same object in the overlapping regions of adjacent tiles.  We find
that the errors must be rescaled by a factor of $2.5$.

\begin{figure*}
  \centering
  \includegraphics[angle=-90, width=0.7\textwidth]{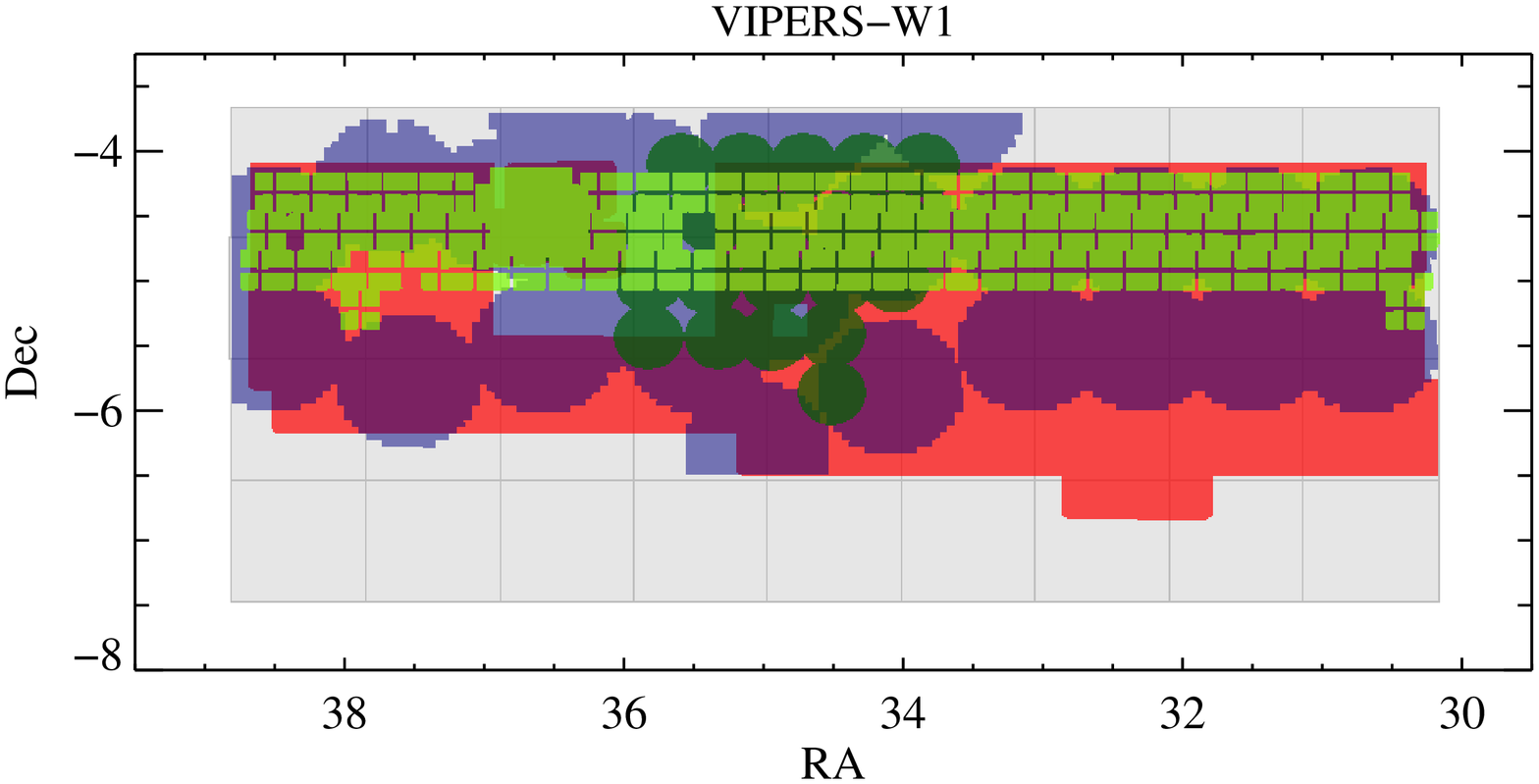}
  \includegraphics[angle=-90, width=0.7\textwidth]{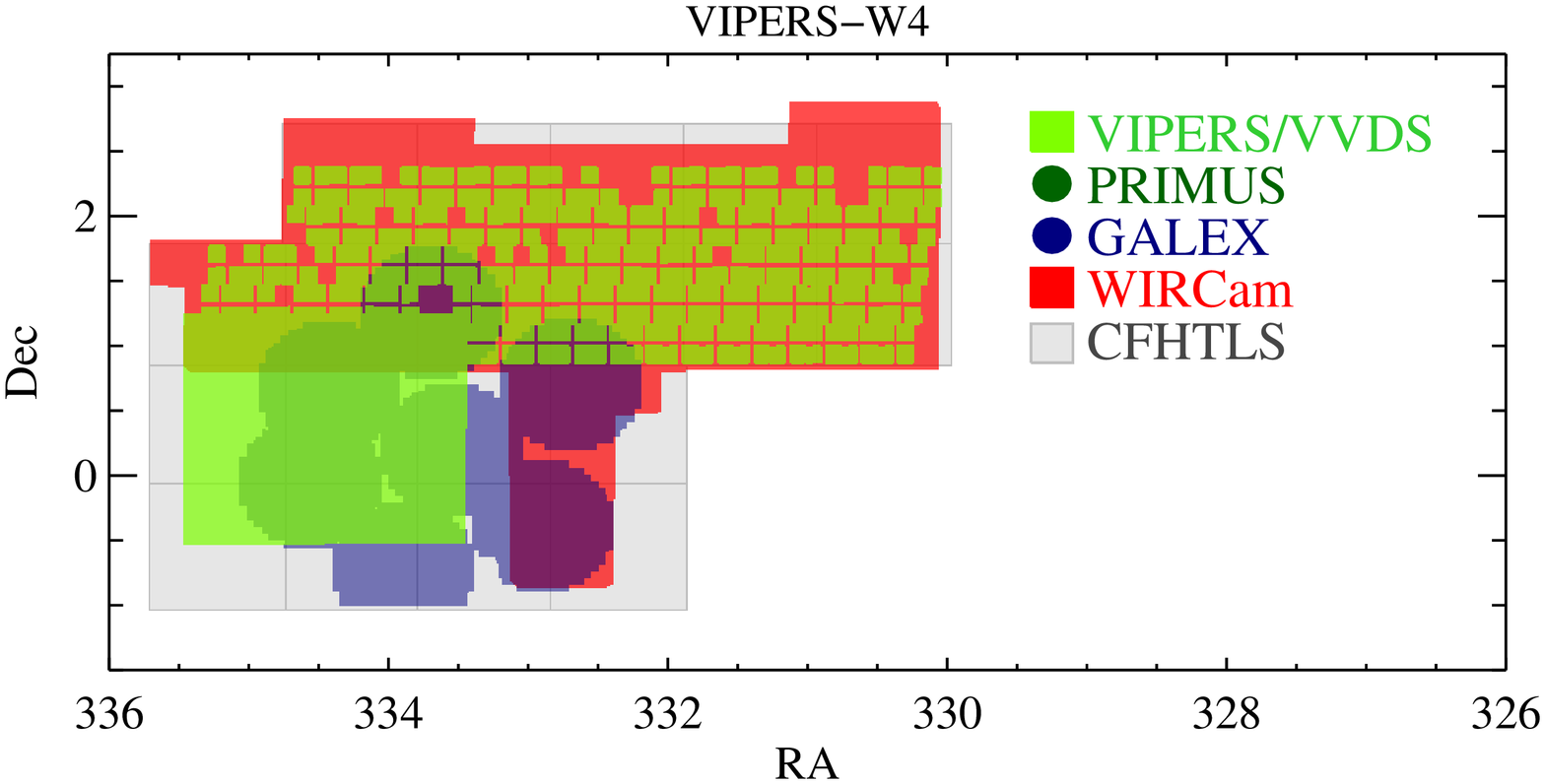}
  \caption{Footprints of the different datasets used in this work. Our
    selection is based on WIRCam data shown in red and covering
    approximately 25 deg$^2$ (23.1 deg$^2$ after masking). The CFHTLS
    MegaCam pointings are shown in grey, the GALEX DIS observations as
    large blue circles (in purple if overlapped with WIRCam), the
    spectroscopic surveys VIPERS/VVDS in light green and PRIMUS in
    dark green. The SDSS/BOSS coverage is almost complete. The data
    outside the WIRCam footprint are not used, and shown here only for
    reference.}
  \label{fig:data}
\end{figure*}

The footprints of the CFHTLS MegaCam tiles overlapping the VIPERS
survey are shown as gray squares in Fig.~\ref{fig:data}.
%

\subsection{The Near-IR observations}

We have conducted a $\Ks$-band follow-up of the VIPERS fields with the
WIRCam instrument at CFHT \citep{Puget:2004aa} for a total allocation
time of $\sim$120 hours.  The integration time per pixel was 1050
seconds and the average seeing of all the individual exposures was
$0.6\arcsec \pm$0.09.  The data have been reduced by the Terapix
team\footnote{http://terapix.iap.fr/}: the images were stacked and
resampled on the pixel grid of the CFHTLS-T0007 release
\citep{Hudelot:2012aa}.  The images reach a depth of $\Ks\ab= 22$ at
$\sim3\sigma$ \citep[][]{Arnouts:2014aa}. The photometry was performed
with \texttt{SExtractor} in dual image mode with a $gri-\chi^2$ image
\citep{Szalay:1999aa} as the detection image. To correct for the noise
correlation introduced by image resampling, we multiply the errors by
a factor 1.5, obtained from the dispersion between the WIRCam
$\Ks$-band magnitudes and the magnitudes measured in the deeper ($K <
24.5$) UKIDSS Ultra Deep Survey \citep[UDS,][]{Lawrence:2007aa}.  We
also used the UDS survey to confirm that our sample completeness based
on $gri-\chi^2$ detections reaches 80\% at $\Ks\ab=22$. Using the
WIRCAM/CFHTLS-Deep data with an $i$-band cut simulating the
CFHTLS-Wide data depth, we have checked that this incompleteness is
caused by red galaxies above $z=1$ and does not affect our sample
selected in the range $0.5 < z < 1$.
\rs{The $\Ks$ \texttt{MAG\_AUTO} estimates are then simply matched to their
  optical counterparts based on position.}

In addition to this dataset, we also use the CFHTLS-D1 WIRDS data
\citep{Bielby:2012aa}, a deep patch of 0.49 deg$^2$ observed with
WIRCam $J$-, $H$- and $\Ks$-bands and centered on 02$^h$26$^m$59$^s$,
$-04^{\circ} 30' 00''$. All three bands reach 50\% completeness at AB
magnitude 24.5.

The WIRCam observations are shown in Fig.~\ref{fig:data} as the red
regions.  After rejecting areas with poor WIRCam photometry and those
with CFHTLenS mask flag larger than 2, the corresponding effective
area used in this work spans over 23.1 deg$^2$, divided into 15 and
8.1 deg$^2$ in the VIPERS-W1 and VIPERS-W4 fields, respectively.

\subsection{The UV-GALEX observations}

When available, we make use of the UV deep imaging photometry from the
GALEX satellite \citep{Martin:2005aa,Morrissey:2005aa}. We only
consider the observations from the Deep Imaging Survey (DIS), which
are shown in Fig.~\ref{fig:data} as blue circles ($\O \sim
1.1^{\circ}$).  All the GALEX pointings were observed with the NUV
channel with exposure times of $T_{\rm exp} \ge 30$~ksec. FUV
observations are available for 10 pointings in the central part of W1.

Due to the large PSF (FWHM$\sim$5\arcsec), source confusion becomes a
major issue in the deep survey. To extract the UV photometry we use a
dedicated photometric code, \texttt{EMphot} \citep{Conseil:2011aa}
which will be described in a separate paper \citep{vibert:inprep}. In
brief, EMphot uses $U$-band (here the CFHTLS $u$-band) detected
objects as a prior on position and flux.  The uncertainties on the
flux account for the residual in the [simulated$-$observed] image.
The images reach a depth of $m_{\rm NUV}\sim 24.5$ at $\sim 5\sigma$.
\rs{As for the WIRCAM data, the GALEX sources are matched to the
  optical counterparts based on position.}

The NUV observations cover only part of the WIRCam area with
$\sim$10.8 and 1.9~deg$^2$ in VIPERS-W1 and VIPERS-W4, respectively.
The UV photometry slightly improves the precision of photometric
redshifts and the stellar mass estimates in the GALEX area. However,
by comparing our measurements inside and outside the GALEX area, we
have checked that the addition of UV photometry does not make a
significant change for the galaxies of interest in this study.
Therefore, in the final sample, we mix galaxies inside the GALEX area
with those outside.

\subsection{Spectroscopic data}

To optimise the calibration and the validation of our photometric
redshifts we make use of all the spectroscopic redshifts available in
the WIRCam area.
 
The largest sample is based on the VIPERS spectroscopic survey
\citep{Guzzo:2014aa, Garilli:2014aa} and its first public data release
PDR1\footnote{http://vipers.inaf.it/rel-pdr1.html}. VIPERS aims to
measure redshift space distortions and explore massive galaxy
properties in the range $0.5 < z < 1.2$.  The survey is located in the
W1 and W4 fields of the CFHTLS-Wide survey and will cover a total area
of 24~deg$^2$ when completed, with a sampling rate of $\sim$40\% down
to $i\ab<22.5$.  In Fig.~\ref{fig:data} we show the layout of the
VIMOS pointings as the light-green squares.  The PDR1 release includes
redshifts for \rs{$\sim 54\,204$} objects. After keeping galaxy spectra
within the WIRCam area \rs{($44\,474$)} and with the highest confidence flags between
2.0 and 9.5 \citep[95\% confidence, see][]{Guzzo:2014aa}, we are left
with \rs{$35\,211$} galaxies\rs{, which corresponds to a spectroscopic
  success rate of $80\%$.}

In addition to VIPERS, we also consider the following spectroscopic
surveys:
\begin{itemize}
\item the VIMOS-VLT Deep Survey (VVDS) F02 and Ultra-Deep Survey
  \citep{Le-Fevre:2005aa,Le-Fevre:2014aa} which consist of $11\,353$
  galaxies down to $i\ab<24$ (Deep) and $1\,125$ galaxies down to
  $i\ab<24.5$ (Ultra-Deep) over a total area of 0.75~deg$^2$ in the
  VIPERS-W1 field. We also use part of the VIMOS-VLT F22 Wide Survey
  with $12\,995$ galaxies over 4~deg$^2$ down to $i\ab<22.5$
  \citep[shown as the large green square in the southern part of the
  VIPERS-W4 field in Fig.~\ref{fig:data}]{Garilli:2008be}.  In total,
  we use \rs{$5\,122$} galaxies with secure flags 3 or 4 from the VVDS
  surveys within the WIRCam area;
\item the PRIMUS survey \citep{Coil:2011el} which consists of low
  resolution spectra ($\lambda/\Delta \lambda\sim40$) for galaxies
  down to $i\ab\sim 23$ and overlapping our VIPERS-W1 field. PRIMUS
  pointings are shown as the dark green circles in
  Fig.~\ref{fig:data}.  We keep \rs{$21\,365$} galaxies with secure flags 3
  or 4;
\item the SDSS-BOSS spectroscopic survey based on data release DR10
  \citep{Ahn:2014aa} down to $i\ab < 19.9$, overlapping both VIPERS-W1
  and VIPERS-W4 fields, totalling \rs{$4\,675$} galaxies with
  \texttt{zWarning=0} (99\% confidence redshift) within our WIRCam
  area.
\end{itemize}

In total, the spectroscopic sample built for this study comprises
\rs{$62\,220$} unique galaxy spectroscopic redshifts with the highest
confidence flag. We use the spectroscopic redshift value, when
available, instead of the photometric redshift value. The galaxies
with a spectroscopic redshift represent \rs{6.5\%} of the full sample, and
12\% after selection in the range $0.5 < z < 1$, where most of the
galaxies are from the VIPERS sample.

\subsection{Photometric redshifts}
\label{sec:photo_z}

To compute the photometric redshifts, we use the template fitting code
\texttt{LePhare}\footnote{http://www.cfht.hawaii.edu/~arnouts/lephare.html}
\citep{Arnouts:1999aa,Ilbert:2006bw}.  We adopt similar extinction
laws and parameters as \cite{Ilbert:2009aa} used in the COSMOS field
\citep{Scoville:2007aa}, and identical priors as in
\cite{Coupon:2009aa} based on the VVDS redshift distribution and
maximum allowed $g$-band absolute magnitude.  We note that the use of
priors is essential for the $z>1$, low signal-to-noise (or no NIR
flux), galaxies used as lensed (background) sources (see
Section~\ref{sec:measurements}).  A probability distribution function
(PDF) in steps of 0.04 in redshift is computed for every galaxy.

We use the full spectroscopic sample to adjust the magnitude relative
zero-points in all the passbands on a MegaCam pointing-to-pointing
basis. For the pointings with no spectroscopic information, we apply a
mean correction obtained from all the pointings with spectra. The mean
zero-point offsets and standard deviations in all passbands are given
in Table~\ref{tab:zeropoints} for the two fields separately.  We
further add the zero-point scatter in quadrature to the magnitude
errors in each band.
%
\begin{table}
\caption{Magnitude zero-point offsets measured per CFHTLS 
MegaCam pointing in VIPERS-W1 and VIPERS-W4 (mean and standard
deviation). $J$ and $H$-band zero-points were computed for the
pointings overlapping WIRDS data.
\label{tab:zeropoints}}
\centering
\begin{tabular}{lcc}
  \hline
  \hline
  Filter & VIPERS-W1 & VIPERS-W4 \\
  \hline 
  FUV & 0.18 $\pm$  0.11 & 0.02 $\pm$  0.16 \\ 
  NUV & 0.11 $\pm$  0.09 & 0.15 $\pm$  0.10 \\ 
  u & 0.10 $\pm$  0.03 & 0.13 $\pm$  0.03 \\ 
  g & $-$0.02 $\pm$  0.01 & $-$0.01 $\pm$  0.01 \\ 
  r & 0.02 $\pm$  0.01 & 0.01 $\pm$  0.01 \\ 
  i & $-$0.01 $\pm$  0.01 & $-$0.00 $\pm$  0.01 \\ 
  z & $-$0.02 $\pm$  0.01 & $-$0.01 $\pm$  0.01 \\ 
  J & 0.08 $\pm$  0.05 & ----- \\ 
  H & 0.02 $\pm$  0.05 & ----- \\
  K & 0.02 $\pm$  0.03 & 0.01 $\pm$  0.05 \\
  \hline
\end{tabular}
%
\end{table}
We recall that these zero-point corrections may not represent absolute
calibration offsets but rather \emph{relative} (i.e. depending on
colours) ones and tied to the adopted spectral energy distribution
(SED) template set.  We come back to the impact of this issue on
stellar mass measurements in Section~\ref{sec:systematic_errors}.

Our SED templates are based on the library used in
\cite{Ilbert:2009aa}, however the fewer bands used in this study
compared to COSMOS necessitate adapting the templates to reduce
redshift-dependent biases. The initial templates are based on the SEDs
from \cite{Polletta:2007ab}, complemented by a number of starburst
SEDs from the \cite{Bruzual:2003aa} SPS library. Using \rs{$35\,211$}
spectroscopic redshifts from VIPERS, we adapt the templates with
\texttt{LePhare} using the following procedure. First, a best-fit
template from the original set is found for each galaxy and normalised
to unity, and the photometry is then corrected into the rest frame
given the spectroscopic redshift value. The rest-frame photometry for
all galaxies with identical best-fit templates is combined and the
adapted template is constructed from the sliding-window median values
as a function of wavelength. The process is repeated iteratively.
Given the high number of galaxies with spectroscopic redshifts, we
found that only two iterations were necessary to reach
convergence. Interestingly, although the improvement in the
photometric redshift bias is significant, the new templates appear
very similar ``by eye'' compared to the original ones, which implies
that small features in the SED templates may lead to large photometric
errors, as also noted by \cite{Ilbert:2006bw}.

\begin{figure*}
  \centering
  \includegraphics[width=0.35\textwidth]{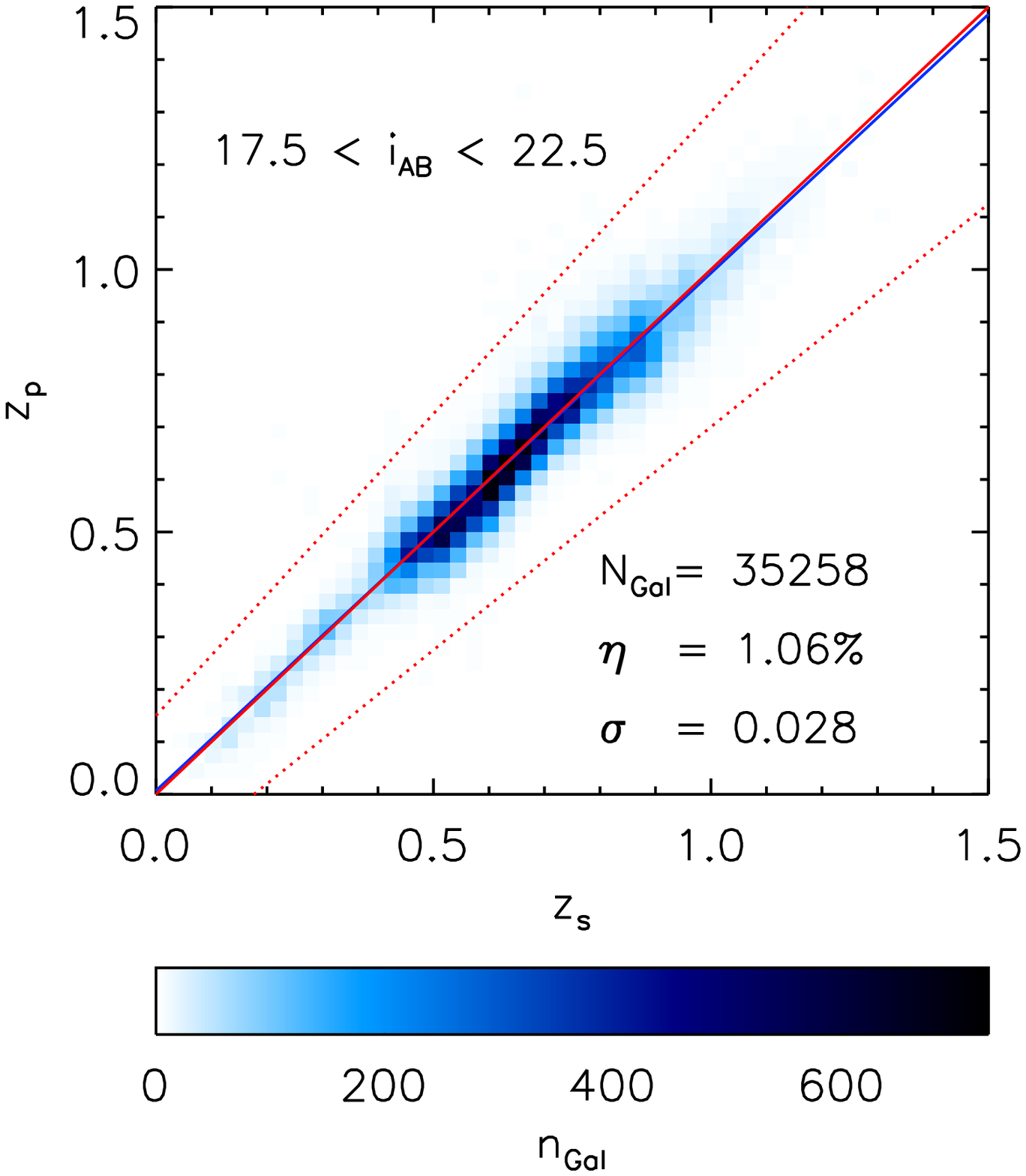}
  \includegraphics[width=0.35\textwidth]{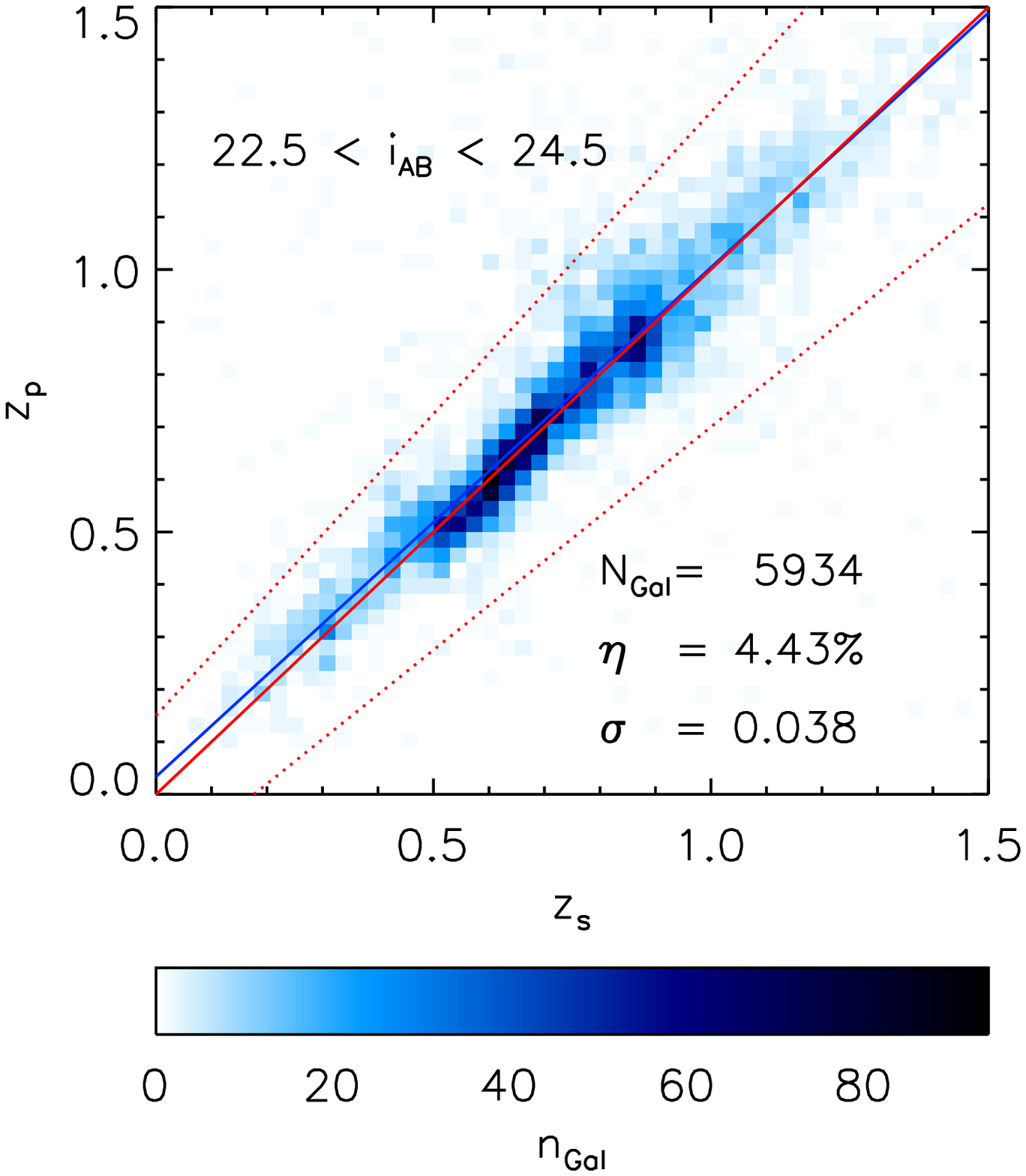}
  \caption{Photometric redshifts measured with $ugrizK$ (left) or
    $ugrizJHK$ (right) photometry versus VIPERS and VVDS spectroscopic
    redshifts. Left: $17.5 < i\ab < 22.5$, where the sample is
    dominated by galaxies between $0.5 < z < 1.2$ due to the VIPERS
    selection. Right: $22.5 < i\ab < 24.5$, from the VVDS Deep and
    Ultra-Deep surveys. \rs{The limits for the outliers are shown as
      red dotted lines.}}
  \label{fig:zp_zs}
\end{figure*}

In Fig.~\ref{fig:zp_zs}, we show the accuracy of the photometric
redshifts by comparing with the spectroscopic redshift sample from
VIPERS ($i\ab < 22.5$, left panel) and VVDS Deep/Ultra-Deep ($22.5 <
i\ab < 24.5$, right panel).  We observe a
\rs{dispersion\footnote{Defined as the normalised median absolute deviation
    \citep{Hoaglin:1983aa}: 1.48$\times$Median($|z_{\rm s}-z_{\rm
      p}|$/($1+z_{\rm s}$)), and robust to outliers.}} of $\sigma/(1+z)
\sim 0.03-0.04$ and a fraction of catastrophic redshifts ($|\Delta
z|\ge 0.15(1+z)$) of $\eta\sim1-4\%$.  The dispersion in both
magnitude ranges is significantly better than previous results in the
CFHTLS-Wide \citep{Coupon:2009aa}, due to the choice of isophotal
magnitudes and PSF homogenisation \citep{Hildebrandt:2012aa} at faint
magnitude, and the contribution of NIR data above $z\sim1$. We note
that the faint sample is compared to the VVDS redshifts where deep NIR
data from WIRDS are available over a small part ($<1$\,deg$^2$) of the
field, \rs{and with a magnitude distribution biased towards bright
galaxies compared to the photometric sample.}
Therefore, we foresee degraded photometric redshift performance
elsewhere, mainly relevant for $z>1$ galaxies.  However, as shown in
Appendix~\ref{sec:details-measurements}, no systematic bias affecting
our lensing measurements is introduced by the use of sources beyond
$z=1$.
%

\subsection{Stellar mass estimates}
\label{sec:mstar}

To compute stellar masses we adopt the same procedure as
\cite{Arnouts:2013gu} and described in detail in their Appendix~A.  In
brief, we use the photometric or spectroscopic (when available)
redshift and perform a $\chi^2$ minimization on a SED library based on
the SPS code from \cite{Bruzual:2003aa}.  The star formation history
is either constant or described with an exponentially declining
function, with e-folding time $0.01\le \tau \le 15$. We use two
metallicities ($Z_{\odot}$, 0.2$Z_{\odot}$) and adopt the
\cite{Chabrier:2003ki} IMF.  As discussed in \cite{Arnouts:2013gu},
the use of various dust extinction laws is critical to derive robust
SFR and stellar mass, and in this work we adopt their choices for
differing attenuation curves: a starburst \citep{Calzetti:2000aa}, a
SMC-like \citep{Prevot:1984aa} and an intermediate slope
($\lambda^{-0.9}$) law. We consider reddening excess in the range $0
\le E(B-V) \le 0.5$. When fixing the redshift, the typical 68\%
stellar mass statistical uncertainty, as derived by marginalising the
likelihood distribution, ranges from $\sigma(M_{\star})\sim 0.05 $ to
$0.15$ for galaxies with $\Ks\ab \le 22$ and $z < 1$. This stellar
mass uncertainty is an underestimate, since we neglect photometric
redshift uncertainties\footnote{We will see in
  Section~\ref{sec:HOD:cen} that our model accounts for such an extra
  source of uncertainty in stellar mass through a stellar-mass
  dependent parameterisation of the stellar-to-halo mass scatter.}.

In addition to statistical errors, in
Section~\ref{sec:systematic_errors} we investigate the different
sources of systematic effects in the stellar mass estimates, arising
from our lack of knowledge of galaxy formation and evolution. The
choice of differing dust treatments (and resulting dust attenuation
laws) is one of them: \cite{Ilbert:2010ee} have measured a shift of
0.14 dex, with a large scatter, between stellar masses estimated with
the \cite{Charlot:2000aa} dust prescription and the
\cite{Calzetti:2000aa} attenuation law.  The dust parameterisation
leads to systematics larger than the statistical errors in the stellar
mass function.  Even more critical is the choice of the SPS model and
the IMF \cite[see more detailed systematic errors analysis
in][]{Behroozi:2010ja,Marchesini:2009ef,Fritz:2014aa}, leading to
systematic differences in stellar mass estimates up to 0.2 dex. One
must keep these limitations in mind when comparing results from
various authors using different methods, and we come back to these
issues when presenting our results.

\section{Measurements}
\label{sec:measurements}

We aim to compute high signal-to-noise measurements of four distinct
observables: the stellar mass function $\phi(\Mstar)$, the projected
galaxy clustering $w(\theta)$, the real-space galaxy clustering
$w_p(r_{\rm p})$, and the galaxy-galaxy lensing $\Delta \Sigma(r)$.

To do so, we select volume-limited samples in the redshift range $0.5
< z < 1$, where the high sampling rate of VIPERS and our NIR data
guarantee both robust photometric redshift and stellar mass estimates.
As for the stellar mass function, we adopt a lower mass limit of
$\Mstar=10^{10}\Msun$ and employ the $V_{\rm max}$ estimator to
correct for galaxy incompleteness near $z=1$. The total volume probed
in this study is $0.06$~Gpc$^3$.

The stellar mass bins for the clustering and lensing measurements are
defined to keep approximately a constant signal-to-noise ratio across
the full mass range (which may lead to differing mass cuts depending
on the observable), and guarantee complete galaxy samples (see
Appendix~\ref{sec:sample_completeness}).  We summarise our samples'
properties in Table~\ref{tab:samples}.
%
\begin{table*}
  \caption{Sample mass definitions in $\log( \Mstar/\Msun)$ and
    number of galaxies in each sample. The parent sample 
    comprises a total of $352\,585$ galaxies.\label{tab:samples}}
  \centering
  \begin{tabular}{ c c c c c c c}
    \hline 
    \hline 
    & \multicolumn{2}{c}{Clustering -- $w(\theta)$} &
    \multicolumn{2}{c}{Clustering -- $w_p(r_{\rm p})$} &  \multicolumn{2}{c}{Lensing} \\
    Sample & Mass cut  & Number & Mass cut & Number &  Mass cut &
    Number\\
    \hline
    1 &  10.00 -- 10.40$^1$  & $23\,886$ & 10.60 -- 10.90$^1$ & $2\,154$& 10.00 -- 10.40$^1$  & $23\,886$   \\
    2 &  10.40 -- 10.60 &$36\,560$ &  10.90 -- 11.20$^2$ & $1\,964$ & 10.40 -- 10.65  & $45\,032$\\
    3 &  10.60 -- 10.80 &$31\,900$  & 11.20 -- 12.00    & $816$ & 10.65 -- 10.80  &$23\,427$ \\
    4 &  10.80 -- 11.00 & $24\,451$ &  -----               &        -----    & 10.80 -- 10.95  & $19\,293$\\
    5 &  11.00 -- 11.20 & $13\,538$  &  -----              &     -----    & 10.95 -- 11.15  & $16\,317$                \\
    6 &  11.20 -- 12.00 &  $6\,326$   &  -----              &      -----        &   11.15 -- 12.00         &  $8\,654$ \\
    \hline
    \multicolumn{7}{l}{$^1$ $0.5 < z < 0.7$}\\
    \multicolumn{7}{l}{$^2$ $0.5 < z < 0.8$}
  \end{tabular}\\
\end{table*}

To measure each of the observables described below, we use the
parallelised code \texttt{SWOT}, a fast tree-code for computing
two-point correlations, histograms, and galaxy-galaxy lensing signals
from large datasets \citep{Coupon:2012aa}. {\rs The stellar mass
  function is expressed in comoving units, whereas the clustering and
  galaxy-galaxy lensing signal are measured in physical units}.  We
estimate statistical covariance matrices from a jackknife resampling
of 64 sub-regions with equal area (0.35 deg$^2$ each), by omitting a
sub-sample at a time and computing the properly normalised standard
deviation \citep[see more details in][]{Coupon:2012aa}.  
\rs{This number was chosen to meet both requirements of using large
  enough sub-regions to capture the statistical variations at large
  scale, while keeping a sufficient number of sub-samples to compute a
  robust covariance matrix. Nevertheless, we expect the projected galaxy
  clustering errors to be slightly underestimated on 
  scales larger than the size of our sub-regions, $\sim0.5$~deg, and the
  noise in the covariance matrix to potentially bias the best-fit
  $\chi^2$ value.}

A random sample with 1 million objects is constructed using our WIRCam
observations layout and the union of the WIRCam and CFHTLenS
photometric masks. For real-space clustering, measured from VIPERS
spectroscopic redshifts, the random sample is constructed using the
layout of the VIPERS PDR1 geometry (and photometric masks) plus a
random redshift drawn in the range $0.5 < z < 1$ from a distribution
following $\ud V/ \ud z$, to match our volume limited samples.  The
sub-regions for the measurements of statistical errors are constructed
by \texttt{SWOT} based on the random catalogue: the field is divided
into 64 areas with an equal number of random objects.

\subsection{The stellar mass function}
\label{sec:measurements-smf}

The stellar mass function $\widetilde{\phi}(\Mstar) =\ud n / \ud\log
\Mstar$ is measured per unit of comoving volume in 10 equally spaced
logarithmic mass bins of width 0.2 dex, centered on the mass mean
weighted by the number of galaxies.  To correct for the incompleteness
in the low-mass galaxy sample ($10^{10} < \Mstar/\Msun < 10^{10.4} $)
occurring near $z=1$ (see Appendix~\ref{sec:sample_completeness}), we
up-weight low-redshift galaxies by a factor $1/V_{\rm max}$ defined
as:
\begin{equation}
  V_{\rm  max} = \Omega \int^{z_{\rm max}}_{0.5} \, \frac{\ud V}{\ud
    z} \ud z \, ,
\end{equation}
where $\Omega$ is the solid angle of the survey, 23.1deg$^2$, $V$ the
comoving volume per unit area, and ${z_{\rm max}}$ the maximum
redshift for a galaxy to be observed given a $\Ks\ab < 22$ magnitude
cut, calculated with \texttt{LePhare}.

\begin{figure}
  \centering
  \includegraphics[width=0.49\textwidth]{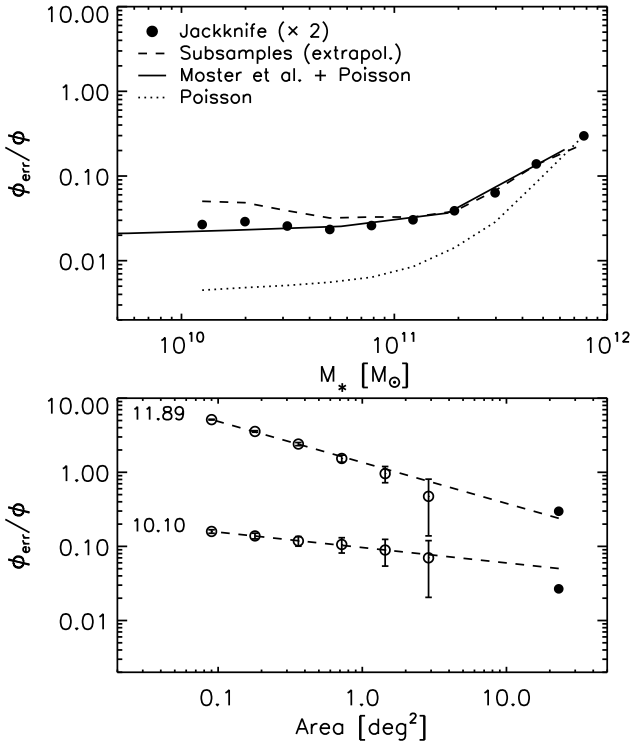}
  \caption{Stellar mass function statistical errors as function of
    stellar mass (top) and area (bottom). In the top panel we show the
    jackknife estimator based on 64 sub-regions and multiplied by a
    factor 2, compared to the theoretical cosmic variance plus Poisson
    error derived from the \protect\cite{Moster:2011ip} \texttt{getcv}
    code (the Poisson error only is shown as the dotted line). The
    bottom panel shows an alternative internal estimate based on the
    standard deviation of sub-regions as a function of their size, in
    two mass bins ($\log \Mstar/\Msun = 10.10$ and $11.89$),
    extrapolated to the size of the full survey (dashed lines in both
    panels).  As in the top panel, the black dots are the jackknife
    estimates, for which the cosmic variance part has been multiplied
    by 2.}
  \label{fig:smf_cv_fig}
\end{figure}

We have performed a number of tests to check our internal error
estimates.  In the top panel of Fig.~\ref{fig:smf_cv_fig} we show our
stellar mass function error estimates (square root of the covariance
matrix diagonal) as a function of stellar mass compared to the
\texttt{getcv} code estimate of \cite{Moster:2011ip} at $z=0.8$.
%
%
The latter code computes the theoretical expectations of cosmic
variance\footnote{We note that the highest mass bin galaxy bias was
  estimated a posteriori from our HOD results, since it was not
  provided by the authors of \texttt{getcv}, although the contribution
  of cosmic variance is negligible compared to the Poisson error in
  this bin, populated by rare massive galaxies.}  assuming a
prediction for dark matter clustering and galaxy biasing
\citep{Bardeen:1986aa}. We add to the \texttt{getcv} cosmic variance
the theoretical Poisson error and show the resulting (total) sample
variance as the thick line in the bottom panel.  Our jackknife
estimate is represented as the black points, for which we find that
the cosmic variance part (after subtracting Poisson noise) needs to be
multiplied by a factor 2 to agree with theoretical expectations (we
then multiply the covariance matrix by a factor 4). We have not found
a definitive explanation for the underestimation of the errors from
the jackknife resampling, however it is likely caused by the strong
correlation between bins (a combined effect of stellar mass scatter
and large scale structure correlations).

In the bottom panel of Fig.~\ref{fig:smf_cv_fig}, we show an
alternative internal estimator as function of area, based on the
standard deviation of subsamples with sizes varying from 0.1~deg$^2$
to 2.9~deg$^2$ (the black dots represent our Jackknife estimates in
the two mass bins $\langle \log \Mstar/\Msun \rangle = 10.10$ and
$11.89$).  We use a power law fit (the amplitudes of the error-bars
are arbitrarily scaled to the square root of the number of subsamples,
ranging from $\sqrt{256}$ to $\sqrt{8}$) to extrapolate to the full
size of the survey. The extrapolated values are shown as the dashed
line in the top panel of Fig.~\ref{fig:smf_cv_fig}.  The bin
correlations between small sub-samples may tilt the slope of the fit
and lead to an overestimate of the extrapolated error estimate, as
observed in the low mass bin. In the high mass bins, characterised by
an uncorrelated sampling variance dominated by Poisson noise, the
extrapolated estimate is consistent with both the jackknife estimate
and the theoretical Poisson noise.

\subsection{Projected galaxy clustering}
\label{sec:wtheta}

We measure the two-point correlation function $\widetilde{w}(\theta)$
in 10 logarithmically spaced bins centered on the pair-number weighted
averaged separation over the range $0.002^{\circ} < \theta <
2^{\circ}$. The modelled $w(\theta)$ is compared to the measured
$\widetilde{w}(\theta)$ by projecting the theoretical spatial
clustering $\xi(r)$ onto the sample redshift distribution computed as
the sum of photometric redshift PDFs (see Section~\ref{sec:photo_z}).

We use the \cite{Landy:1993aa} estimator following a similar procedure
to that described in Section~3.3 of \cite{Coupon:2012aa}.  Owing to
the limited size of the survey, our measurements are affected by the
\emph{integral constraint}, an effect that biases the clustering
signal low.  Here, we adopt a refined way to correct for it: the
correction is calculated directly for every parameter set from the
modelled $w(\theta)$ (instead of a pre-determined power law) and
integrated over the survey area using random pairs as in
\cite{Roche:2002aa}, leading to better agreement between the data and
the model at large scales. \rs{Here the typical values of the integral
  constraint range from $10^{-3}$ to $3\times10^{-3}$.}

We have checked, using the galaxy mocks prepared for the VIPERS sample
\citep{de-la-Torre:2013aa}, that our jackknife error estimates could
reproduce within 20\% the correct sample variance amplitude of
$\widetilde{w}(\theta)$ \citep[this result is in agreement with a
number of tests from the literature,
e.g.][]{Zehavi:2005aa,Norberg:2009by}, and we do not apply any
correction.

\subsection{Real-space galaxy clustering}
\label{sec:wp_rp}

We measure the real-space galaxy clustering for the VIPERS
spectroscopic sample by integrating the weighted redshift-space
correlation function along the line of sight to alleviate
redshift-space distortion effects:
\begin{equation}
  \widetilde{w_p} (r_{\rm p, phys}) = 2\,\int_{0}^{\pi_{\rm max}} \widetilde{\xi}(r_{\rm p, phys}, \pi_{\rm phys}) \ud \pi_{\rm phys} \,,
\end{equation}
where $r_{\rm p, phys}$ and $\pi_{\rm phys}$ are the coordinates
perpendicular and parallel to the line of sight, respectively.
$r_{\rm p, phys}$ is expressed in physical coordinates and divided
into 10 logarithmically spaced bins centered on the pair-weighted
averaged separation over the range $0.2 < r_{\rm p, phys}/{\rm Mpc} <
10$, and $\pi_{\rm phys}$ is divided into linear bins up to $\pi_{\rm
  max} = 40$~Mpc. The value of $\pi_{\rm max}$ is consistently used in
the derivation of the modelled $w_p$. As for $\widetilde{w}(\theta)$,
$\widetilde{\xi}(r_{\rm p, phys}, \pi_{\rm phys})$ is computed using
the \citeauthor{Landy:1993aa} estimator and the covariance matrix
estimated from the jackknife resampling of 64 sub-regions.

Each galaxy is weighted to account for the undersampling of the
spectroscopic sample: we use the global colour sampling rate (CSR),
target sampling rate (TSR), and success sampling rate (SSR), as
described in \cite{Davidzon:2013aa}, to account for the VIPERS colour
selection, the sparse target selection and measurement success as
function of signal-to-noise ratio, respectively. In addition, we also
use number-count normalised (to prevent global CSR, TSR and SSR double
weighting) spatial weights computed for each VIPERS panel by
\cite{de-la-Torre:2013aa} to correct for the position-dependent
sampling. Here, the SSR is the most affected quantity, as a function
of position in the sky, due to the differing observing conditions at
the times of observation.

Small pair incompleteness due to ``slit collision'' is corrected by a
factor $1+\widetilde{w_{\rm A}}$, such that:
\begin{equation}
  1+\widetilde{w}_{\rm p, corr} = \frac{1+\widetilde{w}_{\rm p}}{1+\widetilde{w}_{\rm A}} \, ,
\end{equation}
where
\begin{equation}
  1+\widetilde{w}_{\rm A} = 1-\frac{0.03}{r_{\rm p, phys}} \,
\end{equation}
is derived from the projected correlation as function of angular scale
by \cite{de-la-Torre:2013aa} and translated into physical scales at
$z=0.8$. We note that given our conservative small scale cut of
$r_{\rm p, phys} > 0.2$, the correction remains below 15\%.

\subsection{Galaxy-galaxy lensing}
\label{sec:ggl}

The gravitational lensing signal produced by the foreground matter
overdensity is quantified by the tangential distortion of background
sources behind a sample of stacked ``lens'' galaxies, also known as
the weighted galaxy-galaxy lensing estimator
\citep[e.g.][]{Yoo:2006aa,Mandelbaum:2006cv}. The excess surface
density of the projected dark matter halo relates to the measured
tangential shear through:
\begin{equation}
  \widetilde{\Delta \Sigma}(r_{\rm p, phys})=\Sigma_{\rm crit}\times
  \widetilde{\gamma_t}(r_{\rm p, phys}) \, ,
  \label{dsigma}
\end{equation}
(see also Appendix~\ref{sec:details-observables}).  We measure the
signal in 10 logarithmically spaced bins centered on the
number-weighted averaged separation, in the range $0.02 < r_{\rm p,
  phys}/{\rm Mpc} < 1$. $r_{\rm p, phys}$ is expressed in physical
coordinates\footnote{\rs{Note that the galaxy-galaxy lensing signal is
    measured in physical units, whereas a number of authors assume
    comoving units, which would require multiplying the excess surface
    density by a factor of $(1+z)^{-2}$ compared to our definition.}}.

The critical surface density $\Sigma_{\rm crit}$ is given by:
\begin{equation}
  \Sigma_{\rm crit} = \frac{c^2}{4\pi G_{\rm N}}\,
  \frac{D_{\rm  OS}}{D_{\rm OL}\,D_{\rm  LS}} \, ,
  \label{sigma_crit}
\end{equation}
with $D_{\rm OS}$ the observer-source angular diameter distance,
$D_{\rm OL}$ the observer-lens (foreground galaxy) distance and
$D_{\rm LS}$ the lens-source distance.  $G_{\rm N}$ is the
gravitational constant and $c$ the speed of light. All distances are
computed in physical coordinates using the photometric (spectroscopic
when available) redshift. For photometric redshift values, a cut
$z_{\rm source} - z_{\rm lens} > 0.1 \times (1+z_{\rm lens})$ is
adopted. The background source galaxy sample includes all galaxies
detected in the $i$-band with a non-zero lensing weight
\citep{Miller:2013aa}. Here we do not restrict our redshift sample to
$z_{\rm p} < 1.2$, but consider galaxies at all redshifts, taking
advantage of the improved photometric redshift estimates in our
sample, increasing the background source sample by 30\% compared to
other CFHTLenS lensing studies, without introducing any systematic
bias (see Appendix~\ref{sec:details-measurements}).

The galaxy shape measurement was performed on individual exposures
using the \texttt{lensfit} analysis pipeline
\citep{Miller:2007aa,Kitching:2008aa,Miller:2013aa} and systematics
checks were conducted by \cite{Heymans:2012aa} for cosmic shear (the
projected large scale structure lensing power spectrum).  The lensing
(inverse-variance) weights account for shape measurement uncertainties
\citep{Miller:2013aa}.  Following \cite{Velander:2014aa}, who
performed extensive systematics checks of the CFHTLenS shear catalogue
specifically for galaxy-galaxy lensing (see their Appendix~C), we do
not reject those CFHTLS-Wide pointings that did not pass the
requirements for cosmic shear, and we applied appropriate shape
measurement corrections as described in their Section~3.1.

We compute the boost factor \citep[to account for dilution due to
sources physically associated with the lens, see][]{Sheldon:2004aa,
  Mandelbaum:2006cv} by randomising the source positions, and correct
the final signal for it. On small scales ($<0.1$~Mpc) the boost factor
reaches up to 20\% for the most massive galaxies.

Here, the relatively low source density implies that our errors are
dominated by the source galaxy shape noise, originating from
ellipticity measurement uncertainties and intrinsic shape dispersion,
rather than sample variance. Indeed, when compared to the sum of
inverse-variance lensing weights, we have checked that our jackknife
estimate was similar at all scales (with small off-diagonal
correlation), confirming the negligible impact of cosmic variance (see
Appendix~\ref{sec:details-measurements}).

Nevertheless, a correlation exists \emph{between} the mass bins due to
the re-use of background source galaxies. We neglect this contribution
in the computation of the combined $\chi^2$, but we note that this
correlation is likely to lead to underestimation of our parameter
confidence limits.

\subsection{Systematic errors in stellar mass measurements}
\label{sec:systematic_errors}

In this section we are concerned with systematic errors affecting the
stellar mass measurements caused by the uncertainties in the assumed
cosmology (i.e. volume and distance estimates), the dust modelling,
and potential biases in the photometry.

To assess the impact of systematics on the measurements of the
observables, we propagate the errors affecting the stellar masses by
changing one parameter configuration at a time, then re-computing all
stellar masses and the observables, and finally measuring the
difference with the reference measurements.  We repeat the process for
the three different kinds of systematics listed below:
\begin{itemize}
\item assumed cosmology. We explore three $\Lambda$CDM parameter sets:
  in addition to the WMAP cosmology used in this study with
  $H_0=72$~km~s$^{-1}$ Mpc$^{-1}$, $\Omega_{\rm m}=0.258$,
  $\Omega_\Lambda=0.742$ \citep{Hinshaw:2009jq}, a ``concordance''
  cosmology model with $H_0=70$~km~s$^{-1}$ Mpc$^{-1}$, $\Omega_{\rm
    m}=0.3$, $\Omega_\Lambda=0.7$, and the Planck cosmology with
  $H_0=67$~km~s$^{-1}$ Mpc$^{-1}$, $\Omega_{\rm m}=0.320$,
  $\Omega_\Lambda=0.680$ \citep{Planck-Collaboration:2013aa} are
  tested. In each case, the stellar masses and the observables are
  consistently re-computed with the same cosmology. We note that the
  term ``systematics'' here refers to the choice for one or another
  set of parameters that produces a systematic shift in stellar mass
  and not to systematic errors associated to the measurement of
  cosmological parameters;
\item lens galaxy dust extinction modelling. We compute five different
  stellar masses for each galaxy by varying one aspect at a time: two
  different extinction law configurations (among our choice of three
  laws, see Section~\ref{sec:mstar}) and three different $E(B-V)$
  maximum allowed values (ranging from 0.2 to 0.7);
\item photometric calibration. As zero-point offsets do not correct
  for absolute calibration uncertainties (but do for colours), nor
  correct for photometric measurement biases (e.g. missing flux of
  bright objects), a change in the photometric calibration may cause a
  shift in the best-fit template and further bias the stellar mass
  measurements.  We re-compute stellar masses applying ad-hoc global
  shifts (in all bands) of $-$0.05 and $+$0.05 magnitude, which
  correspond to typical offsets caused by various calibration
  strategies or photometry measurements \citep{Moutard:2014aa}.
\end{itemize}

\begin{figure*}
  \centering
  \includegraphics[width=0.4\textwidth]{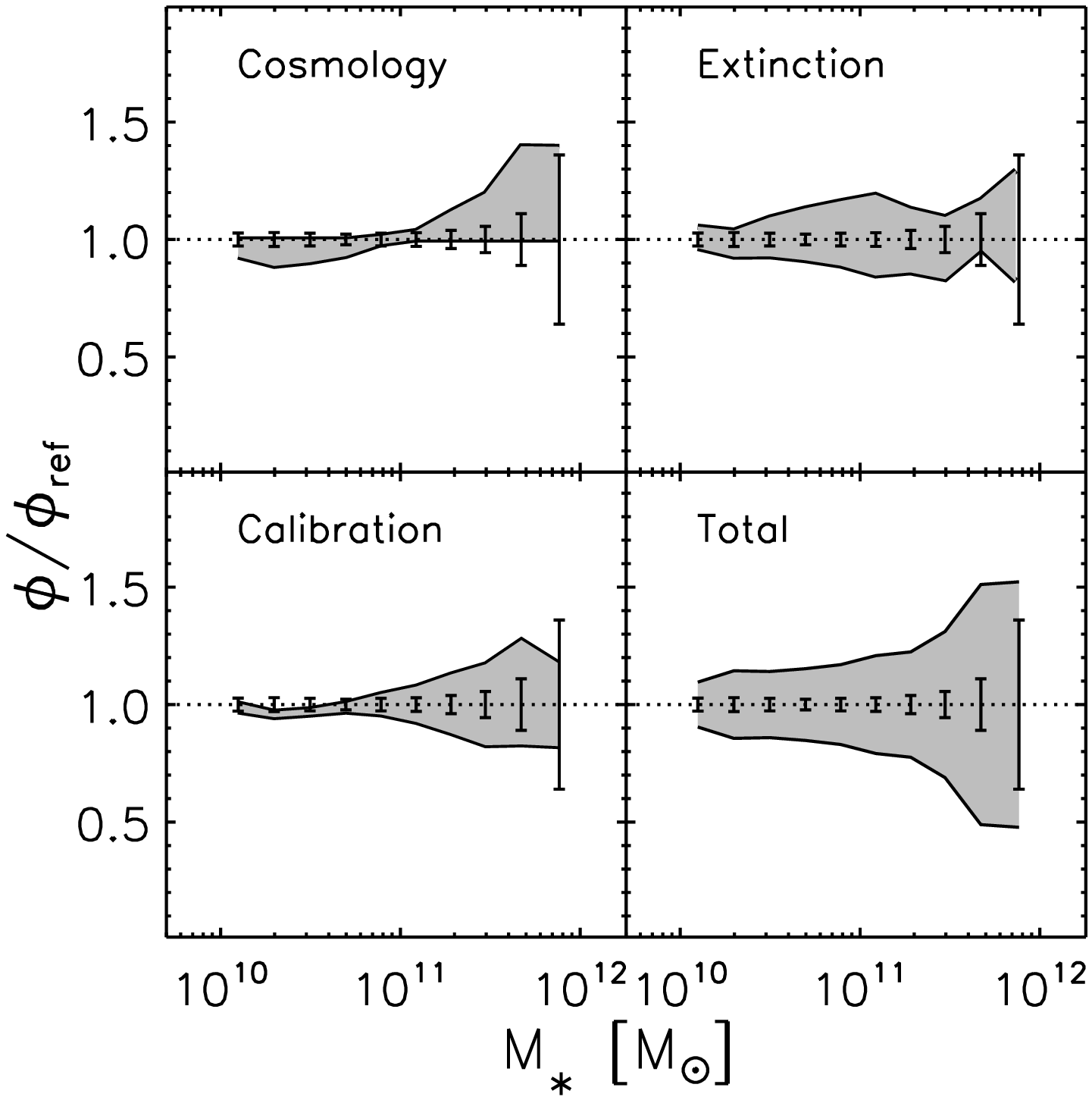}
  \includegraphics[width=0.4\textwidth]{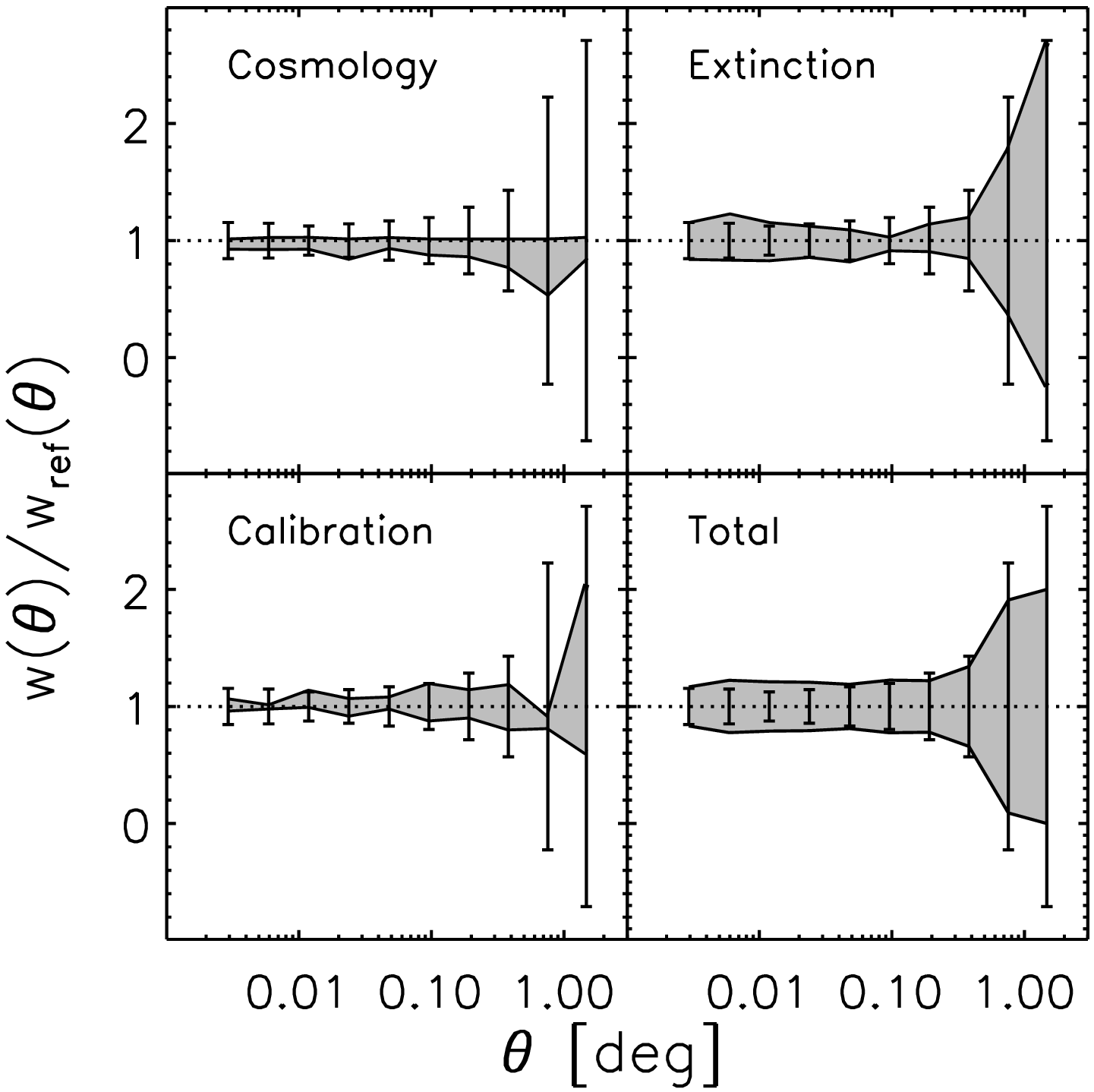}
  \includegraphics[width=0.4\textwidth]{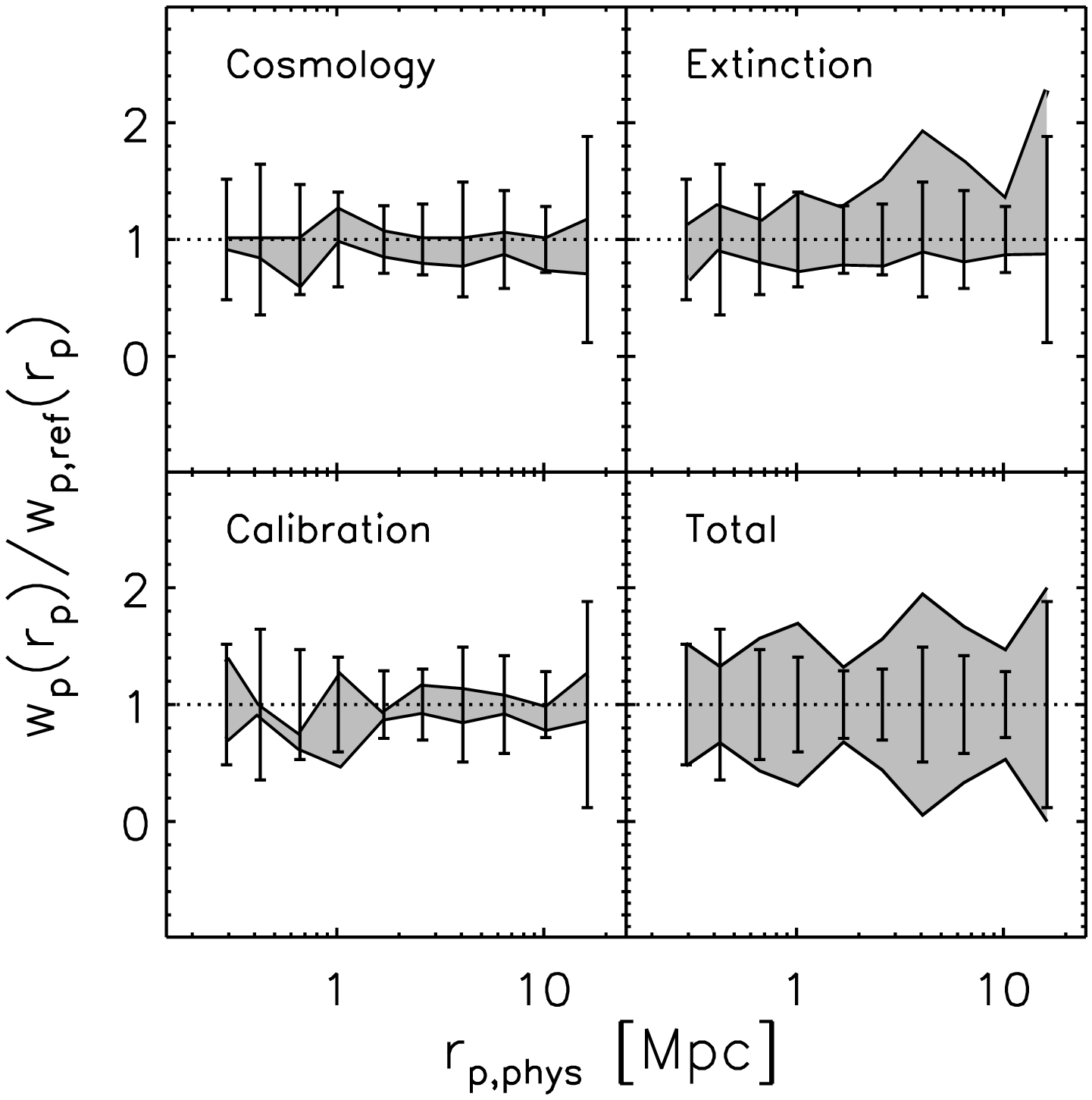}
  \includegraphics[width=0.4\textwidth]{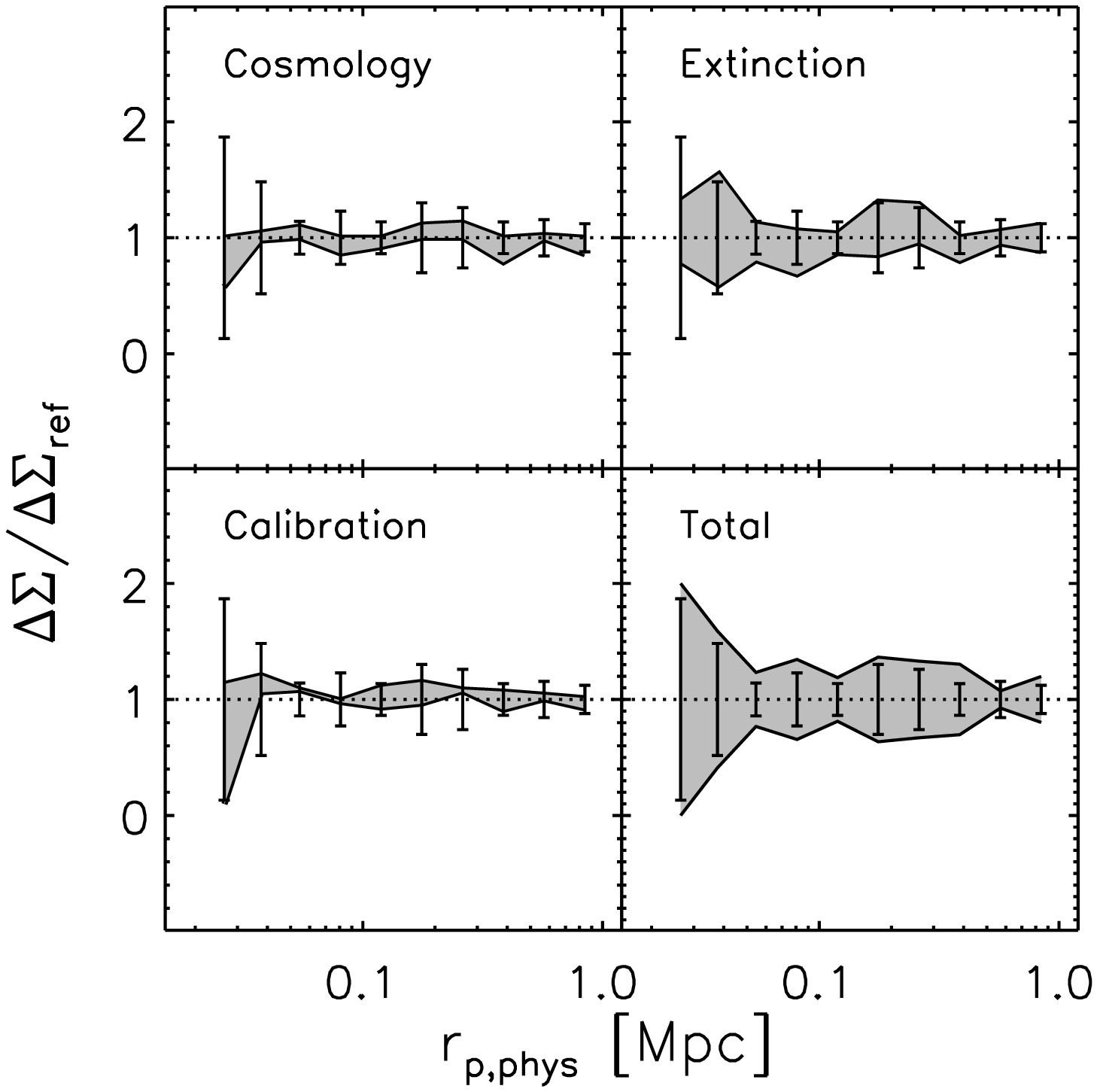}
  \caption{Systematic errors affecting the galaxy stellar mass
    function (top left), the projected correlation function (top
    right), the real-space correlation function (bottom left), and the
    galaxy-galaxy lensing signal (bottom right). In each panel the
    gray area symbolises the envelope (maximum value) of the
    re-computed measurement compared to the reference. The error bars
    are statistical errors from the internal jackknife estimator. The
    ``Total'' panel represents the symmetric sum in quadrature of all
    three contributions. Here we only show the most massive bins for
    the clustering and lensing measurements, however we repeated the
    tests in all mass bins.}
  \label{fig:tot_err}
\end{figure*}

Results are shown in Fig.~\ref{fig:tot_err}.
%
%
For each observable (top left: stellar mass function, top right:
projected clustering, bottom left: real-space clustering, and bottom
right: galaxy-galaxy lensing), we display the re-computed measurements
divided by the reference quantities, in each of the ``Cosmology'',
``Extinction'' and ``Calibration'' panels as well as the sum in
quadrature of all these effects (``total''). The gray area corresponds
to the maximum value among the differing re-computations, not the
standard deviation, as each of the solutions is equally likely to be
opted for. Except for the stellar mass function, we only display the
results in the most massive bins (where we observe the most
significant changes), although the calculations were repeated in all
mass bins.

To allow comparison with the statistical errors, we overplot the error
bars from our jackknife error estimates.  For the stellar mass
function (whose jackknife error estimate is multiplied by a factor 2,
see Section~\ref{sec:measurements-smf}), the systematic errors
compared to the statistical errors are striking, with the former being
larger by one order of magnitude compared to the latter.  The increase
of the systematic errors towards the high mass regime is a direct
consequence of the shift in stellar mass and the steep slope of the
SMF at the massive end.

It is interesting to note that the different cosmologies lead also to
large systematic errors compared to statistical errors. Although many
authors in galaxy evolution studies claim to account for cosmological
parameter uncertainties by presenting $h-$free results, we recall
that, in a flat Universe, both $\Omega_{\rm m}$ and $H_{0}$ enter in
the computation of the comoving volume and luminosity distances and,
in the precision era of WMAP and Planck, happen to contribute equally
to the distance uncertainties. Comparing our results to the recent
literature is therefore not as simple as scaling the different
quantities with respect to $h$, and we must properly account for the
more complex dependence of distances on $\Omega_{\rm m}$ and $H_{0}$.

In comparison, the projected and real-space galaxy clustering as well
as galaxy-galaxy lensing are relatively less prone to systematic
errors. For the effect of cosmology, the measurement of projected
clustering has no dependence on galaxy distances, and the only
difference originates from the modified galaxy selection caused by the
stellar mass shift.  Interestingly, although the real-space clustering
and the galaxy-galaxy lensing do depend on galaxy distance
measurements, the change in cosmology also has little impact at the
level of our statistical errors. We can draw similar conclusions on
the effects of dust extinction modelling and photometric calibration.

Obviously, the stellar mass function is the measured quantity
suffering from the largest systematic error contribution, compared to
the statistical errors. In particular, we will see in
Section~\ref{sec:results} that most of the constraints on the central
galaxy $\Mstar - \Mh$ relationship emanate from the stellar mass
function and taking into account these systematic uncertainties when
comparing our results with the literature is necessary.

Ideally one would like to estimate a best-fit model for each of the
re-computed quantities. Unfortunately this would be computationally
very expensive. Instead, we create two sets of measurements: a
``statistical error'' set based on our jackknife error estimate and a
``total error'' set for which we add in quadrature the systematic
errors \rs{(assuming they are Gaussian distributed)} and the
statistical errors. We present in Section~\ref{sec:results} separate
results for both.

\subsection{Impact of photometric redshift uncertainties}

The dispersion of photometric redshifts may also cause systematic
effects of several kinds, firstly on the stellar mass function, as a
contribution to the stellar mass scatter, which shifts towards higher
masses the high-mass end where the slope is steep, an effect known as
\emph{Eddington bias}.  Secondly, the projected clustering amplitude
is biased low due to the scattering of galaxies falling outside the
mass bins.

\begin{figure*}
  \centering
  \includegraphics[width=0.3\textwidth]{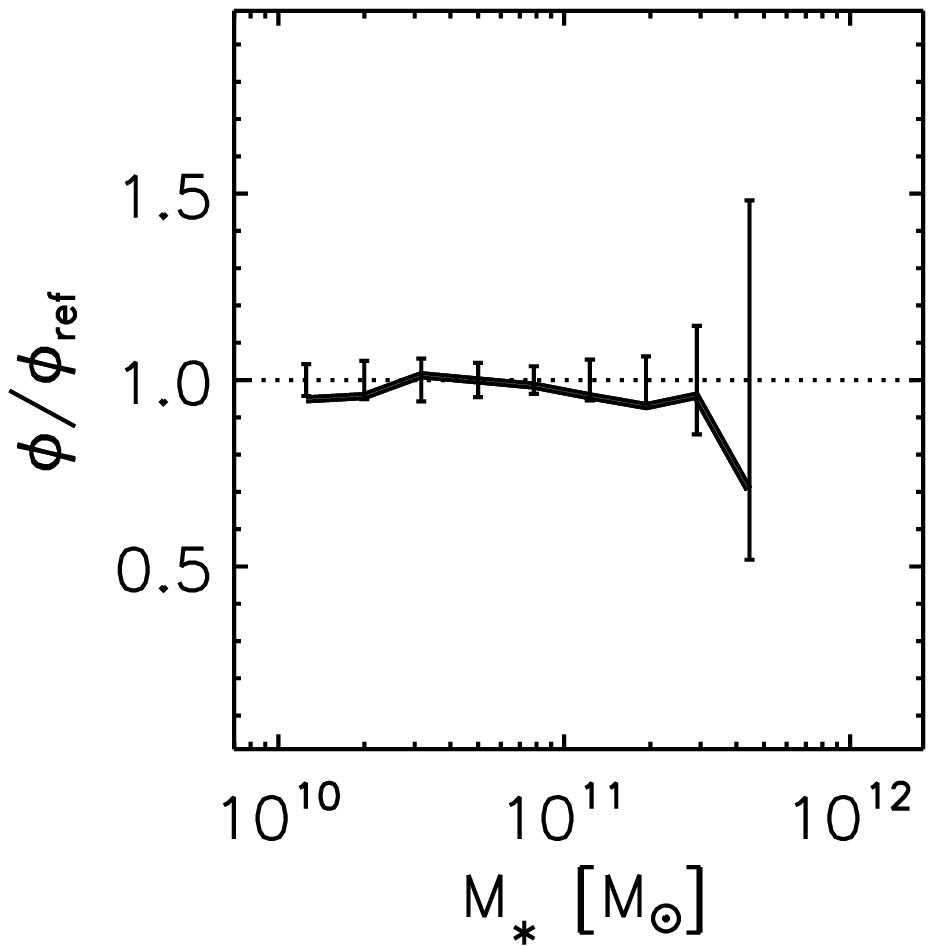}
  \includegraphics[width=0.3\textwidth]{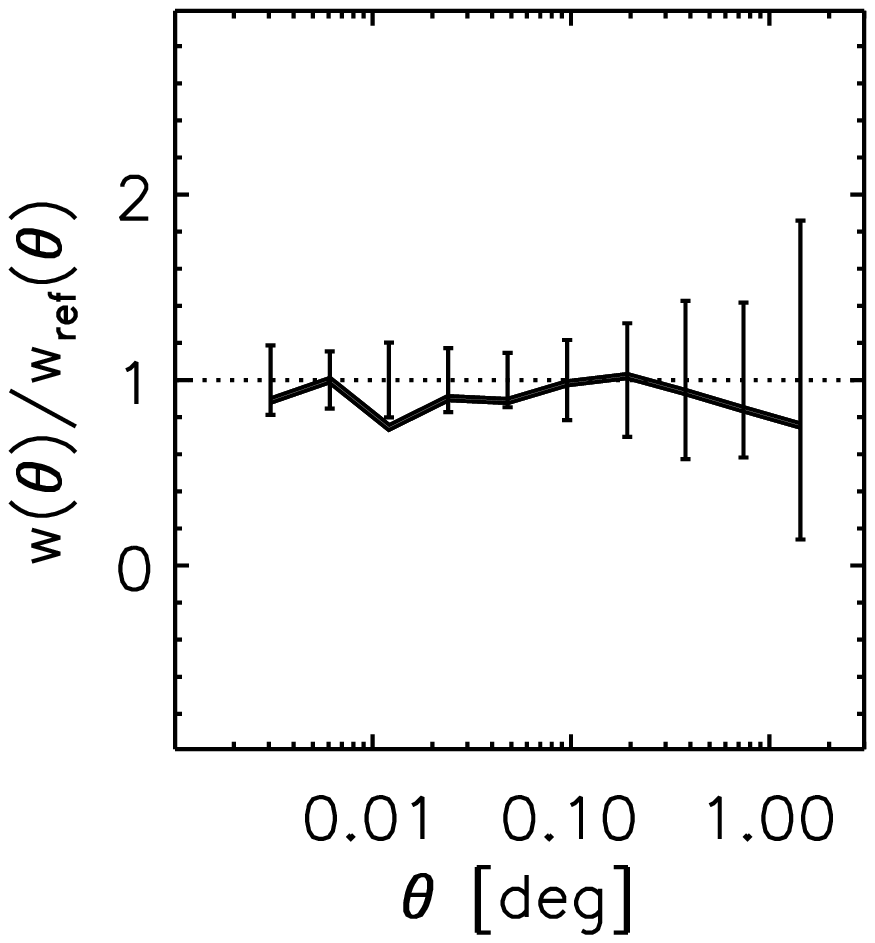}
  \includegraphics[width=0.3\textwidth]{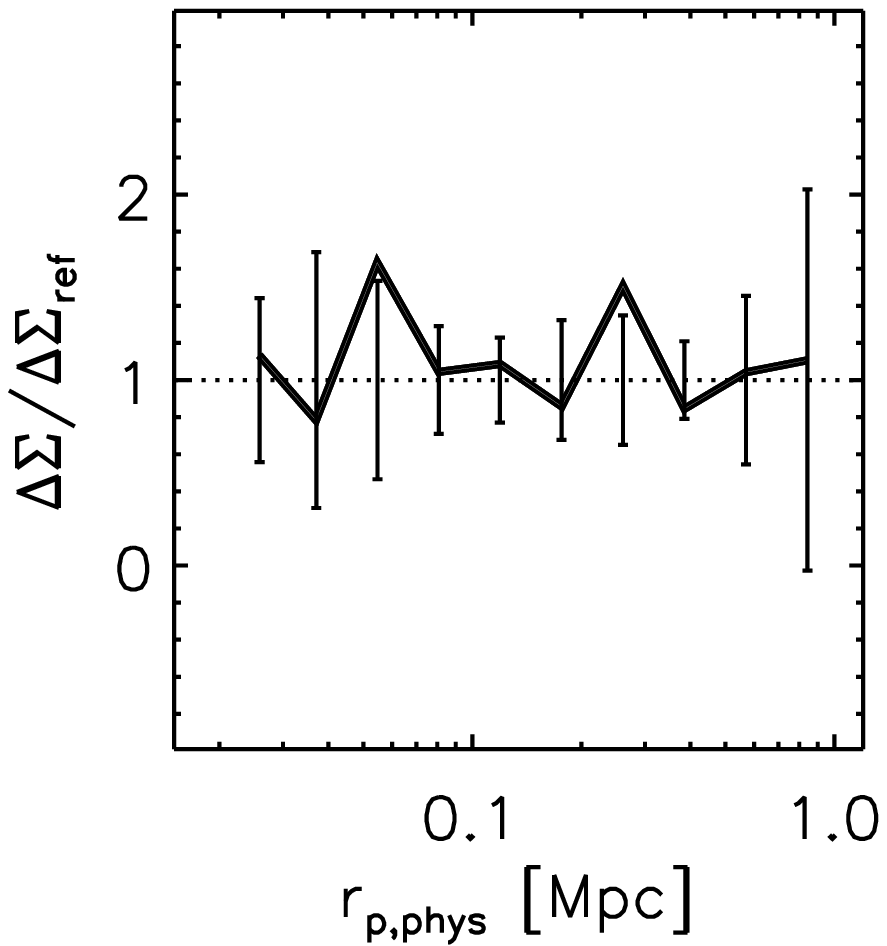}
  \caption{Measurements made with photometric redshifts divided by
    those made with spectroscopic redshifts. From left to right: the
    stellar mass function, the projected clustering and the
    galaxy-galaxy lensing signal, all made with VIPERS galaxies in the
    mass range $10^{10} < \Mstar/\Msun < 10^{12}$ and redshift range
    $0.5 < z < 1$. Error bars represent the statistical error
    estimates from jackknife resampling.}
  \label{fig:photo_z_err}
\end{figure*}

We will see in Section~\ref{sec:model} that our model properly
accounts for these systematic effects caused by photometric redshift
dispersion, through the parameterisation of the stellar mass
scatter. However, catastrophic failures and photometric redshift
biases may be more problematic. We have demonstrated in
Section~\ref{sec:photo_z} that our catastrophic error rate was not
higher than 4\%, and based on results from Section~3.2 of
\cite{Coupon:2012aa}, such a low contamination rate should have no
impact on clustering results at our statistical error level. To check
this statement on the calibration sample (which means the conclusions
are limited to the photometric sample with similar properties to the
spectroscopic sample), we use the VIPERS galaxies with spectroscopic
redshift and re-compute all stellar masses, as well as each
observable, using the corresponding photometric redshift. We show the
measurements in Fig.~\ref{fig:photo_z_err} (solid lines) divided by
the reference measurement made with spectroscopic redshifts and where
the error bars are from the statistical jackknife estimator.  From
left to right we display the results for the stellar mass function,
the projected clustering, and the galaxy-galaxy lensing signal, all in
the mass range $10^{10} < \Mstar/\Msun < 10^{12}$ and redshift range
$0.5 < z < 1$.
%

We conclude that for galaxies with similar properties to VIPERS
galaxies, none of the observables measured with photometric redshifts
display a large bias with respect to the spectroscopic redshift
ones. This represents a reassuring sanity check for the calibration
procedure.  Only the projected clustering presents a slightly low
systematic value, expected from the dispersion of redshifts and
accounted for in the model, through the projection of the modelled 3D
clustering on the redshift distribution constructed from the sum of
photometric redshift PDFs (assuming that estimated PDFs are
representative of the true PDFs).

\section{Model and fitting procedure}
\label{sec:model}

We use the HOD formalism to connect galaxy properties to dark matter
halo masses. Here we assume that the number of galaxies per halo is
solely a function of halo mass, split into central and satellite
contributions. The fitting procedure then consists of finding a set of
parameters to describe the HOD that best reproduces the observables.

A key ingredient of the HOD model is the statistical description of
the spatial distribution of dark matter.  We assume that the matter
power spectrum, the halo mass function, and the dark matter halo
profile are all known quantities over the scales and redshift range
($0.5 < z< 1$) explored in this study. All the technical details about
the halo model are given in Appendix A.1 of \cite{Coupon:2012aa}, with
the exception of the large scale halo bias, for which we use in this
study the fitting formula proposed by \cite{Tinker:2010fz}.

The exact way to parameterise the HOD is often at the origin of the
differences between HOD studies in the literature. In this paper we
follow \cite{Leauthaud:2011bj} who adopted two advanced features:
\begin{itemize}
\item the HOD is a conditional function of the stellar mass given the
  halo mass \citep[this formalism is an extension of the conditional
  luminosity function technique developed by][]{Yang:2003aa}.  In this
  formalism, the central galaxy $\Mstar - \Mh$ relationship is a
  parameterised function representing the mean stellar mass given its
  host halo mass, $\langle\Mstar | \Mh\rangle$;
\item all observables, namely the stellar mass function, the projected
  clustering, the real-space clustering, and the galaxy-galaxy lensing
  signal are fitted jointly.
\end{itemize}

\subsection{The stellar-to-halo mass relationship}

To describe the central galaxy $\Mstar - \Mh$ relationship, we adopt
the parameterised function $f_{\rm SM-HM}$ proposed by
\cite{Behroozi:2010ja}, and defined via its inverse:
\begin{eqnarray}
  & &\log_{10}(f_{\rm SM-HM}^{-1}) =  \log_{10}(\Mh(\Mstar)) = 
  \\
  \label{eq:shmr}
  & & \log_{10} (M_1)+ \beta\,\log_{10}\left(\frac{\Mstar}{\Mstaro}\right) +
  \frac{\left(\frac{\Mstar}{\Mstaro}\right)^\delta}{1 +
    \left(\frac{\Mstar}{\Mstaro}\right)^{-\gamma}} - \frac{1}{2}
  \, . \nonumber
\end{eqnarray}
$M_1$ controls the scaling of the relation along the halo mass
coordinate, whereas $\Mstar_{,0}$ controls the stellar mass
scaling. $\beta$, $\delta$ and $\gamma$ control the low-mass,
high-mass, and curvature of the relation, respectively.

\subsection{The central occupation function}
\label{sec:HOD:cen}

For central galaxies contained in a threshold sample ($\Mstar >
\Mstart{\rm t}$), the HOD is defined as a monotonic function
increasing from 0 to 1, with a smooth transition centered on the halo
mass value corresponding to $\Mstart{\rm t}$ :
\begin{eqnarray}
  \label{eq:ncen}
  && \langle N_{\rm cen}(M_{\rm h}|\Mstart{\rm t}) \rangle = \\
  &&\frac{1}{2}\left[
    1-\mbox{erf}\left(\frac{\log_{10}(\Mstart{\rm t})) -
        \log_{10}(f_{\rm SM-HM}(\Mh)) }{\sqrt{2}\sigma_{\log M_\star} (\Mstart{\rm t})} \right)\right] \, .
  \nonumber
\end{eqnarray}
The parameter $\sigma_{\log M_\star}$, expresses the scatter in
stellar mass at fixed halo mass, which we parameterise as:
\begin{equation}
  \label{eq:sigma_log_m}
  \sigma_{\log M_\star} (\Mstart{\rm t}) = \sigma_{\log M_\star, 0}
  \left ( \frac{\Mstart{\rm t}}{10^{10}\Msun} \right )^{-\lambda}\, ,
\end{equation}
to account for the change in intrinsic stellar mass dispersion as a
function of stellar mass.

\subsection{The satellite occupation function}
\label{sec:HOD:sat}

We describe the satellite HOD for a threshold sample $\Mstart{\rm t}$
with a simple power law as a function of halo mass $\Mh$:
\begin{equation}
  \langle N_{\rm sat}(\Mh|\Mstart{\rm t}) \rangle = \left (
    \frac{\Mh - M_{\rm cut}}{M_{\rm sat}}
  \right )^\alpha \, ,
\end{equation}
for which we fix the cut-off mass scale $M_{\rm cut}$ such that
\begin{equation}
  \label{mcut_eq}
  M_{\rm cut} = f^{-1}_{\rm SM-HM}(\Mstart{\rm t})^{-0.5} \, .
\end{equation}
This assumption is based upon the values reported by
\cite{Coupon:2012aa} for their equivalent parameter ``$M_{0}$''.  We
have checked that the exact parameterisation of $M_{\rm cut} $ had
very little importance compared to the other parameters and did not
change any of our conclusions, in agreement with the loose constraints
observed by \cite{Coupon:2012aa}.

As in \cite{Leauthaud:2011bj}, the normalisation $M_{\rm sat}$ of the
satellite HOD follows the halo mass scaling driven by the central
$\Mstar - \Mh$ relationship, with some degree of freedom controlled by
a power law:
\begin{equation}
  \frac{M_{\rm sat}}{10^{12} M_{\odot}}= B_{\rm sat} \left(\frac{
      f^{-1}_{\rm SM-HM}(\Mstart{\rm t})
    }{10^{12} M_{\odot}}\right)^{\beta_{\rm sat}} \, .
\end{equation}

\subsection{Total occupation functions and observables}

Finally, the total HOD is
\begin{eqnarray}
  &&\langle N_{\rm tot}(\Mh|\Mstart{\rm t}) \rangle =  \nonumber\\
  && 
  \langle N_{\rm
    cen}(\Mh|\Mstart{\rm t}) \rangle + \langle N_{\rm sat}(M_{\rm
    h}|\Mstart{\rm t}) \rangle \, ,
\end{eqnarray}
and since our measurements are made in bins of stellar mass, we
transform the threshold HOD functions into binned functions by
writing:
\begin{eqnarray}
  & & \langle N_{\rm tot}(\Mh|\Mstart{\rm t_1},\Mstart{\rm t_2}) \rangle
  = \nonumber\\
  & &\langle
  N_{\rm tot}(\Mh|\Mstart{\rm t_1})\rangle-\langle N_{\rm
    tot}(\Mh|\Mstart{\rm t_2}\rangle \, .
\end{eqnarray}
Equivalent relations hold for central and satellite binned HODs.

The stellar mass function, the projected two-point correlation
function, the real-space correlation function, and the galaxy-galaxy
lensing signals are computed from the halo model and the HOD as
detailed in Appendix~\ref{sec:details-observables}.
%
\begin{table*}
  \caption{\rs{Estimated systematic errors from the model on the
      central halo mass, $\log_{10} M_1$, and the satellite normalisation, $B_{\rm
        sat}$. The total is the sum in quadrature of the errors.\label{tab:sys_err_mod}.}}
  \centering
  \begin{tabular}{ l c c c c l }
    \hline 
    \hline
    &  \multicolumn{2}{c}{Error on $\log_{10} M_1(\sim12.7)$ } &
    \multicolumn{2}{c}{Error on $B_{\rm sat} (\sim10)$} &  \\
    Assumption &  $\Mstar (\Msun)= 10^{10}$ &  $10^{11.5}$ &  $\Mstar (\Msun) = 10^{10}$ &  $10^{11.5}$ & \multicolumn{1}{c}{Affected quantities} \\
    \hline
   $\sigma_8$ &  0.05 &  0.05 &  1  &  0.5  & SMF, clustering (small \& large scales) \\
    $b(\Mh)$  relation  &  0.08  &  0.1  &  ---  &  --- & clustering  (large scale) \\
    Assembly bias$^1$ &   $<0.04$ & $<0.04$  & $\sim1.5$   & $\sim1.5$ & SMF, clustering  (small \& large scales) \\
    $c(\Mh)$ relation &  0.11  & 0.03  &  0.1 & 0.4   & clustering (small scale), lensing \\
    Satellite concentration &  ---  & --- & 1.1 & 0.9   & clustering (small scale), lensing  \\
    Stripped sub-halos &  0.09 & 0.07  & --- &  ---  & lensing \\
    Total &  0.17 & 0.14  & 2.1 &  1.9  & all \\
    \hline 
    \multicolumn{6}{l}{$^1$ From \protect\cite{Zentner:2014aa}.}
  \end{tabular}
\end{table*}
%

\subsection{Systematic errors in the model}
\label{sec:systematic_errors_model}

\rs{As detailed in the previous sections, the HOD formalism relies on an
accurate description of the dark matter spatial distribution. 
Here we evaluate the impact of our model uncertainties and
assumptions on the best-fit HOD parameters.
Ideally, one would like to
repeat the fitting procedure to test each of the different assumptions of the
model, but to avoid such a time-consuming exercise, 
we take the simple approach of modifying one feature at a time,
and tuning the HOD parameters by hand to reproduce the
modelled quantities derived from the best-fit parameters reported in
Section~\ref{sec:results}. 
We explore two stellar mass bins ($\Mstar = 10^{10}, 10^{11.5}\Msun$)
and we focus on the two parameters $M_1$ and $B_{\rm
  sat}$, controlling the halo-mass scaling of the $\Mstar-\Mh$
relationship, and the normalisation of the satellite HOD, respectively. 
The results are shown in Table~\ref{tab:sys_err_mod}, and we detail
below our calculations for each assumption listed.}

\rs{The power spectrum normalisation parameter, $\sigma_8$, is currently known to a precision of a
few percent. This parameter has a strong impact on the large scale galaxy
clustering, and a larger value
would lead to an increased number of massive
structures, hence shifting the massive end of the halo mass
function towards more massive halos. Choosing Planck over WMAP7
cosmology (as for the tests in Section~\ref{sec:systematic_errors}), would
result in a 5\% increase in $\sigma_8$, leading to relatively small changes
in best-fit HOD parameters, of order of a few percent.}

\rs{Halo bias uncertainties originate from
the measurement of the bias-to-halo mass relation $b(\Mh)$ using
n-body simulations, affected by low-mass resolution,
small volume, or the limitations of halo identification techniques.
In the low-clustering regime, the typical errors on the bias are as small as a few
percent \citep{Tinker:2010fz}, however the rather shallow slope of
bias versus halo mass \cite[see e.g. Fig.~18 of][]{Coupon:2012aa}
translates into a larger uncertainty in the deduced halo mass. In the high-mass regime,
errors are mainly dominated by the sample variance of simulations, up to
$\sim10\%$, but have fewer impact on the deduced halo mass owing to the
steeper slope in this regime.}

\rs{The assembly bias \citep[][and references therein]{Zentner:2014aa}
refers to the correlation between clustering amplitude and time of
halo formation, whereas in our model the bias is assumed to vary only with halo
mass. The effect is stronger when selecting a population of
galaxies based on a parameter correlated with halo formation
history, such as the star formation rate, but moderate when
considering the full galaxy population selected by stellar mass only.
In this case, and in the mass regime explored in this study,
\cite{Zentner:2014aa} found that the systematics caused by assembly
bias on HOD parameters do not exceed $10\%$ to $15\%$.}

\rs{In our model, the dark matter halo profile is assumed to follow a
\cite{Navarro:1997aa} (NFW) profile. While lensing observations tend to
favour NFW profiles \citep{Umetsu:2011fg,Okabe:2013aa,Coupon:2013ke}, the
mass-concentration relation -- driving the slope of the profile -- remains
uncertain. We have
used a simple mass-concentration relation based on theoretical
predictions \citep[updated from][]{Takada:2003aa} and empirical
redshift evolution \citep{Bullock:2001aa}, but more recent relations such
as the work from \cite{Munoz-Cuartas:2010aa} have been measured.
Compared to our concentration values, the difference with
\citeauthor{Munoz-Cuartas:2010aa} rises from 11\% at
$\Mh \sim10^{12} \Msun$ to 30\% at $\sim10^{15} \Msun$ (with a
minimum of 2\% at $\sim10^{13} \Msun$). These systematics affect 
the slope of the small scale clustering and galaxy-galaxy lensing. We
estimate that if all of our constraints came from lensing, this may
result in a 28\% systematic error in $M_1$.}

\rs{We assume that the satellite distribution in the halo follows the dark
matter density profile. However, this assumption may not be always true and
\cite{Budzynski:2012aa} tested this hypothesis from a stacked
analysis of massive clusters from the SDSS. They found a typical factor of 2 (with
$\sim$50\% scatter) lower concentration of the satellite distribution
compared to dark matter, whereas \cite{Muzzin:2007aa} measured a value
closer to dark matter around $z\sim0.3$, and \cite{Burg:2014aa} a relatively
high concentration at $z=1$. These trends may show a redshift evolution of the
concentration or can simply be inherent to the difficulty of observationally measuring the
satellite distribution. In Table~\ref{tab:sys_err_mod} we report
the impact on $B_{\rm sat}$ after setting the satellite concentration
a factor two higher than that of dark matter. The effect on $B_{\rm sat}$ does
not exceed $11\%$.}

\rs{Finally, in our model we neglect the lensing contribution of the sub-halos
hosting the satellite galaxies. This effect, first introduced
by \cite{Mandelbaum:2005ab} under the term ``stripped satellite
central profile'', assumes that a fraction of the satellite halos
survive inside the host halo and further contribute to the lensing
signal at small scales.  As a result, the lensing contribution of
those sub-halos adds up to the central-galaxy halo term in such a way
that the best-fit \emph{host} halo mass gets reduced compared to a
model in which the contribution of sub-halos is neglected.
\cite{Hudson:2015aa} quantify the 
systematic change in best-fit halo mass 
as a systematic decrease by a factor of $\sim(1+f_{\rm sat})$, where $f_{\rm sat}$ is the fraction
of satellites in the sample. Assuming a satellite fraction between $20\%$
and $30\%$, this leads to a systematic error of up to 0.09 in
$\log_{10} M_{1}$. This number must be read as if all the constraints would come from
lensing only. In our study where the stellar mass function and the
clustering signal-to-noise ratio is higher than that of the lensing,
this effect plays relatively little
role, and our results would not significantly change if we
accounted for it.}

\rs{The sum in quadrature of these model systematics is shown 
  as ``Total'' in Table~\ref{tab:sys_err_mod}. Intermediate-stellar mass bins
  ($\sim10^{10} \Msun$) seem to be most affected, with an error of
  0.17 for $\log_{10} M_{1}$ ($\sim50\%$ in $M_1$) and 2.1 ($\sim20\%$) for $B_{\rm
    sat}$. We will see below that these values dominate over the typical
  statistical and systematic errors from the measurements in this mass
  regime. However, as each of these systematic errors affects the
  observables in a different way and we fit all of them jointly, one
  must see these numbers as pessimistic estimates.  
  The high-stellar mass bin ($\sim10^{11.5} \Msun$) is equally affected
  but in a regime where statistical errors are large, hence
  leading to a smaller impact.}

\subsection{MCMC sampling}
\label{sec:mcmc}

We write the combined log-likelihood as the sum of each observable
$\chi^2$:
\begin{equation}
  - 2\,\ln \mathcal{L} = \chi^2_{\phi}  + \sum_{\rm spl} \chi^2_{w(\theta)} +
  \sum_{\rm spl} \chi^2_{w_p(r_{\rm p})} + \sum_{\rm spl}  \chi^2_{\Delta \Sigma} \,,
\end{equation}
where individual $\chi^2$'s are computed as:
\begin{equation}
  \chi^2 = \sum_{i,j}\left[\widetilde{X}_i -
    X_i\right]\left(C^{-1}\right)_{ij}\left[\widetilde{X}_j -
    X_j\right] \,,
\end{equation}
using the covariance matrices evaluated for each measurement as
described in Section~\ref{sec:measurements} ($\widetilde{X}$ and $X$
represent the measured and modelled observables, respectively). 
\rs{Each observable $\chi^2$ is summed over the samples (``spl'')
as described in Table~\ref{tab:samples}. The ``i'' and ``j''
subscripts refer to the stellar mass (stellar mass function) or
transverse separation (clustering and lensing) binning of each
measurement.}
 
We find the best-fit parameters and posterior distribution (assuming
flat priors for all parameters) employing the Markov Chain Monte Carlo
(MCMC) sampling technique with the Metropolis-Hastings sampler from
the software suite \texttt{Cosmopmc} \citep{Wraith:2009jg}.  We check
for individual chain convergence and chain-to-chain mixing using the
\cite{Gelman:1992aa} rule from the R-language CODA
package\footnote{http://cran.r-project.org/web/packages/coda/citation.html}.
We find a typical chain-to-chain mixing coefficient (potential scale
reduction factor) to be equal to 1.01, and the acceptance rate around
30\%.

In practice, we first evaluate a diagonal Fisher matrix at the maximum
likelihood point found using the Amoeba algorithm \citep{Press:2002aa}
and run 10 chains in parallel with the inverse Fisher matrix as the
MCMC sampler covariance matrix. The acceptance rate is usually very
low due to the noisy diagonal Fisher matrix affected by some strong
correlations between parameters.  Once the chains have converged
(after typically $5\,000-10\,000$ steps) we compute the final
likelihood covariance matrix after rejecting the burn-in phase of the
chains (a few thousand steps). This covariance matrix is used as the
input sampler covariance matrix of a second and final MCMC run, in
which 10 chains of $30\,000$ steps each are computed in parallel and
combined together assuming a burn-in phase of $2\,000$ steps and
checking for proper mixing.

We run the full MCMC procedure twice. The first run is performed using
the statistical covariance matrices from the jackknife estimator and
the second MCMC run uses the total error covariance matrices, which
are constructed from the statistical covariance matrices plus the
systematic error estimates added in quadrature to the diagonal, as
described in Section~\ref{sec:systematic_errors}.
%
\begin{table}
  \caption{HOD best-fit parameters and 68\% confidence
    limits (CL) for the statistical errors (top) and total 
    errors (bottom).\label{tab:results}}
  \centering
  \begin{tabular}{ l c c c }
    \multicolumn{4}{c}{Jackknife resampling errors}\\
    \hline 
    \hline
    Parameter & mean &  upper CL & lower CL \\
    \hline 
    $\log_{10} M_1$ & 12.84 &  0.020 &  $-$0.026\\ 
    $\log_{10} \Mstaro$ & 10.98 &  0.015 &  $-$0.019\\ 
    $\beta$ & 0.48 &  0.017 &  $-$0.021\\ 
    $\delta$ &0.63 &  0.094 &  $-$0.073\\ 
    $\gamma$ &1.60 &  0.166 &  $-$0.202\\ 
    $\sigma_{\log M_{\star},0}$ &0.337 &  0.045 &  $-$0.035\\ 
    $\lambda$ &0.21 &  0.047 &  $-$0.044\\  
    $B_{\rm sat}$ & 10.87 &  0.443 &  $-$0.416\\ 
    $\beta_{\rm sat}$ &0.83 &  0.038 &  $-$0.035\\ 
    $\alpha$ &1.17 &  0.020 &  $-$0.021\\ 
    \hline
    \\
  \end{tabular}
  \\
  \begin{tabular}{ l c c c }
    \multicolumn{4}{c}{Total errors}\\
    \hline 
    \hline 
    Parameter & mean &  upper CL & lower CL \\
    \hline 
      $\log_{10} M_1$ & 12.67 &  0.124 &  $-$0.083\\ 
    $\log_{10} \Mstaro$ & 10.90 &  0.082 &  $-$0.067\\ 
    $\beta$ & 0.36 &  0.077 &  $-$0.051\\ 
    $\delta$ &  0.75 &  0.193 &  $-$0.151\\ 
    $\gamma$ &0.81 &  0.477 &  $-$0.386\\ 
    $\sigma_{\log M_{\star},0}$ &0.394 &  0.100 &  $-$0.074\\ 
    $\lambda$ & 0.25 &  0.082 &  $-$0.083\\ 
    $B_{\rm sat}$ & 9.96 &  0.938 &  $-$0.845\\ 
    $\beta_{\rm sat}$ &0.87 &  0.078 &  $-$0.065\\ 
    $\alpha$ & 1.14 &  0.040 &  $-$0.038 \\
    \hline 
  \end{tabular}
%
\end{table}

\section{Results}
\label{sec:results}

Best-fit parameters with 68\% confidence intervals are given in the
top panel of Table~\ref{tab:results} for the statistical- and
total-error MCMC runs. \rs{The 1D and 2D likelihood distributions are
shown in Fig.~\ref{fig:contours}}. The reduced $\chi_{\nu}^2$ for the
statistical-error fit is $\chi^2/(N_{\rm points} - N_{\rm parameters})
= 260/(160-10) = 1.7$, which is an overestimate given the correlations
neglected in the computation of the log-likelihood. \rs{Firstly, we recall that the
lensing and clustering measurements are affected by a sample-to-sample
correlation due to the scatter in stellar mass. The re-use of
background galaxies in the lensing measurements causes an
additional sample-to-sample correlation. Secondly, the projected and
real-space clustering are correlated, as both
observables bring similar information. This mostly affects the satellite
distribution parameter errors, which could be slightly underestimated.
Finally,  the few number of sub-samples (64) used in the computation
of a noisy covariance matrix may
have biased the inverse estimate and contributed to an increase in $\chi_{\nu}^2$.}

\subsection{Measurements and best-fit models}

\begin{figure}
  \centering
  \includegraphics[width=0.49\textwidth]{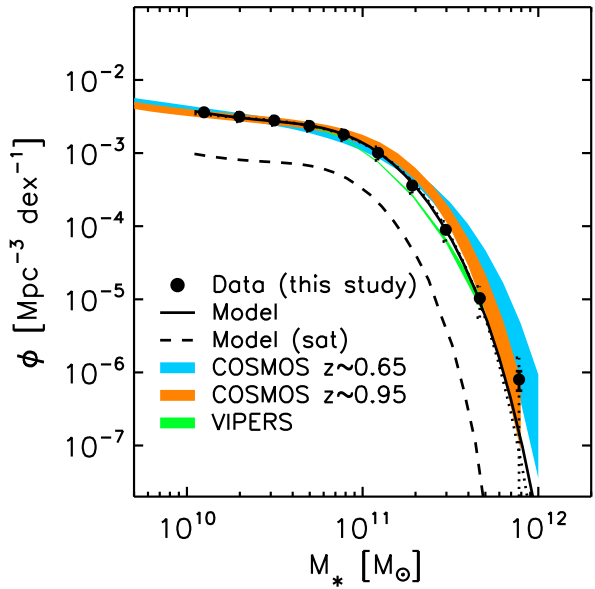}
  \caption{Measured stellar mass function and best-fit model in the
    range $0.5 < z < 1$. The statistical errors from the jackknife
    estimate are shown as black thick lines, whereas the total
    (statistical plus systematic) error bars as dotted lines. The
    COSMOS \citep{Ilbert:2013dq} and VIPERS \citep{Davidzon:2013aa}
    mass functions are displayed with their respective statistical
    errors as shaded areas.}
  \label{fig:smf_fig}
\end{figure}

The measured stellar mass function and best-fit model are displayed in
Fig.~\ref{fig:smf_fig}. Statistical error bars and corresponding
best-fit model are shown as thick black lines, whereas total
(statistical plus systematic) errors and corresponding best-fit model
are represented in dotted lines.
%
%
We compare our measurements with the COSMOS mass function evaluated in
the ranges $0.5 < z < 0.8$ and $0.8 < z < 1.1$ by
\cite{Ilbert:2013dq}, and the VIPERS stellar mass function
\citep{Davidzon:2013aa}, measured in the range $0.5 < z < 1$
(I. Davidzon, private communication).

\begin{figure*}
  \centering
  \includegraphics[width=0.9\textwidth]{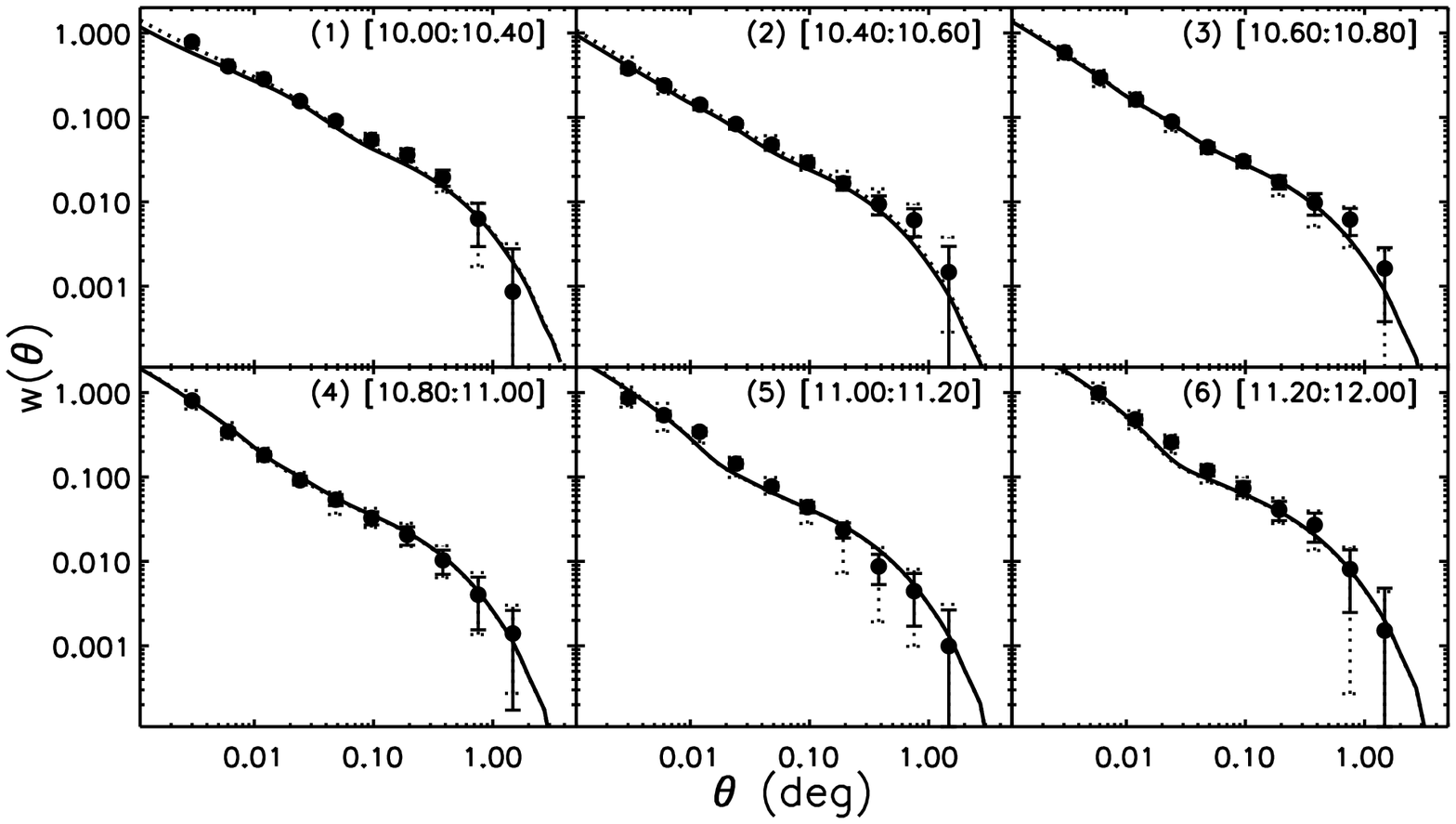}
  \includegraphics[width=0.9\textwidth]{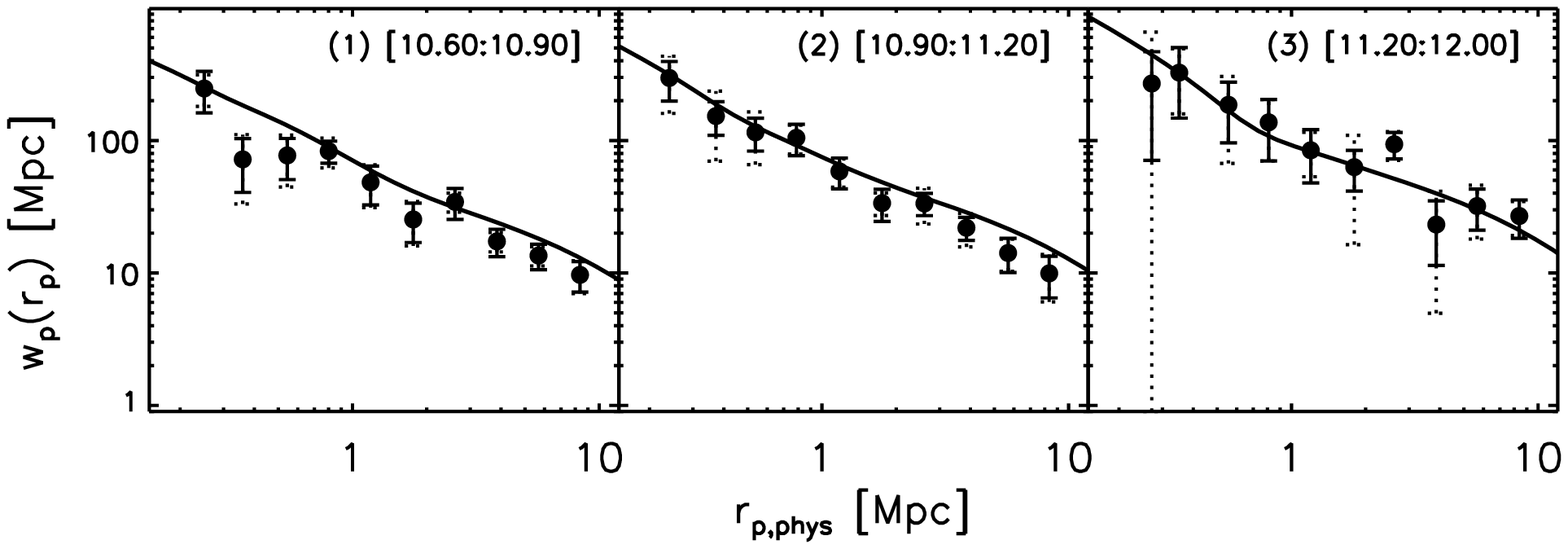}
  \caption{Galaxy clustering measurements (data points with error
    bars) and best-fit models (thick lines). The top panels show the
    projected $w(\theta)$ from the photometric sample (the
    measurements are corrected for the integral constraint), and the
    bottom panels show the spectroscopic real-space $w_p(r_{\rm p})$. The
    thick error bars associated with thick lines represent the
    statistical errors and subsequent best-fit models, whereas dotted
    lines are for total errors.  The mass ranges in the top right
    corner of each panel are given in $\log (\Mstar / \Msun)$.}
  \label{fig:clust_fig}
\end{figure*}

The clustering measurements and best-fit models are shown in
Fig.~\ref{fig:clust_fig}. The projected two-point correlation
functions $w(\theta)$ are displayed in the top panels.
%
%
The mass ranges are given in each top-right panel corner in units of
$\log (\Mstar / \Msun)$. Similarly, the real-space two-point
correlation functions $w_(r_{\rm p})$ are displayed in the bottom panels.

\begin{figure*}
  \centering
  \includegraphics[width=0.9\textwidth]{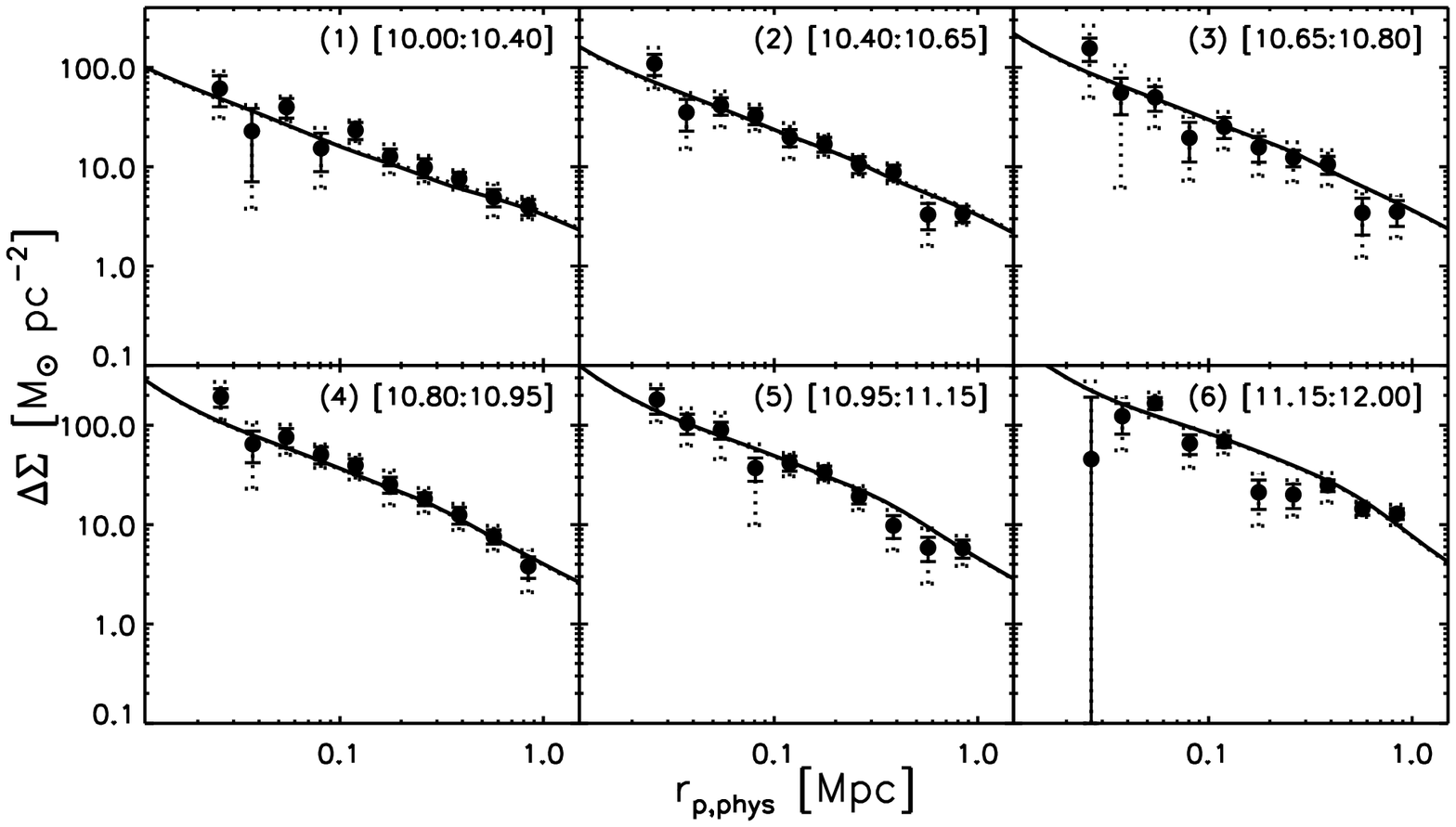}
  \caption{Galaxy-galaxy lensing signal measurements (data points with
    error bars) and best-fit models (thick lines). As in
    Fig.~\ref{fig:clust_fig}, thick and dotted lines are for
    statistical and total error results, respectively.  The mass
    ranges in the top right corner of each panel are given in $\log
    (\Mstar / \Msun)$.}
  \label{fig:ggl_fig}
\end{figure*}

The galaxy-galaxy lensing measurements and best-fit models are shown
in Fig.~\ref{fig:ggl_fig}.
%
%
The most massive lensing bin features a few data points lower than the
model around the transition between the central and the satellites
term. 

For all observables, we report good agreement between the data and the
model.  The constraints on the shape of the central $\Mstar - \Mh$
relationship (parameterised by $\log_{10} M_1$, $\log_{10} M_{\rm
  star0}$, $\beta$, $\gamma$ and $\delta$), are mostly driven by the
high signal-to-noise stellar mass function measurements.  Satellite
HOD parameters ($B_{\rm sat}$, $\beta_{\rm sat}$ and $\alpha$) are
mainly constrained by the clustering and lensing measurements. The
amplitude of clustering at small scale is directly proportional to the
relative number of satellites, hence giving strong leverage on the
satellite galaxy HOD. \rs{Additional information is given on scales
$r\sim0.1$ Mpc from lensing}, through the satellite lensing
signal.  The dispersion in $\Mstar$ at fixed $\Mh$, parameterised in
amplitude by $\sigma_{\log M_{\star},0}$ and in power-law slope by
$\lambda$, is mainly constrained by the high-mass end of the stellar
mass function and the amplitude of the galaxy-galaxy lensing signal in
the most massive bins, resulting in a high-mass ($\Mstar \sim10^{11}
\Msun$) scatter of approximately $\sigma_{\log M_\star} \simeq 0.2$ in
both the jackknife and total error cases, and a medium mass ($\Mstar
\sim10^{10} \Msun$) scatter of $\sigma_{\log M_\star} \simeq 0.35$.

Because the stellar mass function is most affected by the inclusion of
systematics in the error budget, we note a significant increase in
uncertainties associated with the parameters driving the central
$\Mstar - \Mh$ relationship.  From Table~\ref{tab:results}, we report
an increase from a factor $\sim3$ in the error in $\gamma$, up to a
factor $\sim6$ in the error in $\log_{10} M_{1}$. HOD parameters
describing the satellite occupation function such as $B_{\rm sat}$,
$\beta$ or $\alpha$ show substantially less sensitivity to the
addition of systematic errors in the error budget (a maximum of factor
$\sim2$ increase is found).  This is explained by the relatively
smaller contribution of systematic versus statistical errors affecting
the clustering and lensing measurements, compared to the stellar mass
function.

The occasional large differences between best-fit parameters from
statistical alone and total errors, seen in Table~\ref{tab:results},
do not lead to significantly different derived quantities, owing to
the strong correlations between parameters.  This is confirmed by the
almost indistinguishable dotted lines and thick lines in
Figs~\ref{fig:smf_fig}--\ref{fig:ggl_fig}, and is most probably a
consequence of having symmetrically added the systematic errors to the
statistical errors.

\subsection{Central $\Mstar - \Mh$ relationship and the SHMR}

\begin{figure*}
  \centering
  \includegraphics[width=0.9\textwidth]{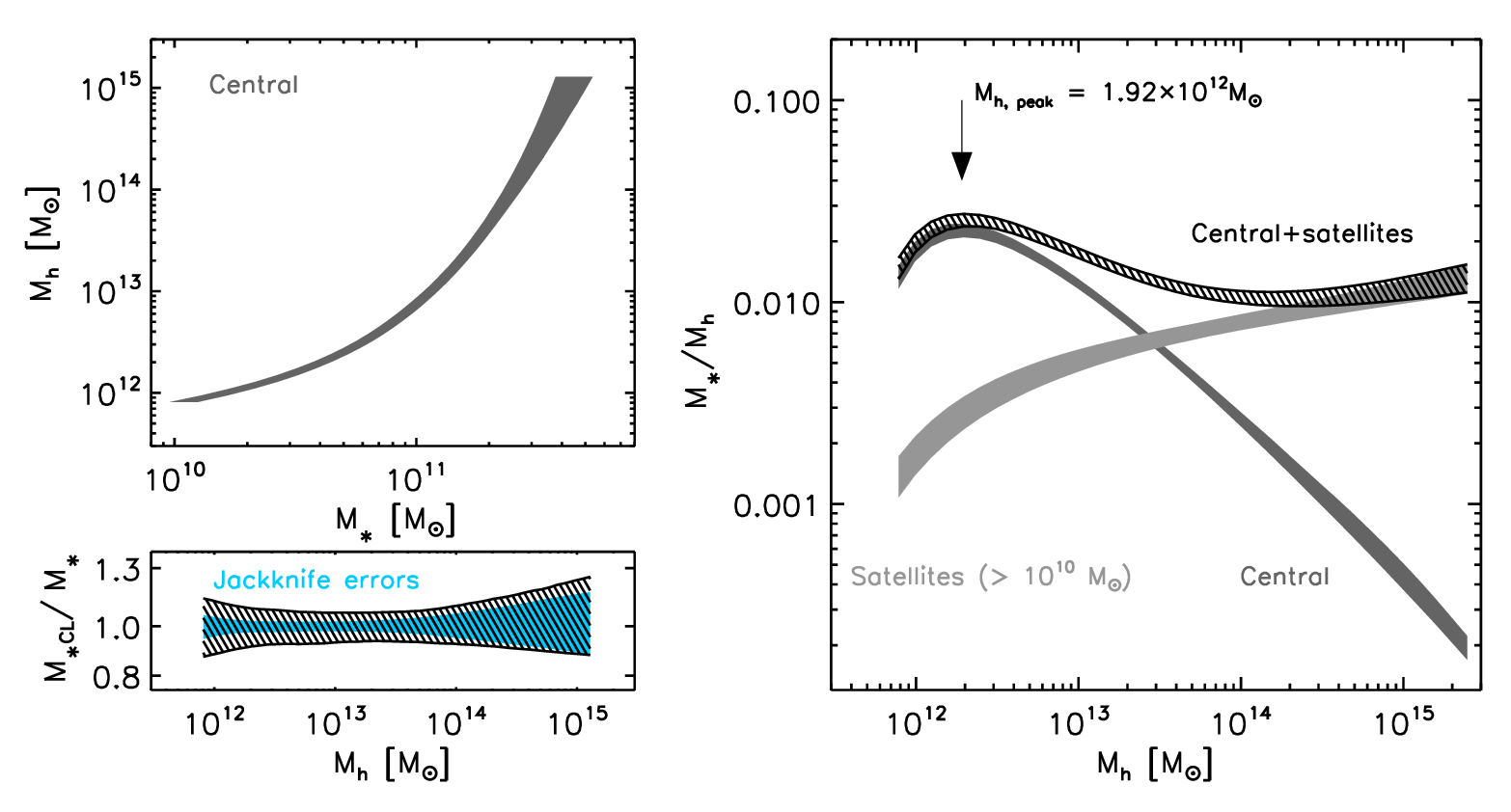}
  \caption{Best-fit $\Mstar-\Mh$ relationship (left) and
    stellar-to-halo-mass ratio (SHMR, right). The black shaded areas
    represent the confidence limits from the total errors. The bottom
    left panel shows the confidence limits interval as a function of
    halo mass in the case of statistical errors (from jackknife
    resampling in light blue) and total errors (in black).  The SHMR
    is derived as function of host halo mass for the central galaxy
    (dark grey), the satellites (light grey) and the sum of both
    (black). The peak value of the central SHMR is indicated by the
    black arrow.}
  \label{fig:shmr_tshmr}
\end{figure*}

In Fig.~\ref{fig:shmr_tshmr} we show the best-fit central galaxy
$\Mstar-\Mh$ relationship (left panel) as parameterised by
Eq.~(\ref{eq:shmr}), and the stellar-to-halo mass ratio (SHMR, right
panel).
%
%
The SHMR is shown as function of host halo mass and is derived for the
central galaxy in dark grey (from the $\Mstar-\Mh$ relationship), the
satellites in light grey (integrated over the galaxies above a mass
threshold of $\Mstar > 10^{10} M_{\odot}$), and the total in black.

The shaded areas represent the 68\% confidence limits, and in the
bottom left panel we have shown the results obtained with statistical
errors in light blue and with total errors in black. As for the
stellar mass function, the statistical uncertainties grow by a factor
$\sim2-4$ in the lower mass regime, when incorporating systematics.

The central SHMR peak position is indicated by a black arrow located
at $M_{\rm h, peak} = 1.92_{-0.14}^{+0.17} \times 10^{12}
M_{\odot}$. The SHMR peak value is ${\rm SHMR}_{\rm peak} = 2.2
_{-0.2}^{+0.2}\times10^{-2}$.  When accounting for satellites, the
peak position and value do not significantly differ from the estimates
for centrals only. However, a remarkable result highlighted in this
figure is the increasing contribution of stellar mass enclosed in
satellites as function of halo mass.  When reaching cluster-size
halos, this contribution reaches over 90\% (and presumably higher when
accounting for satellite galaxies with masses lower than $10^{10}
M_{\odot}$). However we stress that we do not take into account the
intra-cluster light (ICL), which is challenging to quantify using
ground-based photometric data.

\subsection{Comparison with the literature}

In Figs~\ref{fig:shmr_fig_Mstar_Mh} and \ref{fig:shmr_fig_Mh_Mstar} we
compare our best-fit $\Mstar - \Mh$ relationship for central galaxies
with a number of results from the literature. As described in
Section~\ref{sec:model}, our relation describes the mean stellar mass
at fixed halo mass which is, due to the scatter in stellar mass, not
equivalent to the mean halo mass at fixed stellar mass. This issue
becomes particularly important when the slope of the stellar or halo
mass distribution is steep (i.e. at high mass). Therefore, we have
re-computed our results using the latter definition and we
consistently compare our results with the literature in each case.

When required, we convert halo masses to our virial definition using
the recipe given by \cite{Hu:2003aa} in their Appendix~C and,
following \cite{Ilbert:2010ee}, we divide stellar masses by a factor
1.74 and 1.23 to convert from \cite{Salpeter:1955aa} and ``Diet''
Salpeter IMFs, respectively, to our \citeauthor{Chabrier:2003ki} IMF
stellar masses. We apply no correction to \cite{Kroupa:2001aa} IMF
stellar masses.

The mean redshift, measured from the sum of the photometric redshift
PDFs, is found to be $\langle z \rangle=0.82$ for our measurements in
the range $0.5 < z< 1.0$ ($\Mstar > 10^{10.40} \Msun$) and $\langle z
\rangle=0.65$ in the range $0.5 < z< 0.7$ ($10^{10} < \Mstar/\Msun <
10^{10.40}$). We point out that the lensing signal is more sensitive
to lower-redshift lens galaxies characterised by a higher
signal-to-noise (due to the more numerous background sources), and is
likely to be more representative of a lower redshift population, but
this effect is assumed to be small compared to the lensing statistical
errors.

\subsubsection{$\langle \Mstar | \Mh \rangle$ results}

\begin{figure*}
  \centering
  \includegraphics[width=0.9\textwidth]{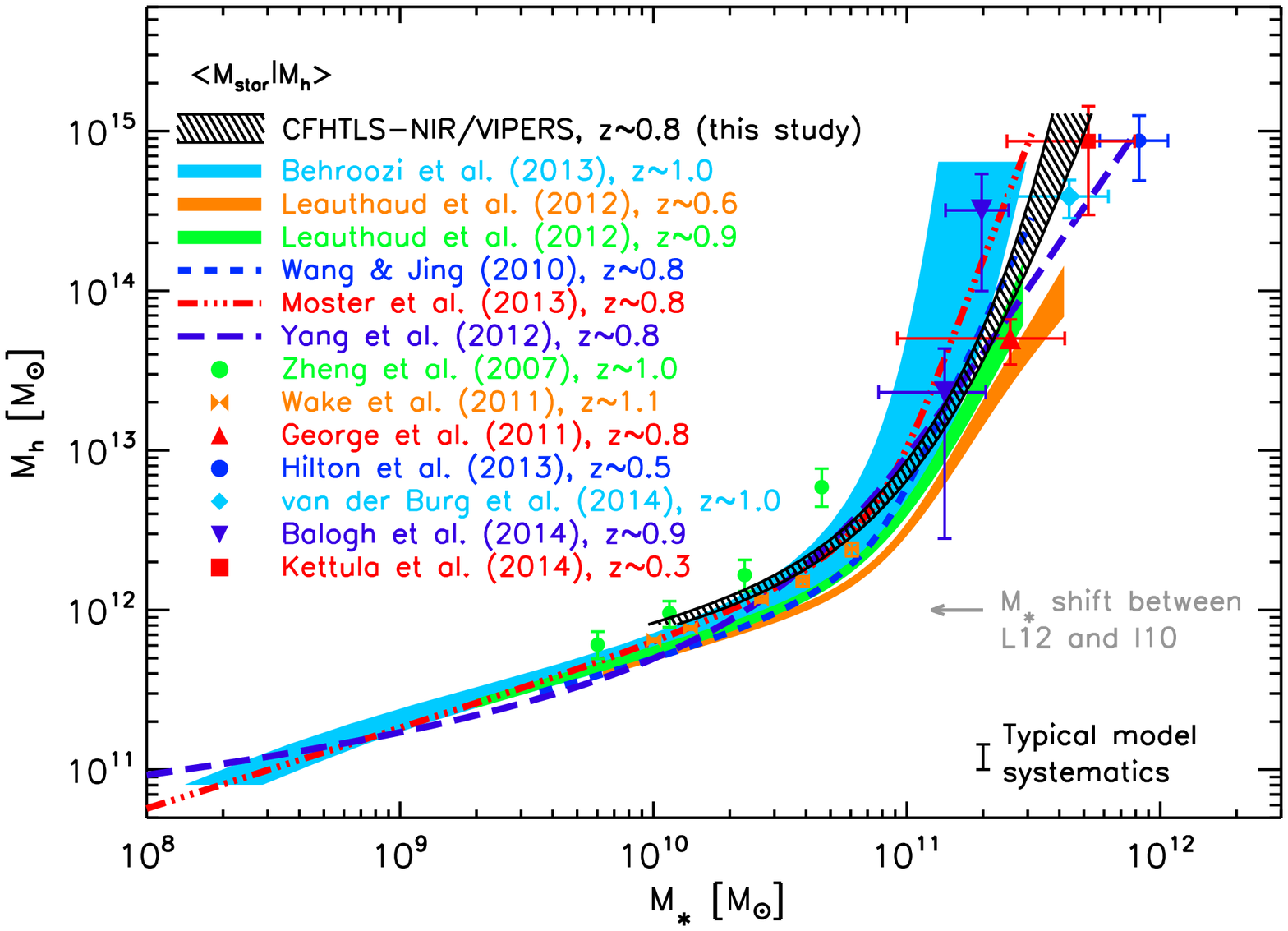}
  \caption{The best-fit $\Mstar - \Mh$ relationship for central
    galaxies, shown in the black shaded area (total-error based 68\%
    confidence limits), compared with a number of results from the
    literature at similar redshifts.  The results shown here represent
    the mean stellar mass at fixed halo mass or halo-mass proxy (X-ray
    temperature or satellite kinematics), $\langle \Mstar |
    \Mh\rangle$, but plotted $\Mh$ as function of $\Mstar$ to ease the
    comparison with the literature.  We perform appropriate halo mass
    conversions and IMF stellar mass corrections when required. The
    length of the grey arrow in the bottom right corner shows the
    shift ($\sim0.2$ dex) measured from the direct comparison between
    stellar masses used in \protect\cite{Leauthaud:2012aa} and
    \protect\cite{George:2011kv}, compared to those in
    \protect\cite{Ilbert:2010ee} which were estimated in a similar way
    to this study. \rs{The error bar on the bottom-right corner indicates the
   typical systematic uncertainty arising from the model.}}
  \label{fig:shmr_fig_Mstar_Mh}
\end{figure*}

We first compare the results for $\langle \Mstar | \Mh \rangle$ in
Fig.~\ref{fig:shmr_fig_Mstar_Mh}.
%
%
The black shaded area shows our results for the central galaxy
relationship with 68\% confidence limits from the total errors. 
The total errors consist of the statistical uncertainties
from jackknife resampling in addition to three sources of systematic
effects \rs{from the measurements}: the cosmology chosen among
widely-used $\Lambda$CDM models,
the fine-tuning of our dust extinction law modelling, and potential
biases in the photometry/calibration.  We recall that this list of
systematic uncertainties is not exhaustive and, for example, ignores
the choice of SPS models, which may be responsible for even larger
systematic effects. \rs{An estimate of the systematic errors from the model,
as detailed in Section~\ref{sec:systematic_errors_model}, is
also shown in the bottom-right corner.}

\cite{Behroozi:2013aa}, shown as the light-blue shaded area, put
constraints on the $\Mstar - \Mh$ relationship by populating dark
matter halos in simulations and comparing abundances using observed
stellar mass functions from a number of surveys. They characterised
the uncertainties affecting stellar mass estimates by accounting for a
number of systematic errors. In particular, unlike in our systematic
errors, the authors had to include uncertainties arising from the
choice of the IMF and the SPS galaxy templates, necessary when
combining the stellar mass functions from several works using
different stellar mass measurement methods.  Here we consider their
results at $z\sim1$. A significant difference with our model resides
in the assumption that satellite galaxies in larger halos are seen as
central galaxies in sub-halos. To circumvent the difficulty of
accurately predicting a sub-halo mass function (e.g. complications
from tidal stripping), the galaxies in sub-halos at the time of
interest are matched to their progenitors at the time of merging onto
the central galaxy halo, under the assumption that the $\Mstar - \Mh$
evolution at a given stellar mass is identical whether the host halo
is isolated or inside a larger halo. In comparison, our model is a
``snapshot'' of the galaxy halo occupation at a given time, where the
satellite distribution is mainly constrained by galaxy clustering.

The results from \cite{Leauthaud:2012aa} in COSMOS are shown in brown
and green at redshifts $z\sim0.6$ and $z\sim0.9$, respectively. We
observe a small discrepancy which, compared to our results, is
unlikely to be explained by differences in the modelling of the HOD
(since the model is essentially identical), nor the sample variance as
confidence limits do not overlap.  A difference in stellar mass
estimates on the other hand is more likely to be at the origin of the
discrepancy. To check this hypothesis, we have compared the stellar
mass estimates from \cite{Ilbert:2010ee}, which were measured in a
similar way to this study, with those used in \cite{Leauthaud:2012aa}
with the method described in \cite{Bundy:2006aa}.  We measured an
offset of $\sim0.2$ dex, illustrated in
Fig.~\ref{fig:shmr_fig_Mstar_Mh} as the grey arrow. Part of the
difference seems to be explained by the separate choice for the dust
extinction law made in each study (which may typically cause a
$\sim0.14$ dex offset, see Section~\ref{sec:mstar}). However we note
that in both cases the same IMF and set of SPS models were used, which
leaves us without a complete understanding of the difference.

The results by \cite{Wang:2010ho} are shown as the blue short-dashed
line. Their model is based on a HOD modelling of the stellar mass
function and real-space galaxy clustering where, as in
\cite{Behroozi:2013aa}, the treatment for satellites is not based on
the distribution of sub-halos in the host halo but on the $\Mstar -
\Mh$ relationship at the time of infall.

\cite{Moster:2013aa}, shown as the red dot-dashed line, also used
abundance matching and provided a redshift-dependent parameterisation
of the central $\Mstar - \Mh$ relationship that we have calculated at
$z=0.8$. As above, the satellites are matched to their halos at the
epoch of merging. Their relation is in good agreement with ours at
intermediate mass, however, it shows a steeper dependence on stellar
masses at higher mass.

The green dots with error bars are from the HOD modelling results of
\cite{Zheng:2007ac}, based on real-space clustering and number density
measurements. Here we show their results for DEEP2, a deep
spectroscopic survey with high density $z=1$ galaxies.  Without deep
NIR data, the authors have computed mean approximate stellar masses
for galaxy samples selected in bins of luminosity. This source of
uncertainty is not shown on the plot, however one may expect a large
scatter and potential biases due to this conversion.

The orange bow-ties with error bars represent the 
\rs{results\footnote{Here we use updated results compared to the
  original publication, estimated with \cite{Bruzual:2003aa} templates
  and with rectified $h$-scaling (D. Wake, private communication)}}
 by \cite{Wake:2011cn} in the NEWFIRM Medium Band Survey at redshift
$z\sim1.1$,
from the combination of NIR-selected galaxy clustering and number
density measurements. Their results are in good agreement with ours.

The five next results were produced using galaxy cluster samples
associated with their brightest cluster galaxies
(BCG). \cite{George:2011kv} built up a catalogue of central versus
satellite galaxies in COSMOS, matched to an X-ray detected
group/cluster sample with robust halo masses from weak lensing
\citep{Leauthaud:2009cg}. From their catalogue we have computed the
mean of stellar mass and halo mass values for clusters in the range
$0.5 < z < 1$, shown as the single red triangle (the error bars show
the standard deviation in halo and stellar masses). As they used
identical stellar masses to \cite{Leauthaud:2012aa}, we also expect a
systematic difference in stellar masses compared with our estimates.

From Sunyaev-Zel'dovich (SZ) detected clusters using the Atacama
Cosmology Telescope, \cite{Hilton:2013eu} presented the measurements
of the galaxy properties between $0.27 < z < 1.07$. Member galaxies
were identified from high density spectroscopic observations, and
stellar masses were measured from Spitzer IRAC1-2 mid-infrared (MIR)
fluxes.  Halo masses were estimated from satellite kinematics. Here we
show the mean halo mass versus mean BCG stellar mass, represented by
the single blue dot with errors bars (standard deviations of both
masses).  Their results appear to be in good agreement with our
$\Mstar - \Mh$ relationship, although our constraints on such
high-mass clusters are extrapolated from the few clusters more massive
than $4-5\times10^{14} \Msun$ expected in our sample.

We show as a single light blue diamond the mean halo mass versus mean
BCG stellar mass from \cite{Burg:2014aa} in the GCLASS/SpARCS cluster
sample at $z\sim1$. Galaxy cluster members were identified from
intensive spectroscopic observations, and halo masses were estimated
from satellite kinematics. We note that stellar masses were measured
from a similar combination of data, redshift range and volume size as
ours, however the methodology used to link halo mass to galaxy stellar
masses was rather different. Thus, the agreement with our high-mass
$\Mstar - \Mh$ relationship within the sample variance is quite
remarkable.

Results from \cite{Balogh:2014aa} are shown as the downward purple
triangles. Halo mass measurements were made using satellite kinematics
for a sample of 11 groups/clusters in the COSMOS field. We show the
mean and standard deviation of their measurements split into two halo
mass bins (the 11 groups are split into 5 and 6 groups below and above
$\Mh=9\times10^{13} \Msun$, respectively). Although their results
suffer from large sample variance, they are in broad agreement with
our results and with the rest of the literature.

Finally, the single red square with error bars shows the mean of halo
mass measurements from a weak lensing analysis of X-ray selected
clusters in the CFHTLenS by \cite{Kettula:2014aa}, versus the mean
stellar mass of associated BCGs \citep{Mirkazemi:2014aa}. We have
re-measured stellar masses of those BCGs in a consistent way to this
study (with the exception of missing NIR data for most of the BCGs,
which may increase the scatter in stellar mass). Despite the lower
redshift range, the identical photometry and lensing catalogue makes
the comparison relevant to our results, where the expected difference
should arise solely from redshift evolution, although the large
statistical uncertainties prevent us from drawing strong conclusions.

\subsubsection{$\langle \Mh | \Mstar \rangle$ results}

\begin{figure*}
  \centering
  \includegraphics[width=0.9\textwidth]{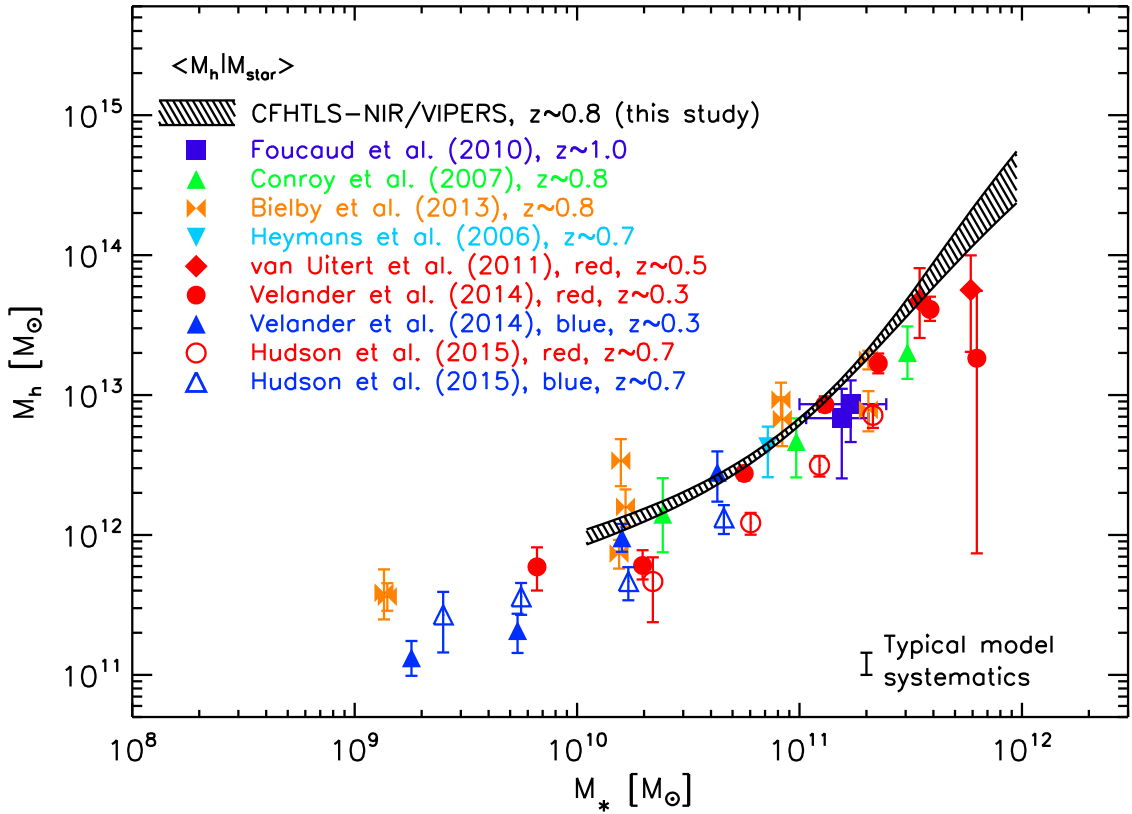}
  \caption{The best-fit $\Mstar - \Mh$ relationship for central
    galaxies, shown in the black shaded area (total-error based 68\%
    confidence limits), compared with a number of results from the
    literature at similar redshifts. Unlike in
    Fig.~\ref{fig:shmr_fig_Mstar_Mh}, the results shown here represent
    the mean \emph{halo mass} at fixed stellar mass $\langle \Mh |
    \Mstar \rangle$.  We perform appropriate halo mass conversions and
    IMF stellar mass corrections when required. The relatively low
    halo masses found by \protect\cite{Hudson:2015aa} is linked to a
    different treatment of the satellite sub-halo contribution to the
    lensing signal at small scale (see text for details).}
  \label{fig:shmr_fig_Mh_Mstar}
\end{figure*}

We compare the results for $\langle \Mh | \Mstar \rangle$ in
Fig.~\ref{fig:shmr_fig_Mh_Mstar}.
%
%
To express the mean halo mass at fixed stellar mass $\langle\Mh|\Mstar
\rangle$ from our results, we derive it from the mean stellar mass at
fixed halo mass $\langle \Mstar | \Mh \rangle$ using the Bayes theorem
relating conditional probability distributions:
\begin{equation}
  P(\Mh|\Mstar) \propto P(\Mstar|\Mh) \times P(\Mh) \, .
\end{equation}
We can then compute $\langle\Mh|\Mstar \rangle$ as the expectation
value of $P(\Mh|\Mstar)$:
\begin{equation}
  \langle\Mh|\Mstar \rangle = \frac{\int  P(\Mstar|\Mh) \, P(\Mh) \Mh \ud \Mh}{\int  P(\Mstar|\Mh) \, P(\Mh) \ud \Mh}
\end{equation}
with
\begin{equation}
  P(\Mstar|\Mh) = \frac{\ud \langle N_{\rm cen}
    (\Mh  | \Mstar) \rangle}{\ud \Mstar} \, ,
\end{equation}
the distribution of central galaxies given a halo mass, and
\begin{equation}
  P(\Mh) =  \frac{\ud n}{\ud \Mh} \, ,
\end{equation}
the halo mass function.

We show the results of \cite{Foucaud:2010gb} at $z\sim1$ from
clustering measurements in the UKIDSS-UDS field as the blue squares
with error bars. The UKIDSS-UDS field is a small patch of $\sim1~{\rm
  deg}^2$ with deep NIR and optical data. They have converted their
clustering amplitude measured in bins of stellar mass into halo
masses, using the analytical galaxy-bias halo-mass relationship from
\cite{Mo:1996aa}. As they do not use any constraints from galaxy
number density, their error bars are dominated by sample variance and
uncertainties on the projected galaxy clustering.

Green upward triangles represent the results by
\cite{Conroy:2007aa}. Halo masses were derived from satellite
kinematics using spectroscopic measurements from the DEEP2
survey. Since the authors have selected their samples based on bins of
stellar masses, we can compare their results with our $\langle \Mh |
\Mstar \rangle$ $\Mstar - \Mh$ relationship. The agreement is found to
be good.

Results from clustering measurements in the CFHTLS-DEEP/WIRDS fields
by \cite{Bielby:2014aa} are displayed by the brown bow-ties with error
bars. We select all mass bin results in the range $0.5 < z < 1$.
Although the total field-of-view is small ($\sim2.4~{\rm deg}^2$), the
combination of four independent fields allowed them to reduce the
cosmic variance. As in \citeauthor{Foucaud:2010gb}, they used an
analytical prescription based on the large-scale clustering amplitude
to estimate halo masses per bin of stellar mass, so that their results
should be compared to our $\langle \Mh | \Mstar \rangle$ results. The
two points well above the other results correspond to the measurements
at $z\sim0.7$ and seem to disagree with our constraints and the rest
of the literature. The authors claim to have observed an unusually
high clustering signal at those redshifts, potentially explained by
cosmic variance effects.

Results by \cite{Heymans:2006ct} in the COMBO-17/GEMS field are shown
as the downward light-blue triangle with error bars. Here we have
picked their unique measurement at $z > 0.5$.  Halo masses were
measured using weak lensing with galaxy shapes from the Hubble Space
Telescope observations.

We show as red diamonds the results for $z\sim0.5$ red galaxies by
\cite{Uitert:2011aa} in the Red Sequence Cluster Survey 2, a
medium-deep CFHT-MegaCam survey in three bands ($gri$) which overlaps
300 deg$^2$ of the SDSS. The authors have measured the galaxy-galaxy
lensing signal for SDSS lens galaxies with a spectroscopic redshift
using background source galaxies from the RCS2 survey. Here the large
area permits a high signal-to-noise measurement for very massive
galaxies from lensing only. Their results are consistent with ours as
this mass bin ($> 3\times10^{11} M_{\odot} $) is dominated by red
galaxies.

We compare our results with those from \cite{Velander:2014aa} at
$z\sim0.3$, shown as filled symbols (red dots and blue triangles for
red and blue galaxies, respectively), and those from
\cite{Hudson:2015aa} at $z\sim0.7$, shown as empty symbols (red dots
and blue triangles for red and blue galaxies, respectively).  In both
studies, halo mass measurements were obtained from galaxy-galaxy
lensing measured using the CFHTLenS lensing catalogue and stellar
masses computed in a similar way to this study\rs{, with the
  exception that, in both cases, no NIR data were available at the time.
  This mostly affects the stellar mass estimates of \citeauthor{Hudson:2015aa} at $z\sim0.7$
 which, unlike \citeauthor{Velander:2014aa} at $z\sim0.3$, do not benefit from the
 leverage of the CFHTLS $z$-band.
We expect the $\Mstar - \Mh$ relationship of the full
galaxy population to lie between those of the red and blue
populations, however the results from 
\citeauthor{Hudson:2015aa} lie below our results for both galaxy
populations. 
The bias caused by the scatter in stellar mass
 partially explains this difference (by shifting their mean stellar mass
 to higher values), but not entirely: \citeauthor{Hudson:2015aa} account for the
contribution of sub-halos around satellites occurring at small scale in
the lensing signal, whereas we do not (see
Section~\ref{sec:systematic_errors_model}\footnote{This point is also investigated in
detail in Appendix~D of \cite{Hudson:2015aa}.}). 
As \citeauthor{Velander:2014aa} also accounted for sub-halos in
their lensing model, we cannot exclude that the
apparent good agreement may result from a redshift evolution going in
the opposite direction, and requires further investigation.}

\subsubsection{The total SHMR}

\begin{figure*}
  \centering
  \includegraphics[width=1.0\textwidth]{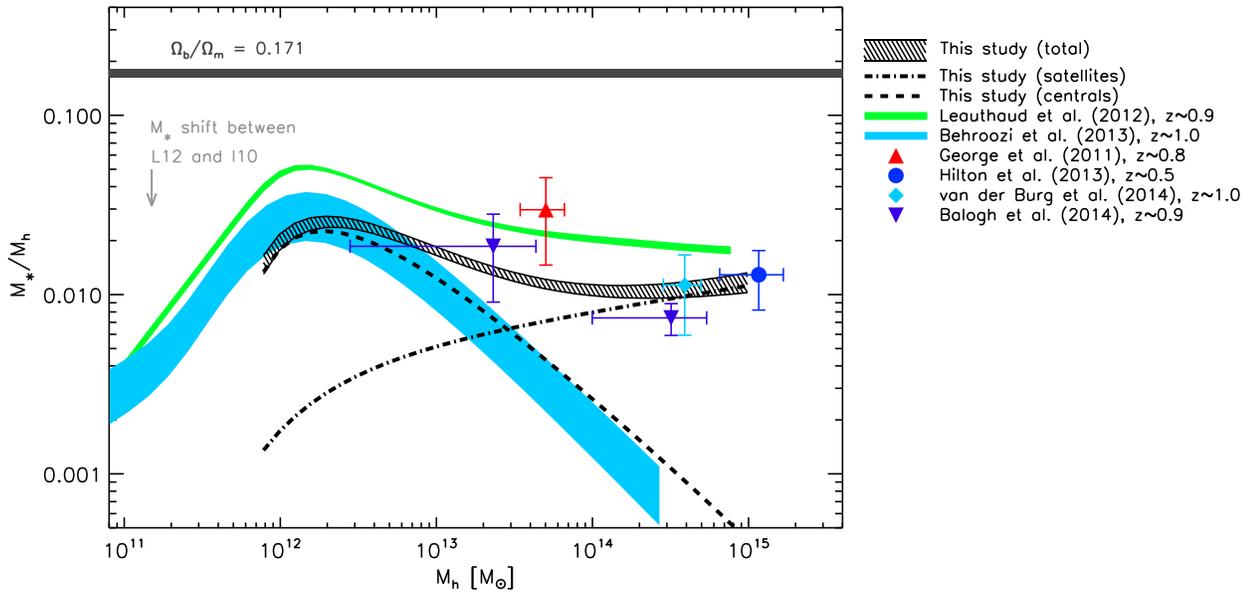}
  \caption{Stellar-to-halo mass ratio (SHMR) as function of halo mass
    compared with observations from the literature. Our best-fit
    result for total (central plus satellites) SHMR is shown as the
    black shaded area. The black dashed line represents the best-fit
    central relationship, whereas the dot-dashed line is for the
    integrated stellar-mass satellite contribution.  For
    \protect\cite{Behroozi:2013fg}, only the central SHMR was
    published and we display it here for comparison with our central
    SHMR and as an illustration of typical stellar mass
    systematics. The length of the grey arrow represents the shift to
    apply to \protect\cite{Leauthaud:2012aa} and
    \protect\cite{George:2011kv} to reconcile their results with ours,
    based on the stellar mass comparison with
    \protect\cite{Ilbert:2010ee}.}
  \label{fig:tshmr_fig}
\end{figure*}

In Fig.~\ref{fig:tshmr_fig} we show the SHMR as function of halo mass
compared with observations from the literature.  The black shaded area
represents the total SHMR as the sum of the central and satellite
contributions. The central SHMR (in dashed line on the figure) is
simply derived from the central $\Mstar - \Mh$ relationship. The
satellite SHMR (in dot-dashed line on the figure) is computed from the
sum of satellite stellar masses over the halo occupation function at
each halo mass, with a lower integration limit of $\Mstar = 10^{10}
\Msun$. The total baryon fraction compared to dark matter in the
Universe is assumed to be 0.171 and represented on the figure by the
grey shaded area on the top \citep[the width of the line represents
the uncertainty]{Dunkley:2009aa}.
%
%

In green we display the total SHMR from \cite{Leauthaud:2012aa}
measured at $z\sim0.9$. The procedure to compute the total SHMR is
identical to ours, i.e. the integrated stellar masses from the
satellite HOD were added to the central stellar mass at each halo
mass. The authors adopted a mass threshold of $10^{9.8} \Msun$, which
does not change the integrated stellar mass from satellites by a large
amount compared with a cut of $> 10^{10} \Msun$.  As shown in
Fig.~\ref{fig:shmr_fig_Mstar_Mh}, part of the vertical shift is
explained by the systematic difference in stellar mass estimates.

We show in light blue the central SHMR from \cite{Behroozi:2013fg}. As
seen in Fig.~\ref{fig:shmr_fig_Mstar_Mh}, the agreement with our
central SHMR is good, although their peak is located at a slightly
lower halo mass value.

The red triangle shows the results by \cite{George:2011kv} in COSMOS
in the redshift range $0.5 < z < 1$.  The point represents the mean
total stellar mass divided by the halo mass versus the halo mass, and
the error bars the standard deviation in each direction. Here we
computed the total stellar mass as the sum of the central galaxy
stellar mass plus the stellar masses of associated group members with
$\Mstar > 10^{10}$. As they used the stellar masses of
\citeauthor{Leauthaud:2012aa}, the agreement is consistently good with
their results, however shifted compared to ours.

The single blue dot with error bars marks the mean and standard
deviation of estimates by \cite{Hilton:2013eu}. Here the total cluster
stellar mass is measured from the background-subtracted sum of galaxy
IRAC fluxes within $R_{500}$ from the BCG. Based on the stellar mass
completeness computed by \cite{Ilbert:2010ee}, a IRAC AB magnitude cut
of $24$ gives a complete passive galaxy sample down to $\Mstar =
10^{9} \Msun$ at $z\sim0.5$. With an IRAC completeness AB magnitude
limit of 22.6, it is therefore safe to assume that
\citeauthor{Hilton:2013eu} are complete above $10^{10} \Msun$ at
$z\sim0.5$, which matches our sample. We then conclude that their
measurements are in good agreement with our results.

Results from \cite{Burg:2014aa} are shown as the single light-blue
diamond, representing the mean SHMR versus halo mass with its standard
deviation. Total stellar masses are computed as the sum of the BCG
stellar mass and the stellar mass from galaxy members
spectroscopically identified and corrected for target sampling rate.
The authors have checked that for $> 10^{10} \Msun$ galaxies, which
contribute the most to the total SHMR (see their Fig.~2), the
spectroscopic success rate reaches 90\%. We note that the median
stellar mass completeness $\sim 10^{10.16} \Msun$ is slightly higher
than ours (limited by their $\Ks$-band data), however the contribution
of satellites compared to a mass limit of $10^{10} \Msun$ will not
significantly change the total SHMR and their measurements can be
fairly compared to our results, and we observe an excellent
agreement. Interestingly, the authors conclude that when comparing
with the literature, no redshift evolution in the total SHMR at high
mass is found below $z\sim1$ and the comparison with our results
($z\sim0.8)$ and those from \citeauthor{Hilton:2013eu} ($z\sim0.5$)
confirm their findings.

The two purple downward triangles represent the results from
\cite{Balogh:2014aa} in the GEEC2 survey in COSMOS.  Here we show the
mean and standard deviation of the SHMR versus halo mass in two halo
mass bins. Galaxy members are identified from the spectroscopic
redshift when available or using the PDF-weighted photometric redshift
computed from the 30-band COSMOS photometric catalogue
\citep{Ilbert:2009aa}. The spectroscopic (photometric) sample is
complete for group members with $\Mstar > 10^{10.3} \Msun$ ($\Mstar >
10^{9} \Msun$). Again, since most of the contribution to the total
SHMR originates from $\Mstar > 10^{10} \Msun$ galaxies, the comparison
with our results is fair. We note a slightly lower value at high mass,
and a good agreement within the error bars at the group-scale halo
mass.

The value of the central SHMR peak may also be compared to that of
\cite{Coupon:2012aa} computed from a clustering and galaxy number
density analysis of the CFHTLS-Wide. In their study the authors have
measured the evolution of the SHMR peak as function of redshift and
have found a lower value compared to ours ($1.1\times10^{12}
M_{\odot}$ at redshift $z\sim0.7$). The difference may not be fully
explained by cosmic variance, firstly because our field significantly
overlaps with the full CFHTLS and secondly because the difference is
larger than our error bars. In fact, due to their selection in the
optical ($i\ab<22.5$), the SHMR peak above $z=0.6$ is much less
constrained than for our $\Ks\ab < 22$ sample, and their peak location
suffers from higher uncertainties than in this study, not properly
accounted for in their published error bars.

\begin{figure*}
  \centering
  \includegraphics[width=0.45\textwidth]{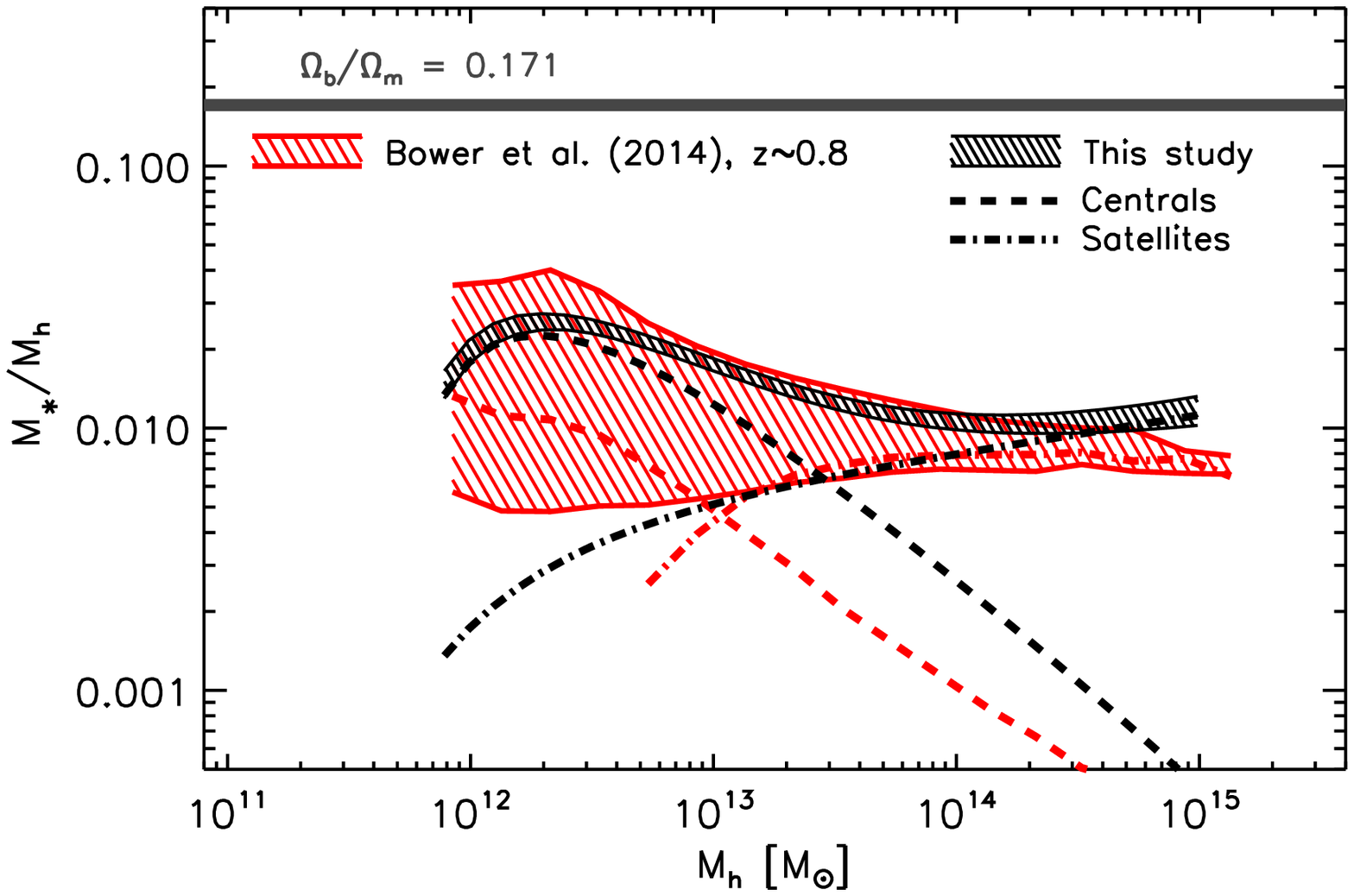}
  \includegraphics[width=0.45\textwidth]{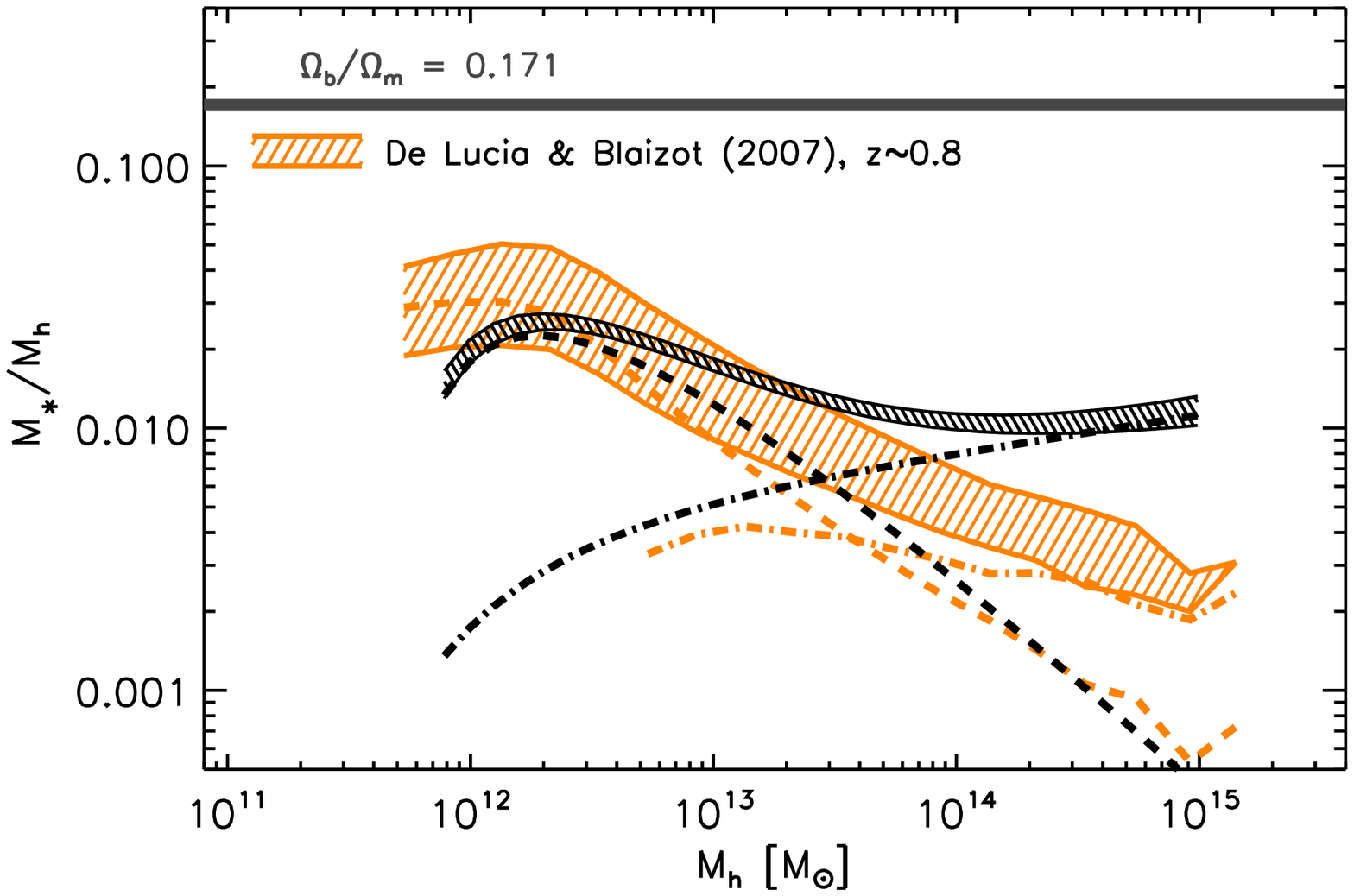}
  \includegraphics[width=0.45\textwidth]{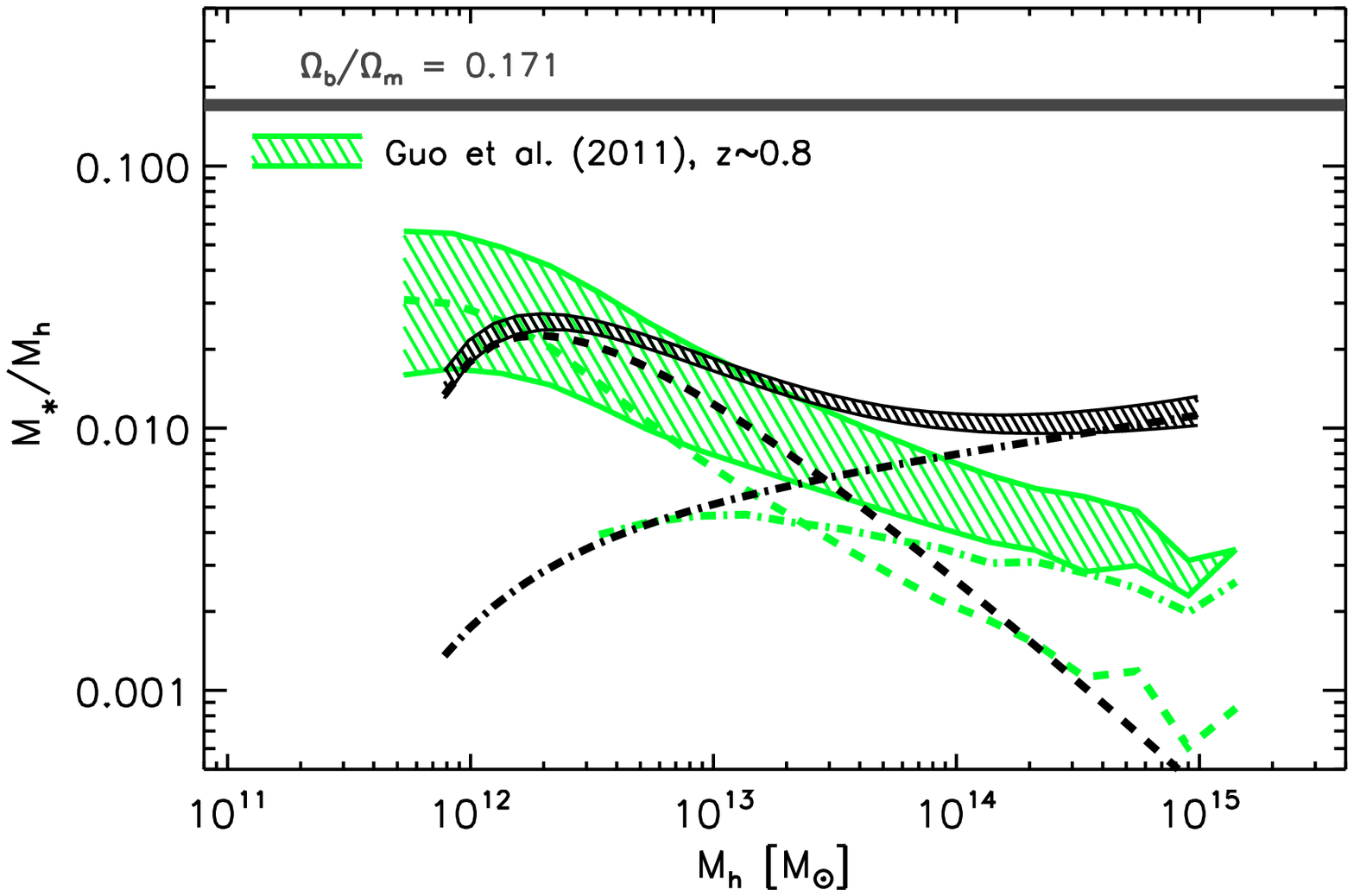}
  \includegraphics[width=0.45\textwidth]{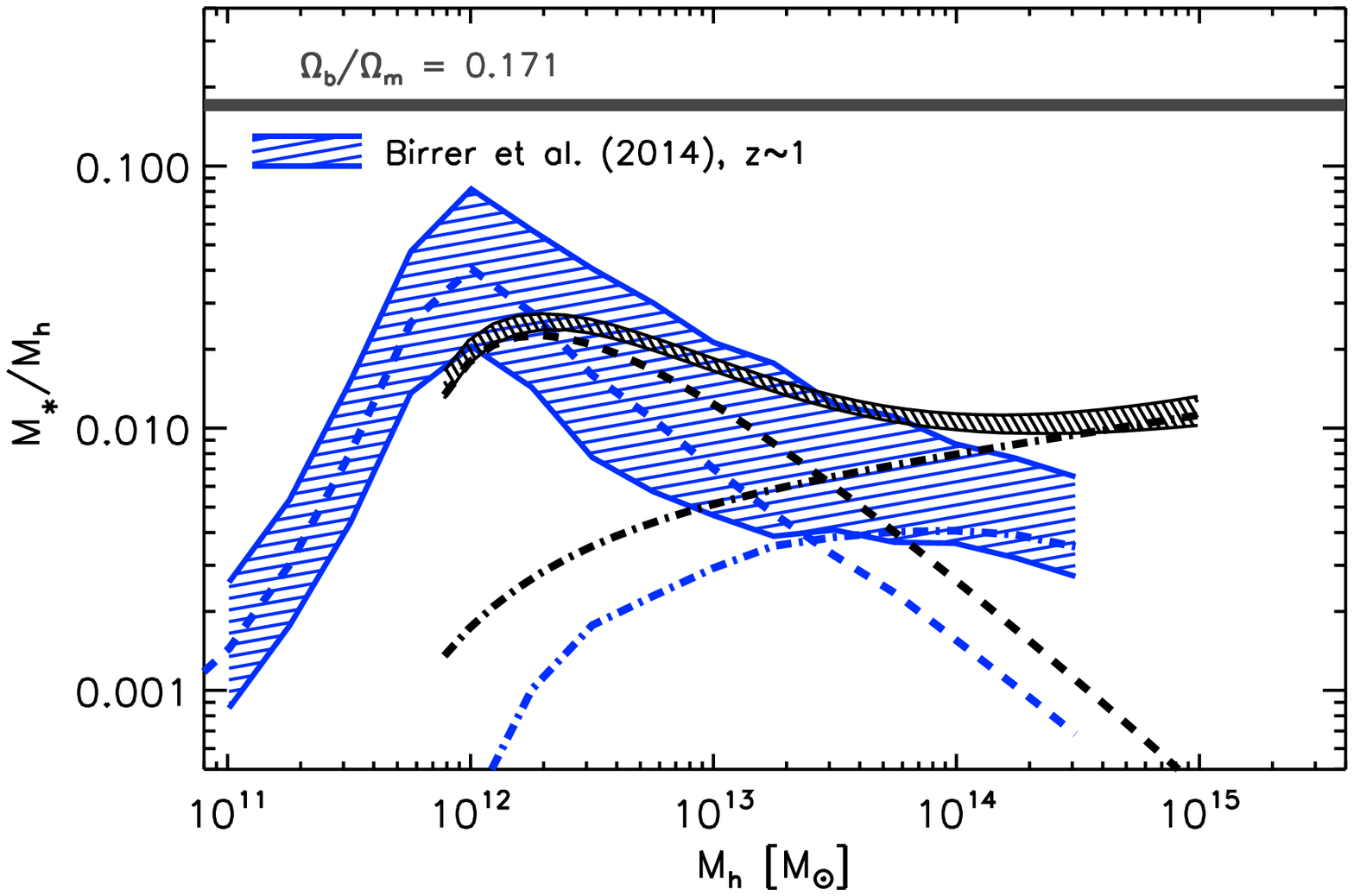}
  \caption{Stellar-to-halo mass ratio (SHMR) as function of halo mass
    compared with simulations from the literature.  We compare our
    total, central and satellite SHMR results with three studies based
    on semi-analytic models applied to the Millennium simulation
    (top-row and bottom-left panels) and one study (bottom right
    panel) based on the ``gas-regulator'' analytical model.  For each
    model we also display the corresponding central (dashed line) and
    satellite (dot-dashed line) SHMR.}
  \label{fig:tshmr_SAM_fig}
\end{figure*}

In Fig.~\ref{fig:tshmr_SAM_fig} we compare our results with a number
of semi-analytic predictions from the Millennium simulation
\citep{Springel:2006fq}.
%
%
In brief, semi-analytic models are anchored to the dark matter halo
merger trees provided by N-body simulations, in which empirical
recipes of physical processes drive the evolution of galaxies. The
fine-tuning of those different processes aim at reproducing the
observed galaxy statistical properties.  In each case, to derive the
total SHMR we compute the sum of the central galaxy stellar mass and
the integrated stellar masses of satellites with $\Mstar > 10^{10}
\Msun$ to match our sample mass completeness limit.  The central SHMR
is represented as a dashed line and the shaded area represents the
total SHMR with 15\% and 85\% percentiles. All quantities were
computed at redshift $z=0.8$. The model of \cite{Bower:2006fj} is
shown in red (top left), the model of \cite{De-Lucia:2007aa} in orange
(top right), and the model of \cite{Guo:2011eb} -- a modified version
of the latter -- in green (bottom right). In both
\citeauthor{De-Lucia:2007aa} and \citeauthor{Guo:2011eb} models, the
contribution from satellites to the total SHMR is significantly below
the observations. Despite a different treatment of satellite galaxies
and the efficiency of stellar feedback in the latter model, compared
to the former, those changes do not show up here. The discrepancy with
our results could not arise from a limitation caused by the simulation
resolution, as we imposed a cut of $\Mstar > 10^{10} \Msun$ to match
our observations.  The model of \citeauthor{Bower:2006fj} better
reproduces the observed satellite SHMR, however it underestimates the
central SHMR and features a significant scatter in the $\Mstar - \Mh$
relationship.

We also show the results from the analytical model proposed by
\cite{Birrer:2014aa} in blue (bottom right).  Their model is an
application of the gas-regulator model \citep{Lilly:2013aa}, in which
the star formation efficiency is driven by the amount of available gas
in the reservoir. In its simplest form, the model describes the
inflows and outflows of the gas in the reservoir by two adjustable
parameters: a star-formation efficiency $\epsilon$, and a mass-loading
factor $\lambda$ that represents the outflows, proportional to the
SFR.  We show their SHMR at $z=1$ from the model ``C''.

\section{Discussion and conclusions}
\label{sec:conclusions}

Using a unique combination of deep optical/NIR data and large area, we
have combined galaxy clustering, lensing and galaxy abundance, to put
constraints on the galaxy occupation function in the range $0.5 < z <
1$ and to link galaxy properties to dark matter halo masses. Our main
result is an accurate measurement of the central galaxy $\Mstar - \Mh$
relationship at $z\sim0.8$ ranging from halo masses at the peak of the
SHMR up to the galaxy cluster mass regime. We also provide separate
measurements of the SHMR for central and satellite galaxies.

We have shown that the statistical errors (computed using a jackknife
estimator) were smaller than systematic errors in the stellar mass
measurements caused by uncertainties in the assumed cosmology, dust
modelling, and photometric calibration. Due to the relatively small
amount of statistical uncertainties, the low- to intermediate-mass
regime of the stellar mass function is most affected by systematic
errors: a factor of $\sim8$ was found between statistical errors and
total errors, increasing the error bars of parameters controlling the
shape of the $\Mstar - \Mh$ relationship by approximately the same
amount (see Table~\ref{tab:results}). Conversely, clustering and
lensing measurements feature relatively higher statistical
uncertainties and only a factor of $\sim2$ increase in error of the
HOD parameters describing the satellite population is observed
compared to statistical errors. By probing such a large volume, nearly
0.1~Gpc$^3$, this study brings unprecedented constraints on the
$\Mstar - \Mh$ relationship from statistical methods in the cluster
mass regime at those redshifts.  As shown in
Fig.~\ref{fig:shmr_fig_Mstar_Mh}, our results make the link between
statistical methods based on HOD applied to deep, small-volume
surveys, with direct measurements of massive clusters from large-scale
surveys.

For central galaxies, we have shown that when properly accounting for
halo mass definition, choice of the IMF and the scatter between
$\Mstar$ and $\Mh$, there is a general agreement among results from
the literature. We find that stellar mass estimates are the main
source of uncertainty, as reflected by the light-blue shaded area from
\cite{Behroozi:2013aa} in Fig.~\ref{fig:shmr_fig_Mstar_Mh}, or the
stellar mass shift measured between \cite{Leauthaud:2012aa} and
\cite{Ilbert:2010ee}.  We stress, however, that if stellar mass
differences may induce a global shift (for instance caused by a
separate choice for the IMF), it may also translate into a
mass-dependent shift in the more general case (e.g. between two sets
of SPS models): hence applying a constant shift may not necessarily
reconcile two measurements.

In Fig.~\ref{fig:shmr_fig_Mh_Mstar}, stellar mass systematics do not
seem to explain all of the observed differences with some results from
the literature for which the stellar mass was measured in a similar
way to this study. 
\rs{To measure the impact of some of the assumptions made in our model, we have
compiled a list of potential systematics propagated through the halo
mass and satellite normalisation best-fit parameters. We quote an estimate
of 50\% error in $M_1$ and 20\% error in $B_{\rm sat}$, respectively.}

For satellite galaxies, the combination of lensing and clustering in
this work represents a significant improvement over studies using only
the stellar mass function. In Fig.~\ref{fig:tshmr_fig} we have shown
the measured total SHMR as function of halo mass, compared with a
number of results from observations and simulations in the
literature. Starting from group-size halos up to the most massive
clusters, we find that the total SHMR is gradually dominated by the
contribution from satellites.

Clearly, most SAMs tend to underestimate the total amount of stellar
mass produced in medium- to high-mass satellites ($ 10^{10} <
\Mstar/\Msun < 10^{11}$) at $z\sim1$ compared to observations.  This
would suggest that, in SAMs, the bulk of star formation occurs in
low-mass galaxies, but is quenched or suppressed at higher
mass. Possible explanations for this include either a too strong
quenching of halos in the mass regime $ 10^{10} < \Mstar/\Msun <
10^{11}$ \citep[e.g. the work by][who argue that the gas could be
later reincorporated into the halos]{Henriques:2012gs}, or that
low-mass sub-halos are too numerous and would ``catch'' the gas in
detriment of high-mass sub-halos. It is interesting to link this
feature to the overabundance of low-mass galaxies found in numerical
simulations compared to observations \citep[see
e.g.][]{Guo:2011eb,Weinmann:2012aa,De-Lucia:2014aa}.  In this context,
\cite{Schive:2014bm} recently proposed that cold dark matter could
behave as a coherent wave and have shown using N-body simulations that
this would suppress a large amount of small-mass halos.

Finally, we can summarise our findings as follows:
\begin{itemize}
\item the HOD model accurately reproduces the four observables within
  the statistical error bars in all mass bins over three orders of
  magnitudes in halo mass and two orders of magnitudes in stellar
  mass;
\item our $\Mstar - \Mh$ relationship shows generally good agreement
  with the literature measurements at $z\sim0.8$ and we have shown
  that, when modelling differences are properly accounted for, we are
  able to make a fair comparison of a number of results derived using
  independent techniques;
\item the systematic errors affecting our measurements were propagated
  through the whole fitting process. For the parameters describing the
  $\Mstar - \Mh$ relationship, we find that including systematic
  errors leads to a factor of 8 increase in error bars, and for the
  parameters describing the satellite HOD a factor of 2 increase in
  error bars, compared to statistical error bars;
\item \rs{the sum of systematic errors from the halo model and
    our model assumptions may be as high (but likely overestimated) as 50\%
    in halo mass and 20\% in the satellite number normalisation;}
\item the central galaxy stellar-to-halo mass ratio peaks at $M_{\rm
    h} = 1.9\times10^{12}\Msun$, a value slightly larger than the
  clustering results from the full CFHTLS from \cite{Coupon:2012aa},
\item the total (central plus satellites) SHMR is dominated by the
  satellite contribution in the most massive halos, in apparent
  contradiction with SAMs in the Millennium simulation.
\end{itemize}

We have demonstrated the power of associating a large and deep area
with a combination of independent observables to constrain the
galaxy-halo relationship with unprecedented accuracy up to $z=1$. The
potential of these data will undoubtedly allow us to extend this
analysis to galaxies split by type in a future work.

Additionally, studying the evolution in redshift of the SHMR above
$z=1$ is one of the greatest challenge in the near future. If
abundance matching already probes the central galaxy-halo relationship
up to high redshift, clustering and lensing are necessary to put
constraints on the satellite HOD and break some of the
degeneracies. Large-scale clustering measurements require wide-field
imaging, whereas high-redshift lensing techniques are yet to be
improved, but on-going projects such as HSC, DES or COSMOS/SPLASH,
which will increase by orders of magnitude the currently available
data, represent the ideal data sets to address those issues.

\section*{Acknowledgments}
\label{sec:acknowledgments}

This work is primarily based on observations obtained with WIRCam, a
joint project of CFHT, Taiwan, Korea, Canada, France, at the
Canada-France-Hawaii Telescope (CFHT) which is operated by the
National Research Council (NRC) of Canada, the Institut National des
Sciences de l'Univers of the Centre National de la Recherche
Scientifique of France, and the University of Hawaii.  The WIRCAM
images have been collected during several semesters and from different
programs. We thank G. Morrison, J. Willis and K. Thanjavur for leading
some of these programmes as PIs, and a special thanks to the canadian
agency who always highly ranked these proposals. We thank the Terapix
team for the reduction of all the WIRCAM images and the preparation of
the catalogues matching the T0007 CFHTLS data release.  The CFHTLenS
project is based on observations obtained with MegaPrime/MegaCam, a
joint project of CFHT and CEA/IRFU, at the Canada-France-Hawaii
Telescope (CFHT). This research used the facilities of the Canadian
Astronomy Data Centre operated by the National Research Council of
Canada with the support of the Canadian Space Agency.  We thank the
CFHT staff for successfully conducting the CFHTLS observations and in
particular J.-C. Cuillandre and E. Magnier for the continuous
improvement of the instrument calibration and the Elixir detrended
data that we used. We also thank TERAPIX for the quality assessment
and validation of individual exposures during the CFHTLS data
acquisition period, and E. Bertin for developing some of the software
used in this study. CFHTLenS data processing was made possible thanks
to significant computing support from the NSERC Research Tools and
Instruments grant program, and to HPC specialist O. Toader.  Part of
the numerical computations were done on the Sciama High Performance
Compute (HPC) cluster which is supported by the ICG, SEPNet and the
University of Portsmouth.

We thank M. Cousin and S. de la Torre for useful discussions related
to this work. We thank S. Birrer and D. Wake for providing us with
their measurements. JC acknowledges the support from the laboratoire
de Marseille during his stay in September 2014.  The early stages of
the CFHTLenS project was made possible thanks to the support of the
European Commission's Marie Curie Research Training Network DUEL
(MRTN-CT-2006-036133) which directly supported six members of the
CFHTLenS team (LF, HH, PS, BR, CB, MV) between 2007 and 2011 in
addition to providing travel support and expenses for team
meetings. BR +CH and HHo+ acknowledge support from the European
Research Council under EC FP7 grant numbers 240672 (BR), 240185 (CH)
and 279396 (HHo). H.Hi. is supported by the DFG Emmy Noether grant Hi
1495/2-1. AF acknowledges support by INAF through VIPERS grants PRIN
2008 and PRIN 2010. MJH acknowledges support from the Natural Sciences
and Engineering Research Council of Canada (NSERC).

{\small Author Contributions: All authors contributed to the
  development and writing of this paper.  The authorship list reflects
  the lead authors of this paper (JC, SA, LVW, TM, OI, EVU) followed
  by two alphabetical groups.  The first alphabetical group includes
  key contributors to the science analysis and interpretation in this
  paper, the founding core team and those whose long-term significant
  effort produced the final CFHTLenS data product, and the VIPERS
  collaboration leaders. The second group covers authors in
  alphabetical order who made a significant contribution to either the
  projects, this paper, or both.  The CFHTLenS collaboration was
  co-led by CH and LVW.}

\appendix

\section{Completeness of the samples}
\label{sec:sample_completeness}

\begin{figure}
  \centering
  \includegraphics[width=0.49\textwidth]{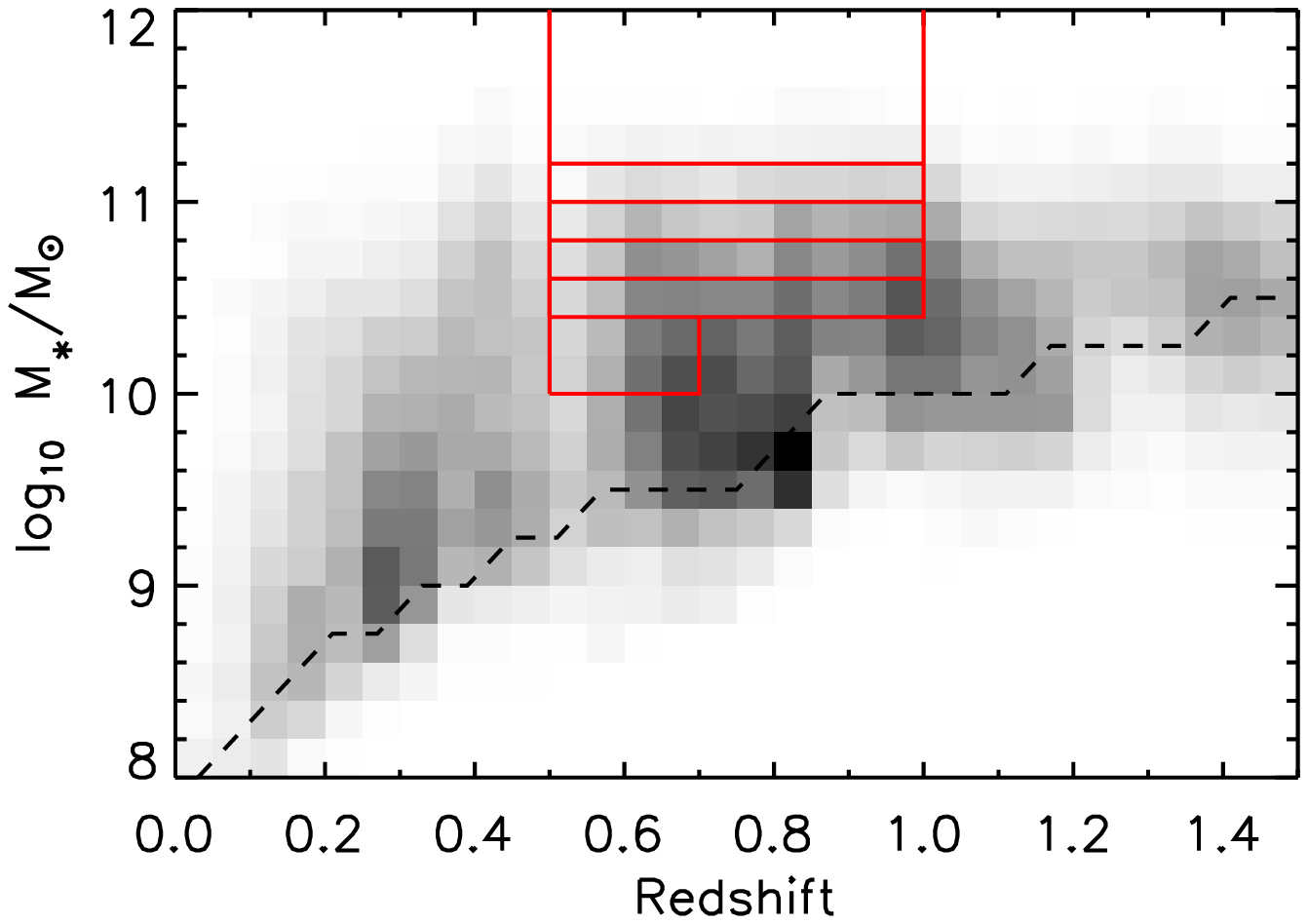}
  \includegraphics[width=0.49\textwidth]{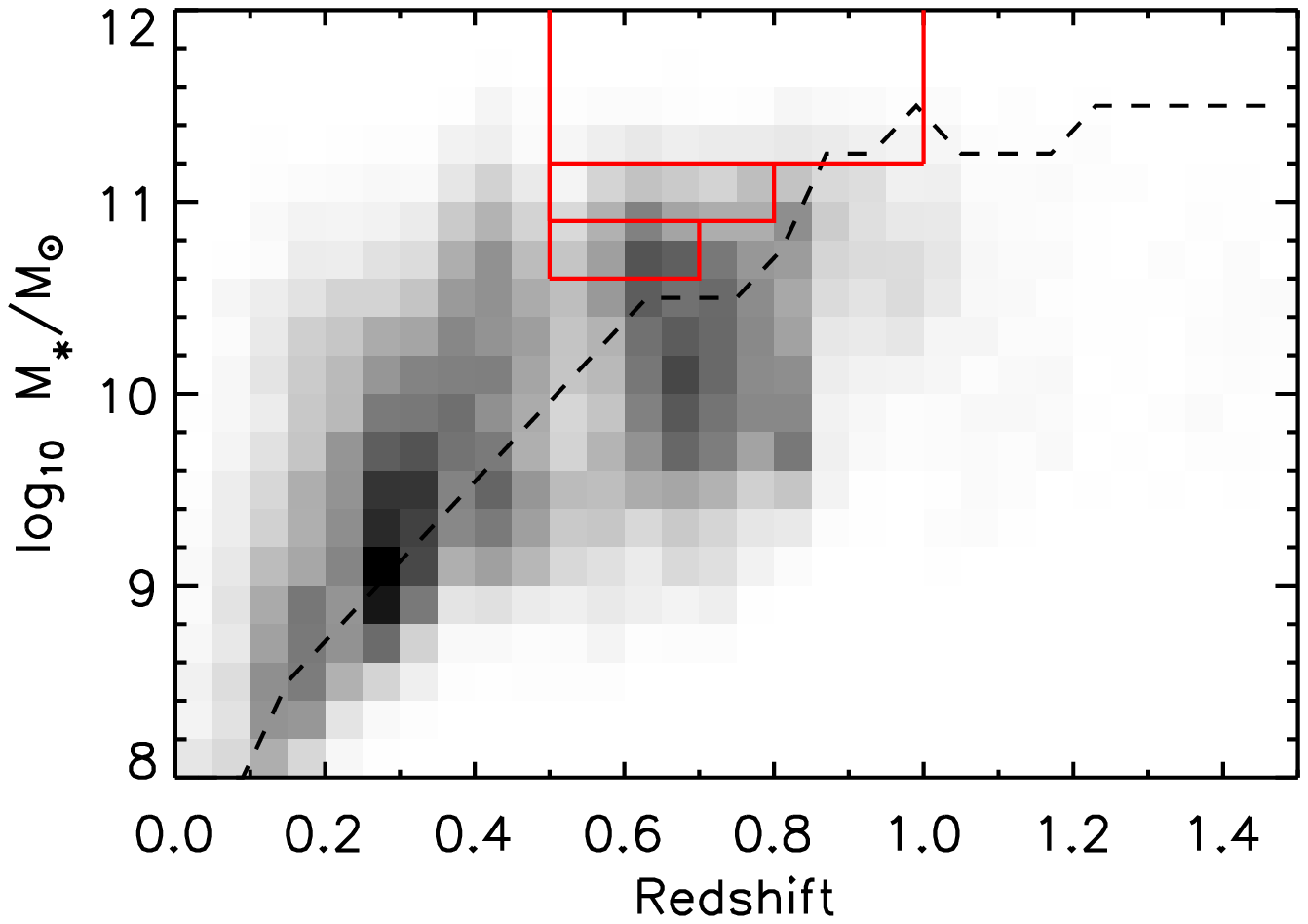}
  \caption{Galaxy distribution as function of stellar mass and
    redshift in WIRDS. Stellar mass 90\% completeness limits of
    $\Ks\ab < 22$ (top) and $i\ab < 22.5$ (bottom) selected samples
    are represented as the dashed black line and the sample selection
    as the thick red line.}
  \label{fig:Mstar_completeness_K22}
\end{figure}

In this section we use the CFHTLS-Deep/WIRDS combined data to test our
samples' mass completeness. The CFHTLS-Deep/WIRDS data are over 2
magnitudes deeper in all bands compared to our CFHTLS-Wide/WIRCam data
and with accurate photometric redshift and stellar mass estimates
computed in a similar fashion to this study.
Fig.~\ref{fig:Mstar_completeness_K22} shows the galaxy distribution in
WIRDS as function of stellar mass and redshift corresponding to our
selection $\Ks\ab < 22$ for the photometric sample (top) and $i\ab <
22.5$ for the spectroscopic sample (bottom).

The density fluctuations seen as function of redshift are due to
cosmic variance (the field of view is smaller than 1~deg$^2$), but we
do not expect any significant impact on our completeness
assessments. In both panels we represent the 90\% completeness limits
as dashed lines, and our samples' selection as red boxes. In the case
of the photometric sample, a conservative $z<0.7$ cut is adopted in
the lower mass sample to prevent missing red galaxies caused by the
optical incompleteness at the CFHTLS-Wide depth.
%
Overall, these verifications show that all of our samples are complete
in mass.

%



\section{Details on the derivation of the observables}
\label{sec:details-observables}

Here we provide detailed calculations of the four observables used in
this study and derived from the HOD model described in
Section~\ref{sec:model}. For the dark matter halo profile and the
distribution of satellites, we assume a \cite{Navarro:1997aa} (NFW)
profile with the theoretical mass-concentration relation from Eq.~(16) of
\cite{Takada:2003aa} with $c_0 = 11$ and $\beta=0.13$\rs{, featuring
the redshift dependence $(1+z)^{-1}$ \citep{Bullock:2001aa}. All
dark matter quantities are derived at the mean redshift of the galaxy sample,
computed from the expectation value of the sum of redshift PDFs. All
quantities are computed in comoving units (``co''). The clustering and
galaxy-galaxy lensing are then converted into physical units (``phys'') to match
the measurements}.

\subsection{Stellar mass function}

The stellar mass function is the integrated HOD over the halo mass
function:
\begin{eqnarray}
  & & \phi_{\rm SMF}(\Mstart{t_1},\Mstart{t_2})\\
  &=&\int_0^{\infty} \langle N_{\rm tot}(\Mh|\Mstar^{t_1},\Mstar^{t2})\rangle
  \frac{{\rm{d}}n}{{\rm{d}}\Mh}{\rm{d}}\Mh \, . \nonumber
\end{eqnarray}

\subsection{Galaxy clustering}

We describe galaxy clustering using the two-point correlation
function, as the sum of the one-halo and two-halo terms:
\begin{equation}
  \xi_{gg}(r_{\rm co}) = 1+\xi_{gg,1}(r_{\rm co}) + \xi_{gg,2}(r_{\rm co}) \, .
\end{equation}
The one-halo term, $\xi_{gg,1}(r_{\rm co})$, expresses the relative
contribution of galaxy pairs within the halo $\langle N_{\rm tot}(\Mh)
(N_{\rm tot} (\Mh)-1) \rangle/2$ and can be decomposed, assuming
Poisson statistics for the satellites, into two terms:
\begin{eqnarray}
  \langle N_{\rm cen}N_{\rm sat} \rangle(\Mh) & = & \langle
  N_{\rm cen}(\Mh) \rangle \langle N_{\rm sat}(\Mh)
  \rangle\, ;
  \nonumber \\
  \langle N_{\rm sat}(N_{\rm sat} - 1) \rangle(\Mh)/2 & = & \langle
  N_{\rm sat}(\Mh) \rangle^2/2  \, .
\end{eqnarray}

The correlation function for central-satellite pairs is given by
\begin{eqnarray}
  & &  1+\xi_{\rm cs}(r_{\rm co}, z) = \nonumber\\
  & & \int_{M_{\rm vir}(r)}^{\infty} \ud \Mh \,
  n(\Mh, z)
  \frac{\langle
    N_{\rm cen} \rangle \langle N_{\rm sat}
    \rangle}{n_{\rm gal}^2/2} \rho_{\rm h}(r_{\rm co}|\Mh) \, ,
\end{eqnarray}
where we assume that the distribution of central-satellite pairs
simply follows that of the dark matter halo profile.  The lower
integration limit $M_{\rm vir}(r_{\rm co})$ accounts for the fact that no halo
with a virial mass corresponding to $r_{\rm co}$ would contribute to the
correlation function.

For the satellite contribution $\xi_{\rm ss}$, the distribution of
satellite pairs is the convolution of the dark matter halo profile
with itself, computed here in Fourier space. The satellite power
spectrum is
\begin{equation}
  P_{\rm ss} ({k}) = \int_{M_{\rm low}}^{M_{\rm high}} \ud \Mh \,  n(\Mh)
  \frac{\langle N_{\rm sat}(\Mh) \rangle^2}{n_{\rm gal}^2}  |u_h({k}|\Mh)|^2\, , 
\end{equation} 
where $u_{\rm h}({k}|\Mh)$ is the Fourier transform of the dark-matter
halo profile $\rho_{\rm h}(r_{\rm co}|\Mh)$. The correlation function $\xi_{\rm
  ss}$ is then obtained via a Fourier transform.

The one-halo correlation function is the sum of the two contributions,
\begin{equation}
  \xi_{gg,1} (r_{\rm co}) = 1+\xi_{\rm cs} (r_{\rm co}) + \xi_{\rm ss} (r_{\rm co}) \, .  
\end{equation}

The two-halo term is computed from the galaxy power spectrum:
\begin{eqnarray}
  \label{eq:p2h}
  && P_2(k,r_{\rm co}) =  P_{\rm m}(k)\times \\
&&  \left [ \int_{M_{\rm low}}^{M_{\rm lim}(r_{\rm co})} \ud \Mh
    n(\Mh) \frac{\langle N_{\rm tot} \rangle}{n'_{\rm gal}(r_{\rm
        co})}b_{\rm h}(\Mh,r_{\rm co}) |u_{\rm h}(k|\Mh)| \right ]^2
  \, ,\nonumber
\end{eqnarray}
where
\begin{equation}
  n'_{\rm gal}(r_{\rm co}) = \int_{M_{\rm low}}^{M_{\rm lim}(r_{\rm co})} n(\Mh) \langle N_{\rm tot}\rangle \, \ud \Mh \, .
\end{equation}
The upper integration limit $M_{\rm lim}(r_{\rm co})$ accounts for halo
exclusion as detailed in \cite{Coupon:2012aa}, and references therein.

Finally, the two-halo term $\xi_{\rm gg,2}$ of the galaxy
auto-correlation function is the Fourier transform of
Eq.~(\ref{eq:p2h}) renormalised to the total number of galaxy pairs:
\begin{equation}
  1+\xi_{\rm gg, 2} (r_{\rm co}) = \left [ \frac{n'_{gal}(r_{\rm co})}{n_{\rm gal}}  \right ]
  [1+\xi_{\rm gg, 2} (r_{\rm co})] \, .
\end{equation}

The projected clustering $w(\theta)$ is derived from the projection of
$\xi_{\rm gg}$ onto the estimated redshift distribution from the sum
of PDFs, assuming the Limber approximation \cite[see details
in][]{Coupon:2012aa}.

The real-space clustering $w_p(r_{\rm p, co})$ is derived from the projection of
the 3D correlation function along the line of sight:
\begin{equation}
  w_p (r_{\rm p, co}) = 2 \, \int_{r_{\rm p, co}}^{\infty} \, r_{\rm co} \,  \ud r_{\rm co} \xi_{\rm gg}(r_{\rm co}) \, (r_{\rm co}^2 -
  r_{\rm p, co}^2)^{-1/2} \, ,
\end{equation}
\rs{converted into physical units as:
\begin{equation}
w_{\rm p, phys}  = w_{\rm p, co} / (1+z) \, .
\end{equation}
}

\subsection{Galaxy-galaxy lensing}

The galaxy-galaxy lensing estimator measures the excess surface
density of the projected dark matter halo profile:
\begin{equation}
  \Delta \Sigma_{\rm co}(r_{\rm p, co})=\overline{\Sigma}_{\rm co}(<
  r_{\rm p, co})-\overline{\Sigma}_{\rm co}(r_{\rm p, co}) \, ,
\end{equation}
where $\overline{\Sigma}_{\rm co}(< r_{\rm p, co})$ is the projected mean surface density
within the comoving radius $r_{\rm p,co}$ and $\overline{\Sigma}_{\rm co}(r_{\rm p,co})$ the mean surface
density at the radius $r_{\rm p,co}$.

To compute the analytical projected dark matter density $\Sigma$, we
write
\begin{eqnarray}
  \Sigma_{\rm co} (r_{\rm p,co}) 
  &=& \int \rho\left(\sqrt{r_{\rm p,co}^2+\pi_{\rm co}^2}\right) \ud \pi_{\rm co} \nonumber\\
  &=& \overline\rho \int \left[1+\xi_{\rm
      gm}\left(\sqrt{r_{\rm p,co}^2+\pi_{\rm co}^2}\right)\right] \ud \pi_{\rm co}, 
\end{eqnarray}
where $r_{\rm p,co}$ is the transverse comoving distance, $\pi_{\rm co}$ the
line-of-sight comoving distance, $\overline\rho $ the mean density of
the Universe, so that $\Delta \Sigma_{\rm co}(r_{\rm p,co})$ is related to the galaxy-dark
matter cross-correlation function $\xi_{\rm gm}$ through
\begin{eqnarray}
  &&  \Delta \Sigma_{\rm co}(r_{\rm p,co}) = \overline{\Sigma}_{\rm co}(< r_{\rm p,co})-\overline{\Sigma}_{\rm co}(r_{\rm p,co}) = \\
  && \overline\rho \left[ \frac{4}{r_{\rm p,co}^2}\int_0^{r_{\rm p,co}} \int_0^{\pi_{\rm max}}  r_{\rm p,co}' \xi_{\rm gm} \left(\sqrt{r_{\rm
          p,co}'^2+\pi_{\rm co}^2}\right) \ud \pi_{\rm co} \ud r_{\rm p,co}' \right. \nonumber\\
  &&  \left. - 2\int_0^{\pi_{\rm max}}  \xi_{\rm gm} \left(\sqrt{r_{\rm
          p, co}^2+\pi_{\rm co}^2}\right) \ud \pi_{\rm co} \right ] \, . \nonumber
\end{eqnarray}
The integration along the line of sight is performed up to the scale
$\pi_{\rm max}=80$~Mpc.

\rs{The excess surface density in physical units writes:
\begin{equation}
\overline{\Delta \Sigma}_{\rm phys} = \overline{\Delta \Sigma}_{\rm co} \times
(1+z_{\rm L})^2 \, ,
\end{equation}
where $z_{\rm L}$ is the redshift of the lens galaxy.}

As for $\xi_{\rm gg}$, $\xi_{\rm gm}$ can be written as the sum of the
one- and two-halo terms:
\begin{equation}
  \xi_{\rm gm} (r) = 1+\xi_{\rm gm, 1} (r) + \xi_{\rm gm, 2} (r) \, .  
\end{equation}
$\xi_{\rm gm, 1} (r)$ is itself decomposed into a contribution from
the cross correlation of the central galaxy-dark matter and from that
of the satellite-dark matter, both assuming a NFW profile.  We write
the former as
\begin{eqnarray}
  & &  1+\xi_{\rm gm, cen}(r, z) = \nonumber\\
  & & \int_{M_{\rm vir}(r)}^{M_{\rm high}} \ud \Mh \,
  n(\Mh, z)
  \frac{\langle
    N_{\rm cen} \rangle}{n_{\rm gal}} \rho_{\rm
    h}(r|\Mh) \frac{\Mh}{\overline\rho} 
\end{eqnarray}
and the latter $\xi_{\rm gm, sat}$ from the Fourier transform of its
power spectrum
\begin{eqnarray}
  && P_{\rm gm, ss} (k) = \nonumber\\
  && \int_{M_{\rm low}}^{M_{\rm high}} \ud \Mh \,  n(\Mh)
  \frac{\langle N_{\rm sat}(\Mh) \rangle}{n_{\rm gal}}
  \frac{\Mh}{\overline\rho}  |u_h(k|\Mh)|^2\, .
\end{eqnarray}

Finally we compute the two-halo term $\xi_{\rm gm, 2}(r)$ from the
Fourier transform of the galaxy-dark matter cross correlation power
spectrum:
\begin{eqnarray}
  \label{eq:2h}
  & & P_{\rm gm, 2}(k,r) =  P_{\rm m}(k)\\
  & &\times  \int_{M_{\rm low}}^{M_{\rm lim}(r)} \ud \Mh
  n(\Mh) \frac{\langle N_{\rm tot}
    (\Mh) \rangle}{n'_{\rm gal}(r)} b_{\rm h}(\Mh,r) |u_{\rm
    h}(k|\Mh)| \, ,\nonumber 
\end{eqnarray}
with a similar treatment of halo exclusion to that of the galaxy power
spectrum.

\section{Systematics checks on lensing and clustering}
\label{sec:details-measurements}

\begin{figure}
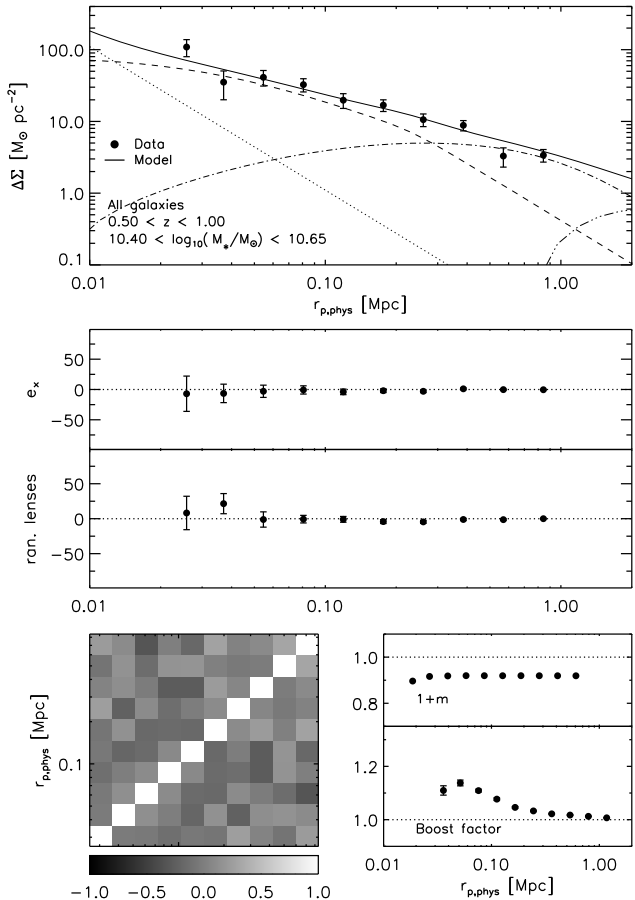

  \centering
  \includegraphics[width=0.49\textwidth]{{{K22_0.50_1.00_Mstar_10.40_10.65_all.ggl}}}
  \caption{Galaxy-galaxy lensing measurements and systematics checks
    for the sample $10.40 < \log (\Mstar/\Msun) < 10.65$. In the top
    panel we show the data (dots with error bars) and the model (thick
    line) split into the stellar term in dotted line, the central term
    in dashed line, the satellite term in dot-dashed line and the
    2-halo term in triple dot-dashed line. The lower panels show the
    systematic tests (rotated-shape signal and random lens positions),
    calibration factor (multiplicative bias correction and boost
    factor), and the lower left corner the correlation coefficients of
    the correlation matrix from the jackknife estimate.}
  \label{fig:ggl_details}
\end{figure}

We have performed systematics checks for the lensing and clustering
measurements. In Fig.~\ref{fig:ggl_details} we detail the
galaxy-galaxy lensing measurement for the sample $10.40 < \log
(\Mstar/\Msun) < 10.65$ as an example. The top panel shows the data
(dots with error bars) and best-fit model (thick line) with the
different components of the model the central galaxy term, the
satellite term, and the 2-halo term. The lower panels show a number of
systematics checks. The ``$e_{\times}$'' panel shows the signal
measured after rotating the ellipticities by 45$^\circ$ and the
``ran. lenses'' panel shows the signal measured by randomising the
lenses positions, both consistent with zero. The ``1+m'' panel shows
the multiplicative bias correction applied to the galaxy-galaxy
lensing measurement, estimated after replacing the ellipticities by
the multiplicative calibration factor $1+m$. The ``boost factor'' was
estimated from randomising the background source positions and
measuring the ratio of the number of real sources over random objects
as a function of distance from the lenses, and applied to the
galaxy-galaxy lensing measurement. The covariance matrix from the
jackknife estimate is shown in the left bottom corner of the
figure. The relatively small off-diagonal values show the low
correlation between data points.  We repeated identical tests for all
mass bins. In all cases, systematics are found to be consistent with
zero.
%
%

\begin{figure}
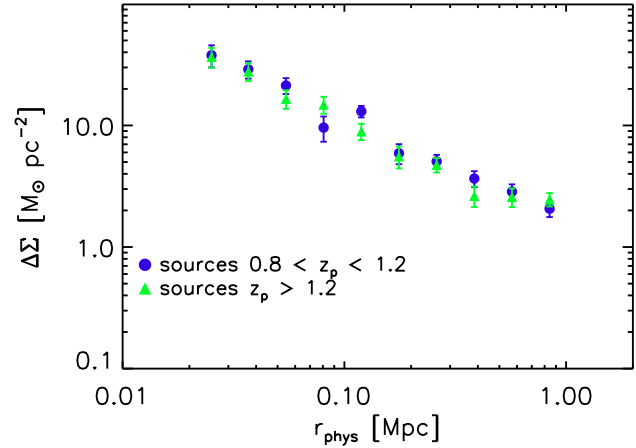

  \centering
  \includegraphics[width=0.49\textwidth]{{{high_z_sources}}}
  \caption{Galaxy-galaxy lensing measurements separating the
    background sample into $0.8 <z_{\rm p}<1.2$ sources (purple dots) and
    $z_{\rm p}>1.2$ sources (green triangles), keeping the same lens galaxy
    foreground sample (low-redshift galaxies with spectroscopic
    redshifts).}
  \label{fig:ggl_photo_z}
\end{figure}

In Fig.~\ref{fig:ggl_photo_z} we test the impact of including high
redshift sources beyond $z>1.2$. To do so, we select an arbitrary
sample of low redshift lens galaxies with a spectroscopic redshift and
we measured the galaxy-galaxy lensing signal using all sources with
$0.8 < z_{\rm p}<1.2$ (purple dots in the figure) and all sources with
$z_{\rm p}>1.2$ (green triangles in the figure). We see no significant
difference between the two signals, meaning that the photometric
redshifts and shape measurements in our catalogue are robust enough
beyond $z_{\rm p}>1.2$.
%

In Fig.~\ref{fig:wtheta_details} we show the projected clustering in
the mass bin $10.60 < \log (\Mstar/\Msun) < 10.80$. The top panel
shows the data points with error bars and the best-fit model, with the
different components of the model: the 1-halo term split into the
central-satellite and satellite-satellite terms, and the 2-halo term.
In the middle panel we show the corresponding HOD, as a dashed line
for the central contribution and as a dot-dashed line for the
satellites' contribution.
%
\begin{figure}
  \centering
  \includegraphics[width=0.49\textwidth]{{{K22_0.50_1.00_Mstar_10.60_10.80_all.wtheta}}}
  \caption{$w(\theta)$ measurements and the corresponding HOD function
    for the sample $10.60 < \log (\Mstar/\Msun) < 10.80$. In the top
    panel, we show the data points with error bars and the best-fit
    model: the dotted line represents the central-satellite
    cross-correlation, the dashed line the satellite-satellite
    auto-correlation, and the dot-dashed line the central-central
    auto-correlation (or 2-halo term). The middle panel displays the
    corresponding HOD, the dashed line shows the central galaxy HOD
    and the dot-dashed line the satellites' HOD. The lower right panel
    shows the corresponding redshift distribution constructed from the
    sum of individual PDFs. The lower left panel shows the correlation
    coefficients of the covariance matrix from the jackknife
    estimate.}
  \label{fig:wtheta_details}
\end{figure}

\section{2D contours}
\label{sec:contours}

We show in Fig.~\ref{fig:contours} the likelihood distributions of the
best-fit HOD parameters. Here the results are shown for the MCMC run
done with total (statistical plus systematic) errors.
%
\begin{figure*}
  \centering
  \includegraphics[width=\textwidth]{{{all_contour2d}}}
  \caption{1D (diagonal) and 2D likelihood distributions of best-fit
    HOD parameters in the case of total errors. The 2D contours
    represent the 68.3\%, 95.5\% and 99.7\% confidence limits. We used
    flat priors within the ranges shown on the figure for all parameters.}
  \label{fig:contours}
\end{figure*}

\bibliographystyle{mn2e_long}
\bibliography{references}

{\noindent\small\it\ignorespaces $^1$ Astronomical Observatory of the  University of Geneva, ch. d'Ecogia  16, 1290 Versoix, Switzerland\\
  $^2$ LAM, Universit\'e d’Aix-Marseille \& CNRS, UMR7326, 38 rue  F. Joliot-Curie, 13388  Marseille Cedex 13, France\\
  $^3$ Department of Physics and Astronomy, University of British  Columbia, 6224 Agricultural Road, Vancouver, V6T 1Z1, BC, Canada\\
  $^4$ Argelander Institute for Astronomy, University of Bonn, Auf dem H{\"u}gel 71, 53121 Bonn, Germany\\
  $^5$ INAF – Istituto di Astrofisica Spaziale e Fisica Cosmica  Milano,  via Bassini 15, 20133 Milano, Italy\\
  $^6$ INAF – Osservatorio Astronomico di Brera, via Brera 28, 20122  Milano, via E. Bianchi 46, 23807 Merate, Italy\\
  $^7$ 3 Dipartimento di Fisica, Universit\`a di Milano-Bicocca, P.zza  della  Scienza 3, 20126 Milano, Italy\\
  $^8$ The Scottish Universities Physics Alliance, Institute for Astronomy, University of Edinburgh, Blackford Hill, Edinburgh EH9 3HJ, UK\\
  $^9$ Leiden Observatory, Leiden University, Niels Bohrweg 2, 2333 CA  Leiden, The Netherlands\\
  $^{10}$ CEA Saclay, Service d'Astrophysique (SAp), Orme des  Merisiers,  B\^at 709, F-91191 Gif-sur-Yvette, France\\
  $^{11}$ Mullard Space Science Laboratory, University College London,  Holmbury St Mary, Dorking, Surrey RH5 6NT, UK\\
  $^{12}$ Institut d'Astrophysique de Paris, Universit\'e Pierre et  Marie  Curie - Paris 6, 98 bis Boulevard Arago, F-75014 Paris, France\\
  $^{13}$ University of Oxford, Department of Physics, Denys Wilkinson Building, Keble Road, Oxford OX1 3RH, UK\\
  $^{14}$ IFAE, Campus UAB, E-08193 Bellaterra, Spain\\
  $^{15}$ Dipartimento di Matematica e Fisica, Universit\`a degli Studi  Roma  Tre, via della Vasca Navale 84, 00146 Roma, Italy\\
  $^{16}$ INFN, Sezione di Roma Tre, via della Vasca Navale 84, 00146  Roma, Italy\\
  $^{17}$ INAF – Osservatorio Astronomico di Roma, via Frascati 33,  00040  Monte Porzio Catone (RM), Italy\\
  $^{18}$ Dipartimento di Fisica e Astronomia - Universit\`a di Bologna,  viale  Berti Pichat 6/2, 40127 Bologna, Italy\\
  $^{19}$ INAF – Osservatorio Astronomico di Bologna, via Ranzani 1,  40127  Bologna, Italy\\
  $^{20}$ INAF – Osservatorio Astronomico di Trieste, via  G. B. Tiepolo 11,  34143 Trieste, Italy\\
  $^{21}$ Shanghai Key Lab for Astrophysics, Shanghai Normal  University,  100 Guilin Road, 200234, Shanghai, China\\
  $^{22}$ Dept. of Physics and Astronomy, University of Waterloo,  Waterloo, ON, N2L 3G1, Canada\\
  $^{23}$ Perimeter Institute for Theoretical Physics, 31 Caroline  Street N, Waterloo, ON, N2L 1Y5, Canada\\
  $^{23}$ INFN, Sezione di Bologna, viale Berti Pichat 6/2, 40127  Bologna,  Italy\\
  $^{24}$ Kavli Institute for the Physics and Mathematics of the Universe, Todai Institutes for Advanced Study, the University of Tokyo, Kashiwa, Japan 277-8583\\
  $^{25}$ Department of Physics and Astronomy, University College London, Gower Street, London WC1E 6BT, U.K\\


  \label{lastpage}

\end{document}